\documentclass{article}
\usepackage{jheppub}
\usepackage{amsmath,amsfonts,amsthm,graphicx}
\usepackage{mathrsfs}

\newcounter{savecounter}

\numberwithin{equation}{section}

\def\cz{{\,\,\bf\hat {\rm *}\,}}
\def\cZ{{\,\,\bf\check {\rm * \rule{0pt}{1ex}}\,}}
\newcommand\secref[1]{section~\ref{#1}}

\newcommand\appref[1]{appendix~\ref{#1}}
\newcommand\Appref[1]{Appendix~\ref{#1}}
\newcommand\figref[1]{figure~\ref{#1}}

\newcommand\IM{{\rm Im}\,}
\newcommand\XmY{S}
\newcommand\XpY{P}
\newcommand\es{\emptyset}
\newcommand\RE{{\rm Re}\,}

\newcommand{\bal}{\begin{align}}
\newcommand{\eal}{\end{align}}
\newcommand{\beq}{\begin{equation}}
\newcommand{\eeq}{\end{equation}}
\newcommand\beqa{\begin{eqnarray}}
\newcommand\eeqa{\end{eqnarray}}
\newcommand\bea{\begin{array}}
\newcommand\eea{\end{array}}

\newcommand{\eq}[1]{(\ref{#1})}

\newcommand{\bB}{{\bf B}}
\newcommand{\bC}{{\bf C}}

\newcommand{\su}{\mathfrak{su}}
\renewcommand{\sl}{\mathfrak{sl}}
\newcommand{\psu}{\mathfrak{psu}}

\newcommand{\Qs}{{\mathsf Q}}

    \newcommand{\neqa}{\nonumber\end{eqnarray}}
    \newcommand{\la}[1]{\label{#1}}

\def\o{{        \omega}}
\def\a{{\alpha}}

\def\[{\left[}
\def\]{\right]}
\def\L{\Lambda}
\def\l{\lambda}
\def\e{\epsilon}

\def\s{\sigma}
\def\a{\alpha}
\def\b{\beta}

\def\<{\langle}
\def\>{\rangle}

\def\sK{\,\slash\!\!\!\! {\cal K}}

\def\i2{\frac{i}{2}}

\def\gas{{\gamma_{\rm as}}}

\def\<{\langle}
\def\>{\rangle}

\def\cF{{\cal F}}

\def\CC{{\cal K}}
\def\cT{{\cal T}}

\def\i2{\frac{i}{2}}

\def\tr{{\rm tr}}
\def\Tr{{\rm Tr}}
\def\det{{\rm det}}

\def\1h{\hat 1}
\def\2h{{\hat 2}}
\def\3h{{\hat 3}}
\def\4h{{\hat 4}}

\def\be{\begin{eqnarray}}
\def\ee{\end{eqnarray}}

    \def\CA{{\cal A}}
    
    \def\CC{{\cal K}}

    \def\CF{{\cal F}}

    \def\CK{{\cal K}}
    
    \def\CM{{M}}
    
    \def\CO{{\cal O}}
    \def\CP{{\cal P}}
    
    \def\CR{{\cal R}}
    
    \def\CT{{\cal T}}

    \def\CZ{{\cal Z}}

    \def\<{\left\langle\,}
    \def\>{\, \right\rangle}
    \def\[{\left[}
    \def\]{\right]}

    \def\xp{ x^{+} }
    \def\xm{ x^{-} }

    \def\D{{\rm D}}
    \def\bT{{\bf T}}
    \def\disc{\,{\rm disc}\,}
    \def\sT{{\mathscr T}}
    \def\wT{{\mathbb{T}}}
    \def\DF{{\CF}}

\newcommand{\ps}{{\bf p}}
\newcommand{\qs}{{\bf q}}
   
   \def\su{{\mathfrak{su}}}
   \def\sl{{\mathfrak{sl}}}
   
   \def\groupn{\mathfrak{n}}

\def\hh{{{\hat h}}}

\def\bZ{{\resizebox{0.28cm}{0.33cm}{$\hspace{0.03cm}\check {\hspace{-0.03cm}\resizebox{0.14cm}{0.18cm}{$Z$}}$}}}
\def\hbZ{{\widehat{ Z}}}
\def\VP{{\tilde Q}}
\def\nU{{U}}
\def\hq{{\hat q}}
\def\hp{{\hat p}}
\def\rrho{{\rho}}
\def\irho{{\eta}}
\def\rhou{{\rho_{_U}}}

\def\Bup{{{\bf B}}}
\def\Bdown{{\overline{\bf B}}}

\renewcommand{\Re}{{\rm Re}\,}
\renewcommand{\Im}{{\rm Im}\,}

\usepackage{amssymb}
\usepackage{subfig}
\usepackage{overpic}

 \title{  Solving the AdS/CFT Y-system }

\author[a,b]{Nikolay Gromov}
\author[c,d]{~~~Vladimir Kazakov%
\note{member of Institut Universitaire de
    France}}
\author[c]{~~~Sebastien Leurent}
\author[e,f]{~~~Dmytro Volin}

\affiliation[a]{Mathematics Department, King's College London,
The Strand, London WC2R 2LS, UK.}
\affiliation[b]{St.Petersburg INP, Gatchina, 188 300, St.Petersburg,
  Russia.}
\affiliation[c]{LPT, École Normale Superieure, 24, rue Lhomond 75005
  Paris, France.}
\affiliation[d]{Universit\'e Paris-VI, Place Jussieu, 75005 Paris,
  France.}
\affiliation[e]{Department of Physics, The Pennsylvania State University,\\
University Park, PA 16802, USA}
\affiliation[f]{Bogolyubov Institute for Theoretical Physics,\\ 14b Metrolohichna str, Kyiv 03680, Ukraine}

\emailAdd{nikolay.gromov$\bullet$kcl.ac.uk}
\emailAdd{kazakov$\bullet$lpt.ens.fr}
\emailAdd{leurent$\bullet$lpt.ens.fr}
\emailAdd{dvolin$\bullet$psu.edu}

\abstract{Using  integrability and analyticity properties of the AdS\(_{5}\)/CFT\(_4\) Y-system
we reduce it to a finite set of nonlinear integral equations.
The \(\mathbb Z_4\) symmetry of the underlying coset sigma model, in its quantum version,
allows for  a deeper insight into the analyticity structure of the corresponding Y-functions and T-functions, as well as for their analyticity friendly parameterization in terms of Wronskian determinants of Q-functions.
As a check for the new equations, we reproduce the numerical results for the Konishi operator previously obtained  from the original infinite Y-system.
}

\subheader{LPTENS-11/24}

\begin{document}

\maketitle

\section{Introduction}
\label{sec:introduction-1}

It is well known that the spectra of integrable two dimensional QFT's in  finite volume can be studied by the \ thermodynamic Bethe ansatz (TBA) approach
 \cite{Zamolodchikov:1991et}. TBA  results in  a system of non-linear
 integral equations which can be always
``miraculously'' recast into the universal Y-system --- a system of  functional equations
on Y-functions,  looking simpler than TBA, but at the price of   loss of a certain analyticity input. Apart from some exceptional  cases (Lee-Yang model, SS-model, Wess-Zumino cosets)
both the TBA system and the Y-system are infinite. However, Destri and de Vega
noticed  \cite{Destri:1992qk} that the infinite TBA system for the
vacuum of the Sine-Gordon model can be  reduced  to only one
non-linear integral equation, and this was later generalized to
excited states \cite{Fioravanti:1996rz}.
For the models with  higher rank groups of symmetries
a similar reduction to a few integral equations is usually possible. DdV-type equations often make possible a systematic study of IR and UV limits.
In general this is
the best one can do with the spectral problem of a finite volume integrable QFT.
  This reduction, still not very well studied and
  understood,  can be  traced down to the fact
  \cite{Gromov:2008gj,Bazhanov:1994ft} that the Y-system itself is a
  gauge invariant version (see \eqref{eq:gaugeT}) of the integrable finite difference bilinear Hirota equation, or T-system:

\begin{equation}\label{eq:Hirota}
T_{a,s}(u+\tfrac{i}{2})T_{a,s}(u-\tfrac{i}{2})=
T_{a+1,s}(u)T_{a-1,s}(u)+T_{a,s+1}(u)T_{a,s-1}(u)\;.
\end{equation}
The equation is essentially the same for all known integrable 2D QFT's\footnote{with slight modifications for the algebras other than \(A_k\).
}, and
what  differs from one model to another is   its boundary conditions w.r.t. the ``representational'' variables \(a,s\)  and the analyticity properties w.r.t. the spectral parameter. In practice, the  Hirota equation is often a very good starting point for attacking the finite volume spectrum problem in any
integrable 2D QFT's (e.g.  for  an important case of the principal chiral field \cite{Gromov:2008gj,Kazakov:2010kf}):
even the  analyticity properties of T-functions appear to be considerably constrained by the structure of Hirota equation.

\begin{figure}[ht]
\begin{center}
\subfloat[lattice for T-functions' (a,s) indices]
{\label{fig:ThookT}
\rlap{\includegraphics[scale=0.7]{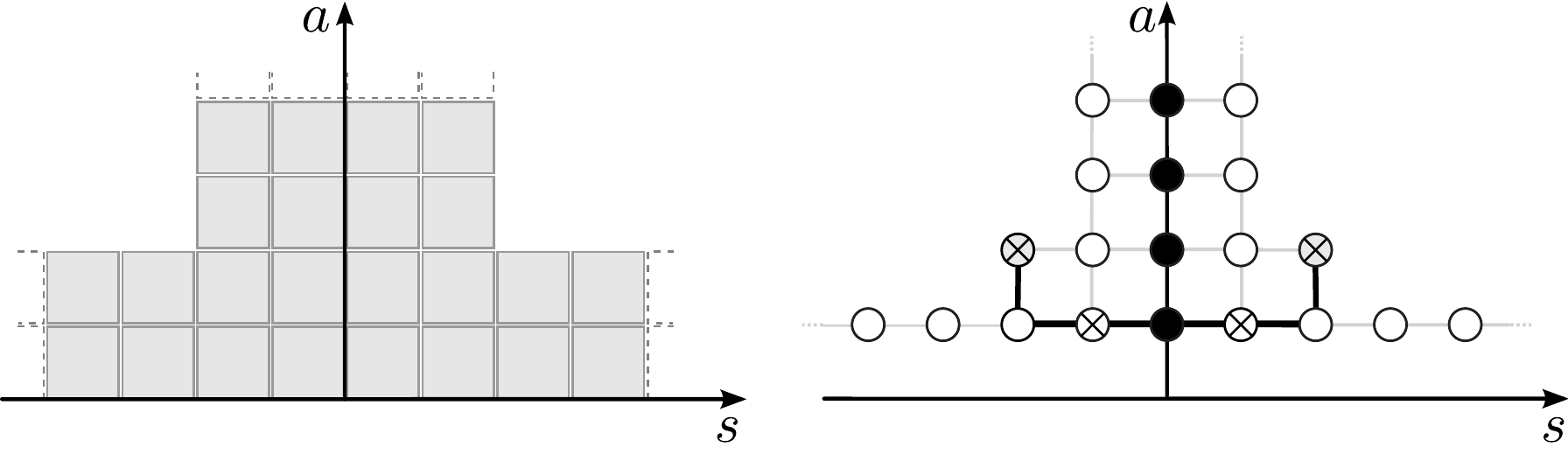}}
~\hspace{6cm}~} 
\subfloat[lattice for Y-functions' (a,s) indices]
{\label{fig:ThookY}
~\hspace{6cm}~}
\end{center}
\caption[Boundary conditions for Y-system (b) and T-system (a) of
  AdS\(_5\)/CFT\(_4\):]{\textbf{Boundary conditions for Y-system (b) and T-system (a) of
  AdS\(_5\)/CFT\(_4\):}   \(\mathbb{T}\)-shaped ``fat hook''
  (\(\mathbb{T}\)-hook). We will often distinguish in the \(\wT\)-hook
  (a)  the  {\it upper band} --  all the nodes  with \(a\ge|s| \), the
  {\it right band} --  all the nodes   with \(a\le s \) and the {\it
    left band} -- all the nodes  with \(a\le|s|, s<0\).
   \label{T-Hook}}
\end{figure}

All these observations should hold for the AdS/CFT Y-systems \cite{Gromov:2009tv}, and in particular for the  most
studied AdS\(_5\)/CFT\(_4\) case, solving the problem of planar spectrum of anomalous dimensions in N=4 SYM theory (see for the recent review \cite{Beisert:2010jr}, and in particular the chapters \cite{Gromov:2010kf,Bajnok:2010ke} therein). The AdS\(_5\)/CFT\(_4\)  Y- and T-systems with \(\mathbb{T}\)-hook boundary conditions
  proposed in \cite{Gromov:2009tv}  and summarized in \figref{T-Hook} were later shown to be equivalent,
 with  certain analyticity requirements \cite{Cavaglia:2010nm,Balog:2011cx,Balog:2011nm},  to the TBA equations \cite{Bombardelli:2009ns,Gromov:2009bc,Arutyunov:2009ur}. It was shown in  \cite{Gromov:2010vb,Gromov:2010km} (see also \cite{Bazhanov:2008yc,Tsuboi:2009ud}) that the T-system, and hence the Y-system, in \(\mathbb{T}\)-hook can be formally solved   in terms of Wronskian determinants of a finite number of  Q-functions
 --- a generalization of Baxter's Q-function. But the corresponding finite system of nonlinear integral equations (FiNLIE), a remote analogue of Destri-de Vega equations,     was still missing.
In this paper we propose such a FiNLIE for the planar  AdS/CFT spectrum.

The difficulty of finding FiNLIE --- possibly the ultimate step in
simplifying the spectral problem in AdS/CFT --- resides in its
formidably complicated analytic structure w.r.t. the spectral
parameter \(u\): the Y-functions live on a Riemann surface
with infinitely many sheets connected by an infinite system of
Zhukovsky cuts (we will call them Z-cuts) parallel to the real axis, with the
fixed branch points at \(u=\pm2g+\frac{i}{2}\mathbb{Z}\).
This cut structure can be already seen in the
asymptotic, infinite length  solution of the    Y-system
\cite{Gromov:2009tv} deduced from the asymptotic Bethe ansatz equations
\cite{Beisert:2005fw} describing the infinite volume spectrum.
This analytic structure is also visible from the dispersion relation of  elementary excitations and their bound states in the string sigma-model in the light-cone gauge, when the energy and momentum are parameterized by the spectral parameter \(u\) (see eqs.\eqref{eps(u)}-\eqref{p(u)}).

One  way to overcome \ these difficulties with  analyticity
is to work in the TBA formulation of the Y-system which implicitly includes all the information
about the analyticity properties of the Y-functions.
This approach made possible the first numerical computation  of the  anomalous dimension of the lowest lying, Konishi state in planar \({\cal N}=4\)  SYM  \cite{Gromov:2009zb}.

However, to write FiNLIE we need  to  explicitly understand  the missing analyticity properties of all Y-functions or T-functions.
In the papers \cite{Cavaglia:2010nm,Balog:2011nm} the basic analytic properties of Y-functions were decrypted from  TBA. Alternatively, together with the Y-system equations, these properties
were shown to be equivalent to the TBA equations. This was an important  progress though  the resulting formulation of the problem contained the same infinite number of Y-functions and it  still remained quite  complex.

We find in this paper that the analytic properties become considerably simpler and more natural
when formulated in terms of T-functions. We managed  to formulate here a simple and sufficient set of analyticity  properties of T-functions, mostly following from  the symmetries of the model. In order to formulate these properties
in a simple way we also have to make a good choice of labeling of the  sheets of the Riemann surface. In this paper, we describe  a
choice which we call ``magic'', additional to the ``physical'' and ``mirror'' labeling already extensively used in the literature. The magic sheet has, unlike the mirror sheet,
``short''  cuts between the pairs of branch points
\(u=\pm2g+\frac{i}{2}\mathbb{Z}\) and coinciding with the mirror sheet
in the vicinity of the real axis.
On this sheet, we discover a new quantum symmetry (already mentioned in \cite{Gromov:2010km}), which generalizes
  the classical  \(\mathbb Z_4\) symmetry  of the string sigma model on the
\(\frac{{\rm PSU}(2,2|4)}{{\rm SO}(1,5)\times {\rm SO}(6)}\) coset
\cite{Beisert:2005bm,Bena:2003wd}.
Using this symmetry, we identified certain Q-functions
having one single short Z-cut, which means that they can be written
in terms of a density with a finite support \([-2g,2g]\).

These   observations  allow us to formulate the first (probably
still perfectible) version of FiNLIE and even to perform its numerical
study for the Konishi dimension. The  results of this numerical study are in perfect
correspondence  with the results obtained in \cite{Gromov:2009zb,Frolov:2010wt} from
TBA and thus confirm the validity of our FiNLIE. We also prove the equivalence between FiNLIE and TBA analytically.

This paper is composed as follows:  \secref{sec:Ysystem} introduces the Y-system and our
notations. The readers familiar with the subject can go directly to
\secref{sec:analyticity} where the analytic properties of
T-functions are described and  the \(\mathbb Z_4\) symmetry  of
the classical string is generalized to a new  symmetry of the full quantum
T-functions. In \secref{sec:wronskian-solution-},  the Wronskian
solution of  T-system in terms of a finite set of
Q-functions is introduced. We work out here a compact and general representation of such solutions in terms of  exterior forms. In \secref{sec:finlie},  the equations constraining these Q-functions are
obtained   with the use of the analyticity and symmetry
constraints identified in \secref{sec:analyticity}.
Finally, \secref{sec:FiNLIEresume} summarizes our construction
and the resulting FiNLIE, while \secref{sec:numer-impl-finl}
presents the results of our first numerical implementation of this FiNLIE.
A systematic study of the analyticity properties, as well
as a recasting of the spectral problem into the  FiNLIE form, are performed explicitly for the \(\sl(2)\) sector's
symmetric states, and in particular \ having only two magnons, containing a relatively large class of operators including
Konishi operator.

\section{Y-system generalities}
\label{sec:Ysystem}
For the detailed description of the  Y-system for AdS/CFT we refer to
\cite{Gromov:2010kf,Beisert:2010jr,Bajnok:2010ke}. In this section we list
the main results and general facts about  Y-systems and Hirota equation.
The reader familiar with the basics of Y-systems can skip this section.

\subsection{AdS/CFT Y-system}
\label{sec:adscft-y-system}
The AdS/CFT Y-system belongs to a class of
functional equations arising in spectral problems of many integrable models
\begin{equation}\label{eq:Y-system}
Y_{a,s}(u+i/2)Y_{a,s}(u-i/2)=\frac{(1+Y_{a,s+1}(u))(1+Y_{a,s-1}(u))}{(1+1/Y_{a+1,s}(u))(1+1/Y_{a-1,s}(u))}\;\;.
\end{equation}
In particular,   the same equations are used to compute the spectrum of \(SU(N)\)
principal chiral  model \cite{Gromov:2008gj,Kazakov:2010kf}         which is an important source of inspiration for the AdS/CFT case.  Y-functions
are associated with certain nodes on a two-dimensional \((a,s)\)
lattice. The main difference from model
to model is
the shape of the domain on this
lattice where the Y-functions are defined.
 In the
present AdS\(_5\)/CFT\(_4\) it
is  the \(\mathbb{T}\)-shaped hook  \cite{Gromov:2009tv}   depicted
in
\figref{fig:ThookY}.~\footnote{More general Y-systems with \(\wT\)-hook shapes are possible for generalizations to other non-compact supersymmetry groups, see \cite{Hegedus:2009ky,Volin:2010xz,Tsuboi:2011iz}.}

In addition to the  shape of the \((a,s) \) domain one should also specify the analyticity
properties of the Y-functions. These properties can be deduced from
the TBA equations
\cite{Bombardelli:2009ns,Gromov:2009bc,Arutyunov:2009ur}  and were explicitly written in
\cite{Cavaglia:2010nm,Gromov:2010kf,Balog:2011nm}.

Y-functions are  natural objects in the TBA approach,
however  not
the most elementary ones.
More fundamental  quantities  are the T-functions defined by
\begin{equation}\label{eq:Y-T0}
Y_{a,s}=\frac{T_{a,s+1}T_{a,s-1}}{T_{a+1,s}T_{a-1,s}}\;\;.
\end{equation}

As it is  the case for many relativistic sigma models, T-functions can sometimes be identified with eigenvalues of the transfer matrices  of an underlying spin chain-type discretization. A similar scenario may exist in the AdS/CFT case, however  the physical interpretation of the AdS/CFT T-functions and their operatorial realization is not yet  established. In the strong coupling limit   the T-functions were identified with the \(\psu(2,2|4)\) characters of the monodromy matrix \cite{Gromov:2010vb}.
This classical monodromy matrix was perturbatively quantized, up to two loops on the world sheet, in  \cite{Benichou:2010ts,Benichou:2011ch}  and the Hirota functional equation \eqref{eq:Hirota} was shown to hold at least in this approximation.

It was noticed for the AdS/CFT system that T-functions are much simpler objects than the Y-functions in all limiting cases
where the general solution of the Y-system was obtained explicitly,
i.e. in the asymptotic large volume limit \cite{Gromov:2009tv} and
in the strong coupling limit \cite{Gromov:2010vb}. It is very natural to assume that the formulation of
analytic properties at the level of T-functions should be
also more revealing and simple. Indeed this is the case, as described in
the next sections.

As one can see  from  \eqref{eq:Hirota}, the T-functions live in an
enlarged  \(\mathbb{T}\)-hook depicted on
\figref{fig:ThookY}.
The meaning of the shape of the \((a,s) \) domain in this case also becomes clear in terms of the T-functions:
 the super-Young diagrams of the unitary highest weight
representations of \(\psu(2,2|4)\) should be entirely contained inside this domain \cite{Gromov:2010vb,Volin:2010xz}.
Furthermore, as we explain in this paper, an important \({\mathbb Z}_4\)
symmetry of the super-coset model noticed in \cite{Gromov:2010km}  has a transparent  realization in terms of the analyticity of T-functions w.r.t. both the spectral parameter \(u\) and the representational variables \((a,s)\).

It is easy to see that the definition
of T-functions by \eq{eq:Y-T0} is ambiguous.
This ambiguity is expressed in the
``gauge'' transformation
\begin{equation}\label{eq:gaugeT}
 T_{a,s}\to g_1^{[+a+s]}g_2^{[+a-s]} g_3^{[-a-s]} g_4^{[-a+s]}T_{a,s}
\end{equation}
which does not affect Y's and the Hirota equation.
Here and everywhere in the text we use a short-hand notation for the shift of the spectral parameter:
\begin{equation}
\label{eq:shifts}
f^{[\pm a]}=f(u\pm ia/2)\;,\qquad f^{[\pm1]}\equiv f^{\pm}.
\end{equation}It is  also useful to notice that due to the Hirota equation
\eq{eq:Hirota} there are several equivalent ways for the Y-functions
to be expressed in terms of the T-functions
\begin{equation}\label{eq:Y-T}
1+Y_{a,s}=\frac{T_{a,s}^+T_{a,s}^-}{T_{a+1,s}T_{a-1,s}},\qquad
1+1/Y_{a,s}=\frac{T_{a,s}^+T_{a,s}^-}{T_{a,s+1}T_{a,s-1}}\;.
\end{equation}

\subsection{Notations}
\label{Notations}
\begin{figure}[ht]
\begin{center}
\subfloat[The functions $x(u)$ and $1/x(u)$.]
{\label{fig:Xriemannl}
{\includegraphics{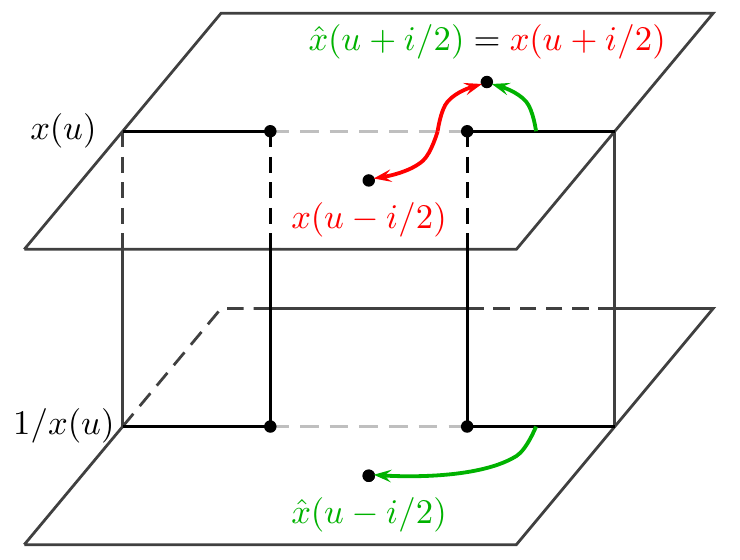}}
}
\subfloat[The functions $\hat x(u)$ and $1/\hat x(u)$.] 
{\label{fig:Xriemanns}
\includegraphics{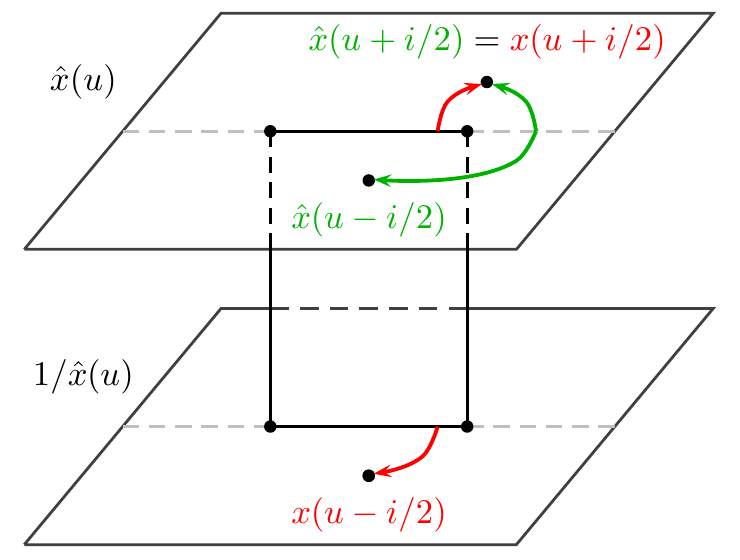}
}
\end{center}
\caption[Riemann sheets of the multivalued function $x(u)$.]{\textbf{Riemann sheets of the multivalued function $x(u)$.}
  The sub\figref{fig:Xriemannl} 
  represents the function $u\mapsto x(u)$ (upper sheet) and $u\mapsto
  1/x(u)$ (lower sheet), whereas the
  sub\figref{fig:Xriemanns} 
  represents the function $u\mapsto \hat x(u)$ (upper sheet) and $u\mapsto
  1/\hat x(u)$ (lower sheet). 
   When $u$ is real, $x^{[\pm 1]}$ is defined by an analytic continuation
   from the real axis, along the red path which avoids the cut
$\bZ_0$, whereas 
$\hat x^{[\pm 1]}$ is defined by a continuation
  along the green path which avoids the cut $\hbZ_0$. We see that \(x^{[+1]}= \hat x^{[+1]}\)
  whereas \(x^{[-1]} = 1/ \hat x^{[-1]}\).
   \label{fig:Xriemann}}
\end{figure}

Let us introduce some important notations used through the text.
We define the Zhukovsky variable \(x+1/x=u/g\), where
\(g=\frac{\sqrt{\l}}{4\pi}\) is related to the `t~Hooft coupling
\(\l\), choosing two different branches\footnote{To define the
  branches, we always use the definition of square roots which has a
  cut on \(\mathbb R^-\) and is positive on \(\mathbb R^+\).}
\begin{eqnarray}
x(u)=\frac{u}{2g}+i\sqrt{1-\frac{u^2}{4g^2}}
\;\;,\;\;
\hat x(u)=\frac{u}{2g}+\sqrt{\frac{u}{2g}-1}\sqrt{\frac{u}{2g}+1}\,.
\label{eq:xphys}
\end{eqnarray}
The first choice $x(u)$ will be referred to as the mirror branch,
whereas $\hat x(u)$ will be called either ``physical'' or
equivalently\footnote{
The function $x$ is quite particular because the magic and
physical sheet coincide. We will see that there exists other functions
(like T- and Y-functions) which do not have this property, and then,
the symbol hat will denote specifically the magic sheet.
} ``magic''.
 $x(u)$ has a long cut, called from now on the \(\bZ\)-cut,
connecting the branch points  \(\pm 2g\) through the  infinity, whereas $\hat x(u)$
has
a short cut \([-2g,2g]\)
called \(\hbZ\)-cut (see \figref{fig:Xriemann}). 

The asymptotic energy  of a single magnon (or of a bound state of \(a\) magnons)  is given by \cite{Santambrogio:2002sb,Beisert:2004hm,Dorey:2006dq}
\begin{equation}
\label{eps(u)}
\hat\epsilon_a(u)=a+\frac{2ig}{\hat x^{[+a]}}-\frac{2ig}{\hat x^{[-a]}}\;.
\end{equation}
Similarly, the asymptotic magnon momentum is
\begin{equation}
\label{p(u)}
\hat p_a(u)=\frac{1}{i}\log\frac{\hat x^{[+a]}}{\hat x^{[-a]}}\;.
\end{equation}

The energy and the momentum of a state are given by
\begin{gather}
  \begin{aligned}
    E
    =&\sum_{k=1}^M \hat \epsilon_1(u_k)+\sum_{a=1}^\infty\int_{-\infty}^\infty \frac{du}{2\pi i}\partial_u\epsilon_a(u)\log(1+Y_{a,0}(u))\;,
    \\
    P
    =&\sum_{k=1}^M \hat
    p_1(u_k)+\sum_{a=1}^\infty\int_{-\infty}^\infty \frac{du}{2\pi
      i}\partial_u p_a(u)\log(1+Y_{a,0}(u))\;,  \label{E:def}
  \end{aligned}\\
\epsilon_a(u)=a+\frac{2ig}{x^{[+a]}}-\frac{2ig}{x^{[-a]}}\;,\qquad\qquad
p_a(u)=\frac{1}{i}\log\frac{x^{[+a]}}{x^{[-a]}}\;.\nonumber
\end{gather}
The anomalous dimension is defined as \(\gamma\equiv E-M\).
The shift of the spectral parameter is always defined by a path which avoids the fixed  \(\hbZ\)-
and \(\bZ\)-cuts   of the functions\footnote{This is a particular case of a general convention
  to avoid the default cuts of a function \cite{Volin:2010cq}. See \cite{Volin:2010cq} for more details about the shift operations.}. For instance, the path  in the shift  \(u\to u+ia/2\) of \(\hat x^{[a]}\)
never crosses the \(\hbZ\)-cut, while the path in the shift of \(x^{[a]}\) never
crosses the \(\bZ\)-cut. As a consequence,   \(x^{[-1]}\neq \hat x^{[-1]}\) whereas
\(x^{[+1]}= \hat x^{[+1]}\) (see
\figref{fig:Xriemann}) and thus the choice of the cuts
affects the functional relations. 
The ``hat''  symbol over a function of the spectral parameter \(u\) (for
instance in the definition \eqref{eq:xphys} of \(\hat x(u)\)) means that
we  choose a Riemann sheet with only short \(\hbZ\)-cuts.
When the ``hat''  appears over a function of \(u\) and of the \((a,s)\) labels, as on
figure~\ref{T-Hook}, then  it denotes also the analytic continuation in
the labels \((a,s)\) performed on the sheet having only
\(\hbZ\)-cuts. This will be introduced in more detail in \secref{sec:genesis} and \secref{sec:z_4-up} (see for instance
\eqref{MHir}).

We will also use  in this paper a short-hand notation for  long  cuts shifted by the distance \(in/2\) from the real axis as \({\bZ}_n\)
 and similarly for the short cuts as \({\hbZ}_n\).    Another important notation which we will
 frequently use is \({\cal A}_{n}\), denoting the class of functions
 analytic\footnote{In this paper,  sometimes what we call the
     ``analyticity'' strip is in fact the ``meromorphicity'' strip since the corresponding function might have poles in that strip.
}
inside the strip \(|\IM u|<\frac{n}{2}\). The class of functions analytic
in the interval \(\frac{-n-m}{2}<\IM u<\frac{n-m}{2}\) is denoted
\({\cal A}_{n}^{[m]}\). Usually it means that we have a pair of
\(\bZ\)-cuts or \(\hbZ\)-cuts at the boundaries of the corresponding analyticity strip.

In addition to the shift operation we will use the fused product for a function:
\begin{equation}
        f^{[s]_\D}\equiv f^{[s-1]}f^{[s-3]}\ldots f^{[1-s]}.
\end{equation}
This definition comes from the \(q\)-number notation: \([s]_{\D}=\frac{\D^s-\D^{-s}}{\D-\D^{-1}}\), where \(\D=e^{\frac i2\partial_u}\). So, e.g. \([2]_\D f=(\D+\D^{-1})f=f^++f^-\) whereas \( f^{[2]_\D}=f^+f^-\).

Finally, for the functions with cuts we define the discontinuity on the cut as follows
\begin{equation}
        \disc(f)=f(u+i0)-f(u-i0)\;.
      \end{equation}

In this paper we use the following Baxter-type polynomials and ``Z-polynomials''\footnote{i.e.,  having a fixed Z-cut  on their Riemann surface, apart from zeros}    :
\begin{equation}
\label{eq:defQB}
Q=\prod_{j=1}^M (u-u_j)\;\;,\;\;
B^{(\pm)}=\prod_{j=1}^{M}\sqrt{\frac {g}{x_j^{\mp}}}\left(\frac {1}{x}-x_j^{\mp}\right)
\;\;,\;\;
R^{(\pm)}=\prod_{j=1}^{M}\sqrt{\frac {g}{x_j^{\mp}}}\left(x-x_j^{\mp}\right)\,.
\end{equation}
For convenience we chose
a normalization which assures the following simple relations:
\begin{equation}
B^{(\pm)} R^{(\pm)}=(-1)^M Q^{\pm}\;.
\end{equation}

The roots \(u_j\) of the Baxter polynomials in \eqref{eq:defQB} are called ``Bethe roots''. They are related to \(x_j\) by \(x_j=\hat
x(u_j)\). Their number and positions completely define  different
states of the   \(\sl(2)\)  sector of the theory to which we restrict
our analysis in this paper. 

In the integral equations that we will write we use the following kernels:

 \begin{equation}
{\cal K}(u)\equiv
-\frac 1{2\pi i u}
\;\;,\;\;
{\cal K}_s(u)\equiv
{\cal K}(u+\tfrac{is}{2})-{\cal K}(u-\tfrac{is}{2})\;,
\end{equation}
\begin{equation}
{\cal Z}(u,v)\equiv
-\frac 1{2\pi i}\frac{\sqrt{4g^2-u^2}}{\sqrt{4g^2-v^2}}
\frac{1}{u-v}\;\;,\;\;
{\cal Z}_s(u,v)\equiv
{\cal Z}(u,v-\tfrac{is}{2})-{\cal Z}(u,v+\tfrac{is}{2})\;,
\end{equation}
\beq
\la{PsiK}
\Psi(u)=-\frac{\psi(-iu)}{2\pi}= \frac{\gamma}{2\pi}+ \sum_{n=0}^\infty\left({\cal
    K}^{[2n]}-\frac 1{2 \pi (n+1)}\right)\,.
\eeq

The convolution is denoted, for arbitrary kernels \(K\) and function \(f\), by
\begin{align}
\nonumber
  \left(K * f\right)(u)&\equiv \int_{-\infty}^{\infty}K(u,v) f(v)dv,
&
\left(\,\slash\!\!\!\! K * f\right)(u)&\equiv -\!\!\!\!\!\!\int_{-\infty}^{\infty}K(u,v) f(v)dv,
\\
  \left(K \cz f\right)(u)&\equiv \int_{-2g}^{2g}K(u,v) f(v)dv,&\left(K
    \check *f\right)(u)&\equiv\left[
    \int_{-\infty}^{-2g}+\int^{\infty}_{2g}\right]K(u,v) f(v)dv\,.
\end{align}

\section{Fundamental properties of T-functions}
\label{sec:analyticity}

The purpose of this section is to find T-functions  in a gauge with
good analyticity properties, different for each of three bands of the
\(\wT\)-hook (see \figref{T-Hook}), and then to relate the three bands by transition functions (gauge transformations).
We describe  and motivate a new structure in  AdS/CFT Y-
and T-systems
which we call \({\mathbb Z}_4\) symmetry. It is tightly related to the analytic properties
of T-functions which we consider first. We will also introduce a concept of the ``magic'' sheet.
Although some analyticity requirements of this section  look a bit speculative, they are verified in
\appref{sec:eqtoTBA} from the TBA form of the Y-system.

It is important to stress that one cannot find a global gauge with good analyticity properties for all T-functions in the whole \(\wT\)-hook.
Therefor we rather describe three different, ``physical'' gauges with a good analyticity for, respectively, the right, left and upper bands of the
\(\wT\)-hook, which are then related by the transitional gauge functions.
Then we show that the analyticity requirements to the functions of these gauge transformations appear to be strong enough\footnote{together with some simple    assumptions about the behavior   at \(u\to\infty\)  and about the structure of poles and zeros   for a given state.} to fix the system of FiNLIE and the corresponding physical solutions of the Y-system.

\subsection{Analyticity strips}
\label{sec:analyticitystrips}

Whereas the Y-system itself  follows to a large extent from the symmetry of the integrable theory
the analytic properties of the Y-functions only partially are constrained by the consistency with
the Y-system. However, they cannot be easily obtained  from general properties of the theory.
For the integrable theories for which the integrable
lattice formulation is known the analytic properties can be
read off from the underlying microscopic Bethe ansatz equations.
However, in the present case such a formulation is  unknown.
Nevertheless, one can make  reasonable assumptions based on the asymptotic
solution of the Y-system proposed for an arbitrary state in \cite{Gromov:2009tv}
and from the integral TBA form of the Y-system for a subclass of \(\sl(2)\)
operators written in \cite{Gromov:2009bc,Gromov:2009zb}.

The analyticity properties of  Y-functions which are expected to be satisfied by the physical solutions of the Y-system
are written in table \ref{tab:AnalY}, in the notations of the
previous section.%
\begin{table}%
  \centering%
\begin{tabular}{|l|}%
  \hline%
 \(Y_{a,0}\in {\cal A}_a \)   \\%
 \(Y_{a,\pm 1}\in {\cal A}_{a-1}\;\;,\;\;a\geq1 \)   \\%
 \(Y_{1,\pm s}\in {\cal A}_{|s|-1}\;\;,\;\;|s|\geq1 \)  \\%
 \(\overline Y_{a,s}=Y_{a,s}\)\;\; (Y-functions are real) \\%
  \hline%
\end{tabular}%
  \caption{Analyticity properties of the Y-functions of the Y-system for AdS/CFT}%
  \label{tab:AnalY}%
\end{table}%
These Y-functions are considered on the mirror sheet which means that we have chosen the long \(Z\)-cuts \cite{Arutyunov:2009ur}.
Moreover, \(Y_{1,\pm 1},\;Y_{2,\pm 2}\)  have a \(\bZ\)-cut on the real axis and  are related on this cut by
\begin{equation}\label{eq:Y11Y22}
Y_{2, \pm 2}(u+ i0)=1/Y_{1,\pm 1}(u- i0)\;\;,\;\; u\in {\bZ}_0\;.
\end{equation}
The last property was discovered from the TBA form of the Y-system  in \cite{Gromov:2009bc}  but it will be considered here as one of the basic analyticity assumptions for our construction of FiNLIE. It means that \(Y_{2,\pm 2}\) and \(Y_{1,\pm 1}\) are not independent functions but rather
the analytic continuation of one into another through the \({\bZ}_0\)-cut.

  The above list of analytic properties for Y-functions also imposes
 certain analytic properties for T-functions. Most of them can be
 inferred from the definition of \(T\)'s \eqref{eq:Y-T0}, the
 analyticity of Y-functions,  and the consequences of Hirota equation \eqref{eq:Y-T}. Recall that  the T-functions are not uniquely defined given the Y-functions, due
 to the gauge freedom \eqref{eq:gaugeT}. This makes the statement about the analyticity
of \(T\)'s a bit more complicated and gauge dependent. A natural assumption, which will be later demonstrated,  is that, in a certain gauge which we denote by the bold font,
\(\bT_{a,s}\)  has good  analyticity properties for the upper band, whether as in another gauge, denoted by the ``blackboard'' font, \(\wT_{a,s}\)
has  good analyticity properties for the right band of the
\(\wT\)-hook.\footnote{\label{ft:mirror}For the left band the gauge is
  also different from the upper and right bands but we will mostly
  consider the right band since the construction for the  left band is
  similar.} 
These analyticity properties are explicitly written in table \ref{tab:AnalT}.

\newcommand{\tabletwo}{
\begin{table}
  \centering
\begin{tabular}{|l|l|}
  \hline
  if \(a\geq |s|\),&   if \(s\geq a\),\\
  \(\bT_{a,0}
\in
  {\cal A}_{a+1}\) & \(\wT_{1,\pm s}
\in {\cal A}_{s}\) \\
  \(\bT_{a,\pm 1}
\in {\cal A}_{a}\) & \(\wT_{2,\pm s}
\in {\cal A}_{s-1}\) \\
\cline{2-2}
  \(\bT_{a,\pm 2}
\in {\cal A}_{a-1}\) & \(\wT_{a,s}\), \(\bT_{a,s}\) are real functions.  \\
  \hline
\end{tabular}
  \caption{Analyticity properties of the T-function of the
    T-hook for AdS/CFT.}
\label{tab:AnalT}
\end{table}
}

Let us notice that there exists still some residual gauge freedom which will not spoil the analyticity properties.
We will fix the remaining gauge ambiguity in the definition of \(\bT_{a,s}\) and \(\wT_{a,s}\) later on.
In particular, we choose 
\begin{equation}
\wT_{0,s}=1\label{eq:wt_0-s=1-}\,.
\end{equation}

Now, with this knowledge of the analyticity strips, we will proceed with fixing the details of the analytic structure of T-functions in each strip.

\subsection{Group theoretical constraints}

Let us remind that there is some evidence \cite{Gromov:2010vb,Benichou:2011ch} for interpretation of the T-functions as  transfer matrices. In particular, at strong coupling the T-functions are reduced to the characters of
classical monodromy matrix, in such  a way that each pair  of indices \((a,s)\)  corresponds to a unitary representation
of the symmetry group PSU\((2,2|4)\) having a rectangular \(a\times s\) Young tableau. This interpretation itself singles
out a particular gauge. Indeed, according to \cite{Gromov:2010vb}
the pairs \((n, 2)\) and \((2, n)  \), or \((n,- 2)\) and \((2,- n)  \), correspond for any \(n\geq 2\)  to the same typical representation and should have equal characters.
For the transfer matrices this also suggests a natural gauge where
\begin{equation}
\label{typical}
\bT_{n,2}=
\bT_{2,n}\;\;,\;\;
\bT_{n,-2}=
\bT_{2,-n}\;\;,\;\;n\geq 2\;.
\end{equation}
This is indeed the gauge in which the Wronskian solution for the \(\wT\)-hook was written
in  \cite{Gromov:2010km}.
\tabletwo{}

Another constraint which we will impose is \(Q_\emptyset=1\). In the  rational (super)-spin chains
\(Q_{\emptyset}\) is the Baxter Q-function on the final step of the
B\"acklund procedure
\cite{Krichever:1996qd,Kazakov:2007fy},
where it corresponds to the trivial \(\mathfrak{gl}(0|0)\) sub-algebra
and therefore it is naturally represented by a constant that
does not depend on the spectral parameter. The same observation holds when \(Q\)-functions are constructed from the first principles, i.e. as eigenvalues of \(Q\)-operators \cite{Bazhanov:2010jq,Kazakov:2010iu,Chicherin:2011sm}. Even though the construction of Q- and T- operators in AdS/CFT is still in a preliminary stage (see
 \cite{Benichou:2011ch} for the first nontrivial quantum computation with T-operators), we believe that \(Q_{\emptyset}\) will play the same role, i.e. it will be an identity operator.

The  quantum super-determinant should be also set to \(1\).
In \cite{Gromov:2010km}\footnote{see the eq.(4.6) and the beginning of sec.5.3 there} the quantum determinant was identified
with the following combination of the \(\Qs\)-functions: \(\frac{\Qs_{\emptyset}^+\Qs_{\bar\emptyset}^-}{\Qs_{\emptyset}^-\Qs_{\bar\emptyset}^+}\).
Setting it to \(1\) and recalling that \(\bT_{0,s}=\Qs^{[-s]}_{\bar\emptyset}\)
 gives the following ``unimodularity'' constraint\footnote{A closer look reveals an intimate connection of \({\bT}_{0,0}\) with the
dressing kernel in TBA. This is similar to the case  of relativistic sigma models with rational S-matrices, where the scalar factor of \(R\)-matrix needed to set the quantum determinant to one is equal to the dressing factor of the   factorized S-matrix in the  integrable field theory with the same symmetry group \cite{Wiegmann:1997zg}.}
\begin{align}
\label{Tper}
\bT_{0,0}^+&=\bT_{0,0}^-,&
\bT_{0,s}&=\bT_{0,0}^{[+s]}=\bT_{0,0}^{[-s]}\;.
\end{align}
This tells that
\(\bT_{0,0}\) should be an \(i-\)periodic function\footnote{We recall that  since the original Y-system is defined on the mirror sheet, the corresponding \(\bT_{a,s}\) are defined as functions with
long cuts. The periodicity condition is sensible to the choice of cuts.
Periodicity of \({\bT}_{0,0}\) can be also obtained from the TBA equations, see \appref{sec:tba-y11-y22}.
}.
 Due to the importance of this quantity, in what follows we introduce a special notation
\begin{equation}
\label{DefF}
\CF\equiv\sqrt{\bT_{0,0}}\;.
\end{equation}

 We expect \(\bT\) to be a kind of a physical gauge.
 In this gauge  the T-functions exhibit all  natural symmetries
 of the characters, and for the  left-right wing symmetric states (in particular, the \(\sl(2)\) or \(\su(2)\) states; we will call them LR-symmetric from now on) it can be shown to be essentially unique
 when supplemented by  a set of analyticity conditions   summarized in (\ref{propbT})
  (see \Appref{sec:uniqueness}).
 Moreover, we will show that in addition to the properties listed above   there is one more very important symmetry
 of the \(\bT\)-gauge which we call \({\mathbb Z}_4\) symmetry.

 Finally, let us define \(\wT\) in terms of \(\bT\). A natural candidate for that is the following, very particular gauge transformation   \eqref{eq:gaugeT} from one to another:
 \begin{equation}
\label{TwTb}
 \wT_{a,s}=(-1)^{a(s+1)}\bT_{a,s}(\CF^{[a+s]})^{a-2}\,.
\end{equation}
This transformation does not change indeed the gauge invariant Y-functions. The relation  \eqref{TwTb} can  be considered as a
   minimal gauge transformation from \(\bT\) (where \(\bT_{0,s}\), for
   instance, has an infinite number of branch points) to a gauge with the
   analyticity bands of table \ref{tab:AnalT}.
 Indeed, we see that  \(\wT_{0,s}=1\). The property
\(\wT_{2,\pm s}\in {\cal A}_{s-1}\)
 is also obviously true since  \(\wT_{2,a}=\bT_{2,a}\).
Then \(\wT_{1,s}=\pm \bT_{1,s}/\CF^{[1+s]}\) seems to be a natural choice
since both \(\wT_{1,1}\) and \(\bT_{1,1}\) should be regular functions\footnote{we frequently use the words ``regular'' and ``analytic'' in the paper for the functions without cuts but potentially with poles}
with the same analyticity strip, according to table \ref{tab:AnalT},
and the factor \(1/\CF\) will not spoil it. Moreover, as it is shown
in \appref{sec:tba-y11-y22}, the property
\(\wT_{1,s}\in {\cal A}_{s}\) and \eqref{TwTb} itself, for which we gave a motivation above, rigorously follow from the TBA
equations. 
The overall sign $(-1)^{a(s+1)}$ does not affect the analyticity strips and  can be chosen arbitrarily, by adjusting  definition of the $\wT$-gauge. Our choice will eventually provide a simpler behavior of Q-functions at $u\to\infty$.

\subsection{Magic sheet  }
\label{subsec:glueingQQbar}

 Let us consider T-functions in the right band of the \(\wT\)-hook, i.e.  for \(T_{a,s}\) with \(s\ge a\).
In this case the T-functions satisfy precisely the same Hirota equations as for the
\(\su(2)\) principal chiral model in \cite{Gromov:2008gj}.
Moreover, at least for the \(\sl(2)\) states
the Y-functions obey a similar asymptotics
\(Y_{1,s}\to s^2-1,\ u\to\infty\). This property is well seen in the ABA
limit, see app.~\ref{app:as}, and also holds at finite size, see app.~\ref{sec:large-u-behavior}. Therefore, to construct the T-functions with a good analyticity in this band   we  choose a gauge\footnote{which will be later related to \(\wT\)-gauge by a gauge transformation conserving its analyticity properties} \(\CT\)
with \(\CT_{1,s}\to s,\ u\to\infty\) (i.e. it tends to the dimension of representation in this limit), preserving the analyticity strip \({\cal A}_s\) and the property \(\CT_{0,s}=1\).

Following the logic of  \cite{Gromov:2008gj,Kazakov:2010kf} we  find the general solution of  Hirota equation satisfying all these properties in the right band of the \(\mathbb{T}\)-hook in terms of the Q-functions\footnote{in the notations of the paper \cite{Gromov:2010km}} \(Q_1,Q_{\bar 1},Q_{2},Q_{\bar 2}\)
in  a gauge where
\begin{equation}\label{Q12b1b2} Q_2=Q_{\bar 1}=1\,,\qquad  Q_1= -iu+G(u),\quad
Q_{\bar 2}=- \overline{Q_{1}}\,=-iu-\bar G(u), \end{equation}  namely,

\begin{equation}
\CT_{1,s}=\begin{vmatrix}Q_1^{[s]} & Q_2^{[s]} \\
Q_{\bar 2}^{[-s]} & Q_{\bar 1}^{[-s]} \\
\end{vmatrix}=Q_1^{[s]}+\overline{Q_{1}}^{[-s]}\;,
\label{eq:ct1s}
\end{equation}where we introduced two  mutually complex conjugate resolvents:

\begin{equation}\label{Gdef}
\begin{array}{ccc}
        G(u)&=&-\frac 1{2\pi i}\int_{-\infty}^{\infty}dv\frac {\rrho(v)}{u-v},\ \ \ {\rm Im}(u)>0,\\
        \bar G(u)&=&+\frac 1{2\pi i}\int_{-\infty}^{\infty}dv\frac {\rrho(v)}{u-v},\ \ \ {\rm Im}(u)<0.
\end{array}
\end{equation}
The function \(G~\)(function \(\bar G\))
is analytic everywhere above(below)
the real axis and should be defined by analytic continuation
in the lower(upper) half-plane.
In general, one would  expect \(G\) to have infinitely many \(\bZ\)-cuts
\({\bZ}_0,\;{\bZ}_{-2},\;{\bZ}_{-4},\dots\).
Also  notice that \(G^{[+0]}+\bar G^{[-0]}=\rrho\).
In these notations we get
\begin{eqnarray}
\label{TGrepr}
\CT_{0,1}&=&1\nonumber\\
      \CT_{1,s}&=&s+G^{[+s]}+\bar G^{[-s]},\\
\nonumber      \CT_{2,s}&=&(1+G^{[s+1]}-G^{[s-1]})(1+\bar G^{[-s-1]}-\bar G^{[-s+1]}).
\end{eqnarray}
In this way we write all T-functions, and thus Y-functions  of the right band
in terms of one single real function \(\rrho\). This result was first time presented at the talk \cite{GroKazIGST}. Later an alternative way to reduce the right band to a finite set of functions was proposed in \cite{Suzuki:2011dj}.

The function \(\rho\) can be fixed  by gluing
this right band  of the \(\mathbb{T}\)-hook to the rest of it, into  the single T-system.
A gauge-invariant way of doing this is to consider the following combination
of \(Y_{1,1}\) and \(Y_{2,2}\):
\begin{equation}
r=\frac{1+1/Y_{2,2}}{1+Y_{1,1}}\;.
\end{equation}
A nice feature of this combination is that it can be  expressed entirely in terms
of T-functions from the right band:
\begin{equation}
\label{rdefinition}
        r=
        \frac{\CT_{2,2}^+\CT_{2,2}^-\CT_{0,1}}{\CT_{1,1}^+\CT_{1,1}^{-}\CT_{2,3}}\;.
      \end{equation}
Using \eq{TGrepr}  we find
\begin{equation}\label{eqr}
r=\frac{(1+G^{[+2]}-G)(1+\bar G^{[-2]}-\bar G)}{(1+G^{[+2]}+\bar G)(1+\bar G^{[-2]}+G)}.
\end{equation}
Due to the property \eqref{eq:Y11Y22} we should get
\begin{equation}\label{rru2g}
r(u+i0)=1/r(u-i0),\ \ u\in \bZ_0.
\end{equation}

Since on the real axis the r.h.s. of (\ref{eqr}) has  branch points due to \(G\) and \(\bar G\), whereas \(G^{[+2]}\) and \(G^{[-2]}\)
are regular there, the relation \eq{rru2g} can be satisfied by imposing the following simple condition:
\begin{equation}\label{QbarQright}
        G^{[+0]}=-\bar G^{[-0]}\;,\ \ u\in \bZ_0.
\end{equation}
This simple observation has some dramatic consequences.
First, due to \(\rrho=G^{[+0]}+\bar G^{[-0]}\) we see that the density \(\rrho\) has a finite support on
the cut \([-2g,2g]\). This allows us  to clarify the definition of the  resolvent as follows:
\begin{equation}\label{Ghat}
\hat G(u)\equiv
-\frac 1{2\pi i}\int_{-2g}^{2g}dv\frac {\rrho(v)}{u-v}=
\left\{\begin{array}{ll}
+G(u)&\;\;,\;\;\IM u>0\\
-\bar G(u)&\;\;,\;\;\IM u<0
\end{array}\right.\;.
\end{equation}
We clearly see the advantage of the choice of the Riemann sheet  with only ``short'' Z-cuts \(\hbZ_n\) over the ``long'' cuts \({\bZ}_n\):
the same function \(G(u),\) when being
defined as a function with the short cuts,
 appears to be   \(\hat G(u)\) --- a function with only one single cut!
\begin{figure}[t]
\centering
\includegraphics[height=3.8cm]{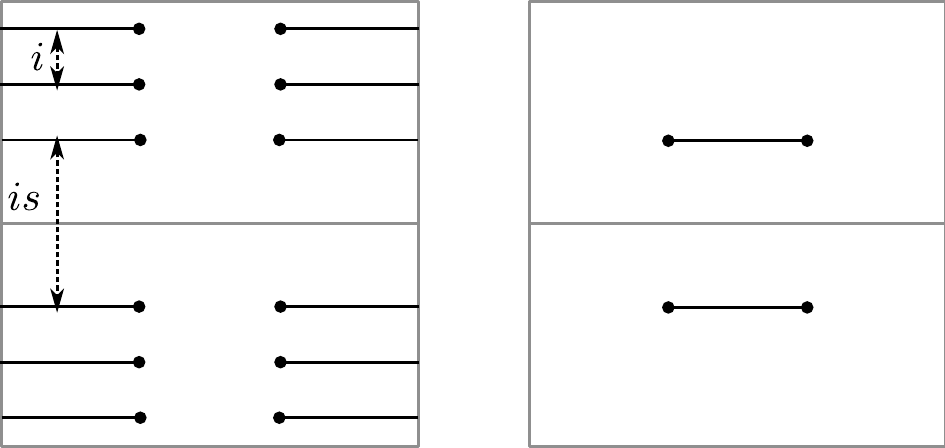}
\caption[Comparison of the mirror and the magic sheets.]{\label{fig:mirrorvsmagic}\textbf{Comparison of the mirror (on the left) and the magic sheets (on the right).} \(\CT_{1,s}\) coincide on the real axis of both sheets. }
\end{figure}

We can go further and convert any function with long \(\bZ\)-cuts into a function
with short cuts by the appropriate choice of the section of the Riemann surface (a priori of an infinite genus). As we shall see, this strategy leads indeed to an essential simplification of analytic
properties of various functions involved in the TBA equations. By this reason we call this section of the Riemann surface the
``magic'' sheet.
We systematically denote
various functions defined on that sheet by a hat over them leaving the
same quantities on the mirror sheet without the hat.
\subsection{\texorpdfstring{${\mathbb Z}_4$}{Z4} symmetry}
\label{sec:genesis}
In this subsection we study the properties of the magic version of   T-functions in the
right band of the T-hook which we denote by \(\hat \CT_{a,s}\). To produce this function we should simply systematically
replace all \(G\) and \(\bar G\) in \eq{TGrepr} by \(\hat G\):
\begin{eqnarray}\label{MHir}
\nonumber\hat\CT_{0,s}&=&1\;,\\
      \hat\CT_{1,s}&=&s+\hat G^{[+s]}-\hat
      G^{[-s]},\\
\nonumber      \hat\CT_{2,s}&=&\hat\CT_{1,1}^{[+s]}\hat\CT_{1,1}^{[-s]}.
\end{eqnarray}

Now we present some important observations. Firstly, \(\hat\CT_{1,s}\) coincide with \(\CT_{1,s}\) everywhere inside the strip \({\cal A}_{s}\). This in particular means that
\begin{equation}
\hat\CT_{1,s}^+\hat\CT_{1,s}^-=\CT_{1,s}^+\CT_{1,s}^-\;\;,\;\;s>1,
\end{equation}
thus the Hirota equation for \(\CT_{1,s}\) also implies the Hirota equation for \(\hat\CT_{1,s>1}\).
However, for \(s=1\) this Hirota equation  should be different since \(\hat\CT_{1,1}^+\hat\CT_{1,1}^-\neq \CT_{1,1}^+\CT_{1,1}^-\).
\(s=1\) is precisely the point where the T-functions from the upper band start entering  the Hirota equation.

Secondly, although the initial parameterization  \eq{TGrepr} of  \(\CT_{a,s}\) (and thus of \(\hat\CT_{a,s}\))
makes sense\footnote{We remind here that the
determinant expression \eqref{eq:ct1s} only holds when \(s\ge a\). As
explained in \cite{Gromov:2010km}, the T-functions on the \(\wT\)-Hook are
given by different determinants in each of the domains \(s\ge a\), \(s\le
-a\) and \(a\ge |s|\).} only for \(s\geq a\), we can formally define \(\hat\CT_{a,s}\) for any \(-\infty< s<\infty\) by \eq{MHir}. Interestingly,
for \(s=1\) we get then
\begin{equation}\label{T11^+T11^-=...}
\hat\CT_{1,1}^+\hat\CT_{1,1}^- = \hat\CT_{0,1}\hat\CT_{2,1}+0\;,
\end{equation}
which is consistent with
 \begin{equation}\hat\CT_{1,0}=0\,.\end{equation}
  In fact, it is easy to see that \(\hat\CT_{a,s}\)
defined by \eq{MHir} satisfy Hirota equation for any  \(s\) if we define it for \(-\infty<s\leq 1\) as an analytic continuation in \(s\).\footnote{Let us point out that  this analytic continuation of \(\CT\)-functions to  negative \(s\) has nothing to do with the physical solution of Hirota equation on the \(\wT\)-hook at negative \(s\).}

Finally, the magic T-functions for the right band defined in this way satisfy
\begin{equation}\label{eq:Z4s-s}
\hat \CT_{a,-s}=(-1)^{a}\hat \CT_{a,s}\;.
\end{equation}
We call this new symmetry of the AdS/CFT T-functions the  \({\mathbb Z}_4\)-symmetry.
In the \appref{app:QCanal} we motivate its relation  to the
\({\mathbb Z}_4\) symmetry of the classical monodromy matrix of the string sigma model
(inherited from \({\mathbb Z}_4\) symmetry in the super coset action construction).

 It is evident that \(\CT\)  is related to \(\wT\) by a gauge transformation. Since in both
gauges \(\wT_{0,s}=\CT_{0,s}=1\) and T-functions are real we have to take in  \eqref{eq:gaugeT} \(g_4=1/g_1,\,\,g_3=1/g_2\),  \(\frac{g_1^{+}}{g_1^{-}}=
\overline {\left(\frac{g_1^{-}}{g_1^{+}}\right)}\equiv h\). Then,
since in both gauges the analyticity strips are the same and given by
table \ref{tab:AnalT}, \(h\) should be analytic in the upper half-plane. In addition, in \appref{sec:zfode} we show that  \(h\) should be real on the magic sheet\footnote{The function \(h\) on the magic sheet, denoted by \(\hat h\), is defined by equality \(h=\hat h\) in the upper half-plane and taking all the Z-cuts being short.}.

Therefore \(\wT\)'s can be found from the  relation \eq{MHir}  as
\begin{equation}\label{wT-->h*calT}
\wT_{0,s}=\hat\CT_{0,s}=1\;,\quad\hat\wT_{1,s}=\hat h^{[+s]} \hat h^{[-s]}\hat\CT_{1,s}\;\;,\quad
\hat\wT_{2,s}=\hat h^{[+s+1]}\hat h^{[+s-1]}\hat h^{[-s+1]}\hat h^{[-s-1]}\hat\CT_{2,s},
\end{equation}
from were it follows that \(\hat\wT_{a,s}\) also obeys \({\mathbb Z}_4\) symmetry property given by \eqref{eq:Z4s-s}.

Note that reality of \(\hat h\) together with its analyticity in the upper half-plane means that \(\hat h\) has only one single cut \(\hbZ_0\), similarly to \(\hat G\). Therefore \(\wT\)-gauge inherits another
``magic'' property of the \(\cT\)-gauge --- finite number of short Z-cuts. In particular, \(\hat\wT_{1,s}\) has only two of them: \(\hbZ_s\) and \(\hbZ_{-s}\).

\subsection{\texorpdfstring{${\mathbb Z}_4$}{Z4} invariance of the upper band}
\label{sec:z_4-up}
In this section we argue that the \({\mathbb Z}_4\) symmetry also leads to  an additional symmetry transformation in
the \(\bT\)-gauge having good analyticity properties in the upper band. In order to see it we have to pass
to the magic sheet and then analytically continue the solution
of  Hirota equation in the variable \(a\)  from  \(a\geq |s|\)
 to  \(a<|s|\), defining  in this way \(\hat\bT_{a,s}\) for an arbitrary integer  \(a\), positive or negative,
providing that \(-2\leq s\leq2\). Our  claim, stemming from the \(\mathbb{Z}_4\) symmetry, is  that \(\hat\bT_{a,s}\)  defined in this way obeys the property similar to \eqref{eq:Z4s-s} of the right band:
\begin{equation}\label{Z4up}
\hat\bT_{a,s}=(-1)^s \hat\bT_{-a,s}\;.
\end{equation}
Let us demonstrate it.
Consider a combination of Y-functions \(Y_{1,1}Y_{2,2}\) which reads in terms of \(T\)-functions as follows:
\begin{equation}\label{Y11Y22inT}
Y_{1,1}Y_{2,2}=\frac{\bT_{1,0}\bT_{2,3}}{\bT_{0,1}\bT_{3,2}}
=\frac{\bT_{1,0}}{\bT_{0,0}^-}
=\frac{\bT_{1,0}}{\bT_{0,0}^+}\,.
\end{equation}
Here we use that, by definition of the bold gauge, \(\bT_{2,3}=\bT_{3,2}\)
and  \(\bT_{0,1}=\bT_{0,0}^+=\bT_{0,0}^-\).
Next, from \eq{eq:Y11Y22} we notice that \(Y_{1,1}Y_{2,2}\)
gets inverted when passing through the cut \({\bf Z}_0\):
\begin{equation}\label{RR}
Y_{1,1}(u+i0)Y_{2,2}(u+i0)=1/(Y_{1,1}(u-i0)Y_{2,2}(u-i0))\;\;,\;\;|u|>2g\;.
\end{equation}
If we simply substitute expression (\ref{Y11Y22inT}) into \eq{RR} we get\footnote{we use a natural notation \(f^{[n\pm 0]}=f(u+in/2\pm i0)\)}
\begin{equation}
\frac{\bT_{1,0}}{\bT_{0,0}^{[-1+0]}}=\frac{\bT_{0,0}^{[+1-0]}}{\bT_{1,0}}\,.
\end{equation}
This relation takes a nice form in terms of \(\hat \bT\)'s.
Employing the definition of the hatted functions we have \(\bT^{[-1+0]}_{0,0}=\hat\bT_{0,0}^{[-1]}\)
(for \(|u|>2g\)) from where we  obtain on the magic sheet simply
\begin{equation}\label{eq:magic00}
\hat\bT^+_{0,0}\hat\bT^-_{0,0}=\hat\bT_{1,0}^2\,.
\end{equation}
On the other hand,  Hirota equation on the magic sheet reads as follows:
\begin{equation}
\hat\bT^+_{0,0}\hat\bT^-_{0,0}=\hat\bT_{1,0}\hat\bT_{-1,0}+\hat\bT_{0,1}\hat\bT_{0,-1}\,,
\end{equation}
which coincides with  \eqref{eq:magic00}  if \eq{Z4up} is satisfied! Indeed, from \eq{Z4up} we have
\begin{equation}{\hat \bT}_{0,\pm 1}=0\end{equation}
 and
\begin{equation}
{\hat \bT}_{1,0}={\hat \bT}_{-1,0}\,.
\end{equation}
To give some more evidence to these results  let us
consider a bit more complicated example.
Take another expression built from \(Y_{1,1}\) and \(Y_{2,2}\) which is rewritten entirely in terms of the T-functions of the upper band of the \(\wT\)-hook (\(a\ge |s|)\)
\begin{eqnarray}\label{sdefinition}
s=\frac{1+Y_{2,2}}{1+1/Y_{1,1}}=
\frac{\bT_{2,2}^+\bT_{2,2}^-\bT_{1,0}}{\bT_{1,1}^+\bT_{1,1}^{-}\bT_{3,2}}\,.
\end{eqnarray}
From the property \eqref{eq:Y11Y22}   of  \(Y\)-functions  it follows again that
\begin{equation}\label{eq:rru2g}
s(u+ i0)=1/s(u- i0),\ \ |u|>2g.
\end{equation}

If we calculate \(s\) slightly bellow the mirror \(\bZ_0\) cut  we can represent it as a product of the following two ratios:
\begin{equation}\label{eq:tilder}
s(u-i0) =
\frac{\bT_{1,1}^{[-1+0]}\bT_{2,2}^{[-1-0]}}{\bT_{1,1}^{[-1-0]}\bT_{2,2}^{[-1+0]}}\ \frac{\hat\bT_{2,2}^+\hat\bT_{2,2}^{-}\hat\bT_{1,0}}{
 \hat\bT_{1,1}^{+} \hat\bT_{1,1}^{-}\hat\bT_{3,2}}\;\;,\;\;{\rm for}\;\;|u|>2g.
\end{equation}
Now we notice that the second multiplier is simply \(1\) if \(\mathbb Z_4\) symmetry holds.
Indeed, the magic Hirota relations give  \(\hat \bT_{2,2}^+ \hat \bT_{2,2}^-=\hat \bT_{3,2}\hat \bT_{1,2}\)
and, if \(\hat \bT_{0,1}=0\) then \(\hat \bT_{1,1}^+ \hat \bT_{1,1}^-=\hat \bT_{1,2}\hat \bT_{1,0}\;\). Thus
\begin{equation}
s(u-i0) =
\frac{\bT_{1,1}^{[-1+0]}\bT_{2,2}^{[-1-0]}}{\bT_{1,1}^{[-1-0]}\bT_{2,2}^{[-1+0]}}\;\;,\;\;{\rm for}\;\;u\in \bZ_0,
\end{equation}
and we see that (\ref{eq:rru2g})
is satisfied assuming, as usual for the AdS/CFT integrability, that all branch points are of a square root type (in fact only Z-cuts are present).
This is an additional argument for the hypothesis that \(\hat \bT_{0,1}=0\) .

In  \appref{sec:eqtoTBA} the \({\mathbb Z}_4\)-symmetry of the upper band is derived rigorously from the TBA equations.
However, from our point of view, the \({\mathbb Z}_4\)-symmetry should be rather included into the list of
fundamental properties of the AdS/CFT Y-system, out of which the TBA
equations can be derived.

\subsection{Bethe equations}

 In the study of integrable spin chains a very convenient way of writing the Bethe equations is the Baxter equation supplemented with the condition that the eigenvalues of transfer matrices are analytic. Analyticity is an obvious fact from the very definition of the transfer matrix,
but it is not immediately clear from the explicit expression for its eigenvalues in terms of the Bethe roots and is only true once
the Bethe equations are satisfied.

Provided that the interpretation of \cite{Tsuboi:2009ud} is correct and the T-functions of AdS/CFT are indeed the transfer matrix eigenvalues one should
expect that the auxiliary Bethe roots (carrying no momentum and energy) can be found by requiring some
good analytic properties for the physical transfer matrices.
In this paper we restrict ourselves to the
\(sl(2)\) sector, which has no
auxiliary roots, and thus the analyticity should be simpler.

As it is shown in \appref{sec:uniqueness}, one can make gauge transformations by means of \(i\)-periodic functions
without Z-cuts preserving all the properties of \(\bT\)- and \(\wT\)-gauge mentioned before in this section.
We claim that it is possible to show that the requirement of
absence of poles
in the T-functions in \(\bT\)- and \(\wT\)-gauges,
at least in their analyticity strips,
together with the requirement that \(\bT_{0,0}\)
has the minimal possible number of zeroes in its analyticity strip,
removes the residual gauge ambiguity. This fixes the \(\bT\)- and \(\wT\)- gauges completely up to an inessential constant factor.

Now suppose that \(\bT_{0,0}\) has some number of zeroes in the strip \(-1/2<\Im(u)<1/2\). Since \(\CF=\sqrt{\bT_{0,0}}\)
defines the gauge transformation between \(\bT\)- and \(\wT\)-gauges, the only possibility to avoid the appearance of
branch points different of those of standard Z-cuts, is to have only the double zeroes in \(\bT_{0,0}\), so that \(\CF\)
 has only simple zeroes.
We denote these zeros as \({u_j}\) and assume that there are \(M\) such zeroes. 

\begin{figure}[t]
\centering
\includegraphics[width=.84\textwidth]{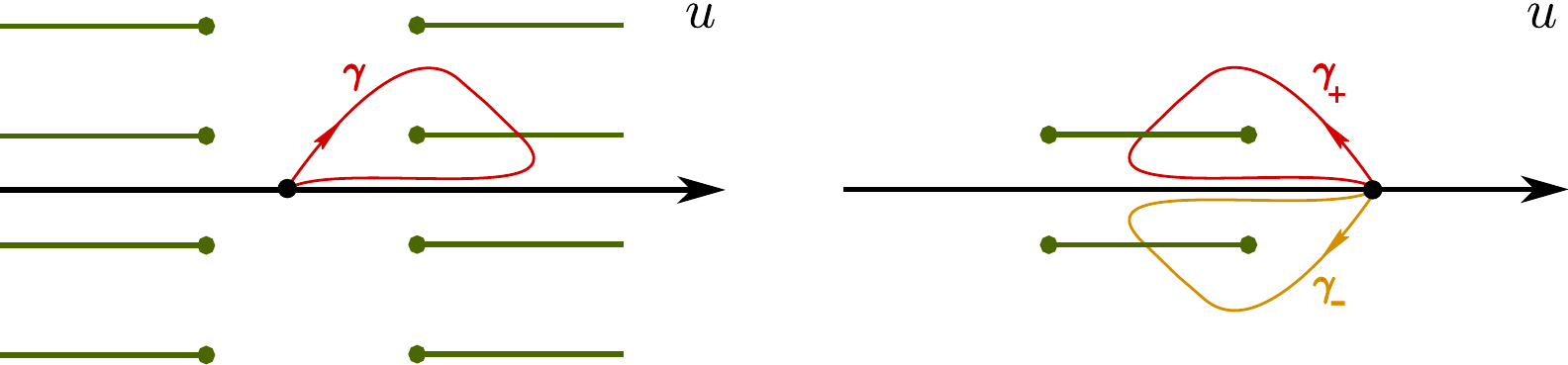}
\caption[Paths used for analytical continuation.]{\label{fig:contourbethe}\textbf{Paths used for analytical continuation.} To get the Bethe equations used in the TBA approach, we continue  \(Y_{1,0}\)  using the path \(\gamma\) (on the left). Alternatively we can formulate Bethe equations using \(\wT_{1,1}\) continued over \(\gamma_+\) and \(\gamma_-\) (on the right).}
\end{figure}

Comparison with the TBA equations shows that  \({u_j}\) are nothing but the Bethe roots and thus we should also satisfy
\be\label{BetheExact}
Y_{1,0}^{\gamma}(u_j)=-1\;,
\ee
where \(Y_{1,0}^{\gamma}\) denotes the analytic continuation along the contour \(\gamma\) defined on  \figref{fig:contourbethe}.

Let us mention a curious  observation. Since \(\wT_{1,2}=\bT_{1,2}/\cF^+\), the absence of poles in \(\wT_{1,2}\) is only possible if \(\bT_{1,2}\) has zeroes at positions \(u_j\pm i/2\).
Assuming that \(\bT_{2,1}\) does not have zeroes at \(u_j\pm i/2\),  \(Y_{2,2}=\bT_{2,1}/\bT_{1,2}\) should have poles at \(u_j\pm i/2\). On the other hand, let us consider analytic continuation along the  contour \(\gamma\) of the Y-system equation at \(a=1,s=1\). Using (\ref{eq:Y11Y22}), one gets
\begin{equation*}
       (1+Y_{1,0}^\gamma){Y_{2,2}^{-}}=\left(\frac{Y_{1,1}^+(1+1/Y_{2,1})}{(1+Y_{1,2})}\right)^{\gamma}\,.
\end{equation*}
On the l.h.s., the poles of the \(Y_{2,2}^-\) at 
zeroes of the Bethe roots are conveniently canceled  with zeroes of \((1+Y_{1,0}^\gamma)\), due to (\ref{BetheExact}).  Therefore we see that the exact Bethe equations (\ref{BetheExact}) can be replaced by the condition that the r.h.s. should be regular at  \(u=u_j\). We believe that it reduces to the condition of regularity of
\(
(Y_{1,1}^+)^\gamma
\)
 at \(u=u_j\). This is an interesting statement deserving a further study.

In \appref{app:BetheRoots} we show that the Bethe equations \eq{BetheExact} can be alternatively written as
\be\label{beqint11t11}
        \frac{\hat\wT_{1,1}^{\gamma_+}}{\hat\wT_{1,1}^{\gamma_-}}=-1,\ \ u=u_j,
\ee
where the analytic continuation along the contours \(\gamma_{\pm}\) is defined on the right of \figref{fig:contourbethe}.

Eq.~(\ref{beqint11t11}) looks more natural than \eqref{BetheExact}  since \(\wT_{1,1}\) has only one \(\hbZ\)-cut in each of the half-planes of the magic sheet, therefore the analytic continuation along \(\gamma_+\) or \(\gamma_-\) is uniquely defined.
On the contrary, \(Y_{1,0}\) has infinitely many \(\bZ\)-cuts in  the mirror sheet and one should  impose an additional  condition in  (\ref{BetheExact}) that the contour \(\gamma\) goes only through the closest cut.
 In \appref{app:BetheRoots} we analyze (\ref{beqint11t11}) and represent it in terms of the quantities suitable for numerics.

\subsection{Expression for the energy}
The transfer matrices of integrable models form a commutative set of
operators. Their expansion with respect to the spectral parameter
produces a family of conserved charges, one of them being
energy. Hence we expect the energy to be encoded into certain
asymptotics of the \(\bT\)-functions. To see how it happens in AdS/CFT
case  let us first consider the 
TBA equation for \(Y_{1,1}Y_{2,2}\) given in \cite{Gromov:2009bc} (the
sum of eqs.(40-41) therein) 
\begin{equation}
Y_{1,1}Y_{2,2}=
\prod_{k=1}^M\frac{\left(\frac 1x-x_k^+\right)(x-x_k^-)}{\left(\frac 1x-x_k^-\right)(x-x_k^+)}
\exp\[\sum_{a=1}^\infty {\cal Z}_a*\log(1+Y_{a,0})\]
\,,\qquad  x_k=\hat x(u_k).
\end{equation}
The kernel \({\cal Z}_a\) is defined in \secref{Notations}
and for  \(u\to \infty+i0\) can be written as\footnote{the next terms contains the
next ``local'' charges. For example the \(1/u^2\) term is
\(\frac{2g^2 \partial_v q_{_3,a}(v)+a}{2\pi u^2}\). Note that also a ``non-local'' part \(\sim a\) is present. Here
\(q_{n,a}=\frac{i}{n-1}\left(\frac{1}{(x^{[+a]})^{n-1}}-\frac{1}{(x^{[-a]})^{n-1}}\right)\) are ``local'' charges \cite{Beisert:2004hm}.
}
\beq
{\cal Z}_a(u,v)=
\frac{\partial_v p_a(v)}{2\pi}+\frac{\partial_v \epsilon_a(v)}{2\pi u}+{\cal O}(1/u^2).
\eeq
The product in the pre-exponent has a similar expansion:
\beq
\prod_{k=1}^M\frac{\left(\frac 1x-x_k^+\right)(x-x_k^-)}{\left(\frac 1x-x_k^-\right)(x-x_k^+)}
=\exp\left[
\sum_{k=1}^M\left(i \hat p_1(u_k)+\frac{i\hat \epsilon_1(u_k)}{u}+{\cal O}(1/u^2)\right)\right].
\eeq
Therefore the exact finite volume expression for \(Y_{1,1}Y_{2,2}\) has the following large \(u\)
expansion:
\begin{equation}\label{Ylauexpan}
 Y_{1,1}Y_{2,2}=\exp\left[ iP+\frac{iE}{u}+{\cal O}(1/u^{2})\,\right],
\end{equation}
where we used \eq{E:def}. Note that for all physical states \(P=0\).

Now let us recall that \(Y_{1,1}Y_{2,2}=\frac{\bT_{1,0}}{\bT_{0,0}^+}\).
Since \(\bT_{1,0}\) is regular on the real axis   we get due to \eqref{eq:Y11Y22}\be\label{eq:discf}
\log\frac{\bT_{0,0}(u+\tfrac{i}{2}+i0)}{\bT_{0,0}(u+\tfrac{i}{2}-i0)}=-2\log Y_{1,1}Y_{2,2}\;\;,\;\;|u|>2g.
\ee

This important property essentially defines the function \(\bT_{0,0}\) in terms of the product of Y-functions
 (see \appref{sec:alteqnonCF}), and it will be often used in this paper.
 Since \(\bT_{0,0}\) is an \(i\)-periodic function, one gets:
\be\label{eq:discf2}
\log\frac{\bT_{0,0}(u-\tfrac{i}{2}+i0)}{\bT_{0,0}(u+\tfrac{i}{2}-i0)}\simeq-2\frac{i E}{u}\;\;,\;\;u\to\infty\;,
\ee
so that both \(\bT_{0,0}\)'s are inside the analyticity strip.
We can see that this expression implies
\begin{equation}\label{T00largeuexpansion}
      E=\frac{1}{2}\lim_{u\to\infty} u\partial_u\log\bT_{0,0}\;.
\end{equation}
Thus  \(\bT_{0,0}\) renders indeed the value of energy of the state when being expanded in \(u\).

\subsection{Summary of properties of \texorpdfstring{$\bT_{a,s}$}{bold T} and \texorpdfstring{${\mathbb T}_{a,s}$}{`blackboard-bold' T}}
\label{sec:prop}
In this subsection, for convenience of the reader we summarize the   properties of the
T-functions   encountered above.

First, the \(\bT\)-functions satisfy the following analyticity properties:

\begin{subequations}
\label{propbT}
\begin{eqnarray}
&\text{\(\bT\)'s are real functions analytic in the upper band: }&
\begin{array}{l}
\bT_{a,0}\in {\cal A}_{a+1}\\
\bT_{a,\pm 1}\in {\cal A}_{a}\\
\bT_{a,\pm 2}\in {\cal A}_{a-1}
\end{array}\\ \hline
&\label{groupPR}\text{\(\bT\)'s satisfy ``group-theoretical'' properties: }&
\bea{l}
\bT_{n,2}=
\bT_{2,n}\;\;,\;n\geq 2\\
\bT_{n,-2}=
\bT_{2,-n}\;\;,\;n\geq 2
\\
\bT_{0,0}^+=\bT_{0,0}^-\\
\bT_{0,s}=\bT_{0,0}^{[-s]}
\eea\\
\hline
\nonumber\\
&\text{\(\bT_{a,s}\) obey \({\mathbb Z}_4\) symmetry
on the magic sheet: }&\label{Z4up2}
\bea{c}
\hat\bT_{a,s}=(-1)^s \hat\bT_{-a,s}\\
\eea\\
\hline
\nonumber\\
&\text{\(\bT\)'s  have no poles in the analyticity strip}&\\
\hline
\nonumber\\
&\text{\(\bT_{0,0}\) has a minimal possible amount of zeroes}&\\
&\text{The double zeroes of \(\bT_{0,0}\) are the Bethe roots \(u_j\)}&
\end{eqnarray}
\end{subequations}
Then,  the \(\wT\)-functions  are related to the \(\bT\)-functions by a gauge transformation involving \(\bT_{0,0}\):
\begin{equation}
\label{eq:Fnice}
 \wT_{a,s}=(-1)^{a(s+1)}
\bT_{a,s}(\CF^{[a+s]})^{{a-2}},\ \ \   \CF\equiv \sqrt{\bT_{0,0}}\,.
\end{equation}
They satisfy the properties:
\begin{subequations}
\label{propwT}
\begin{eqnarray}
&\text{\parbox{18em}{\begin{flushright}
\(\wT\)'s are real functions analytic in the right(left) band:
\end{flushright}}}&
\bea{l}
\wT_{0,\pm s}=1\\
\wT_{1,\pm s}\in {\cal A}_{s}\\
\wT_{2,\pm s}\in {\cal A}_{s-1}
\eea \\ \hline
\nonumber\\&\text{\(\hat\wT_{1,s}\) has only two magic cuts \(\hbZ_s\) and \(\hbZ_{-s}\)}&
\\ \hline
\nonumber\\
&\text{\(\wT_{a,s}\) obey \({\mathbb Z}_4\) symmetry
on the magic sheet: }&\label{Z4right2}
\bea{c}
\hat\wT_{a,s}=(-1)^a \hat\wT_{a,-s}
\eea
\\ \hline
\nonumber\\
&\text{\(\wT\)'s have no poles in the analyticity strip}&
\end{eqnarray}
\end{subequations}

In this paper, restricted to the states of \(sl_2\) sector, the symmetry
between the right and left wing implies an extra relation \(\bT_{a,s}=\bT_{a,-s}\).

All the listed properties of T-functions can be derived from the TBA
equations, as we show in \appref{sec:eqtoTBA}. However, in this
paper we take a different point of view and consider them to be fundamental.
Below we fix the solution using only these properties and taking some extra information
about the large \(u\) behavior of various functions from the  large volume
asymptotic solution.

In \appref{sec:finliedetails} we show that the conditions (\ref{propbT}) assure
that the \(\bT\)-gauge is  unique up to a normalization constant.

\section{Wronskian solution   }
\label{sec:wronskian-solution-}

In  \secref{subsec:glueingQQbar} we managed   to
 express all T-functions    \(\CT_{a,s}\)
in the right  band of the \(\mathbb{T}\)-hook in terms of a single function --- the spectral density \(\rrho\).
In this section we will find a similar representation for the upper band of the  \(\mathbb{T}\)-hook by using the   so-called Wronskian determinant solution of Hirota equation in the infinite band    \(-2\le s\le 2,\,\,-\infty \le a \le \infty \).
The right (left) and upper bands of the \(\mathbb{T}\)-hook   will be represented  by \(2\times 2\) and \(4\times 4\) Wronskian determinants, respectively. In the next section, the full finite set of equations, FiNLIE, will be found by gluing these three bands together into the full \(\mathbb{T}\)-hook. Each of these steps has to be done by respecting the structure of analyticity strips of T-functions.

The Wronskian solution allows to parameterize
the infinite set of T-functions satisfying Hirota equation in a band, in terms of a finite number of Q-functions.
It is thus important to understand  the analyticity properties  of the
underlying Q-functions in virtue of the analyticity conditions and the
symmetries, such as \(\mathbb Z_4\) symmetry, established in the previous section.
The representation in terms of Q-functions will be the base for construction of the FiNLIE system for AdS\(_5\)/CFT\(_4\) formulated in sections \ref{sec:finlie},\ref{sec:FiNLIEresume}
and solved then numerically in \secref{sec:numeric-results}.

\subsection{General Wronskian solution}

\begin{figure}[t]
\centering
\parbox{5cm}{\includegraphics[width=5cm]{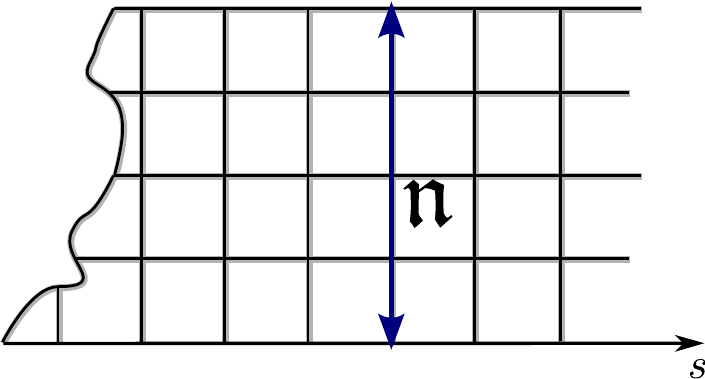}}
\caption{\label{fig:n-strip}Infinite band \(B^{(\groupn)}\) for the
  Hirota equation.
}
\end{figure}

Here we describe the Wronskian solution for    Hirota equation \eqref{eq:Hirota} on an arbitrary
infinite   band of width \(\groupn\) shown in \figref{fig:n-strip}.
Let us denote this band as \(B^{(\groupn)}\). This kind of Wronskian
solutions was used in \cite{Krichever:1996qd} for the analysis of the
\(\su(\groupn)\) quantum spin chains and in \cite{Kazakov:2010kf} for
the solution of the  \(\su(\groupn)\) principal chiral field (PCF)
model in a finite volume.\footnote{In spite of this similarity in the Wronskian representations, the analytic properties of Q-functions are totally different in AdS/CFT and PCF models, especially due to the  difference in position of the momentum carrying nodes in the band. }
At this point, we do not specify any particular boundary conditions
at the end (\(s=0\))  of the band (we need three semi-infinite bands, one \(B^{(4)}\) and two \(B^{(2)}\),  for the construction of the \(\mathbb{T}\)-hook).

A possible basis of the Wronskian ansatz  is a set \(2\groupn+2\)
Q-functions: \(q_{\emptyset}\), \(q_i\) and \(p_{\emptyset}\), \(p_i\)
where \(i=1,2,\dots,\groupn\). 
To get rid of numerous indices it is very convenient to use the formalism of exterior forms. Namely,  introducing
an auxiliary  basis  of \(\groupn\) vectors \(e^i\) we define the 1-forms:
\begin{equation*}
        q\equiv \, q_i\,e^i\;\;,\;\;
        p\equiv \, p_i\,e^i .
\end{equation*}
Then, introducing the k-forms\footnote{
\( q_{(k)}= \frac{1}{k!}q_{i_1\ldots i_k}\, e^{i_1}\wedge\ldots\wedge e^{i_k}\,.\)
}
\begin{equation}\label{eq:WrSol}
       q_{(k)}\equiv\frac{{q^{[k-1]}\wedge q^{[k-3]}\ldots \wedge q^{[1-k]}}}{q^{[k-2]}_{\emptyset}q^{[k-4]}_{\emptyset}\dots q^{[2-k]}_{\emptyset}}\;\;,\;\;
       q_{(0)}\equiv q_\emptyset\;,
\end{equation}
(and similar definitions for \(p_{(k)}\)), we write the general Wronskian solution for the \(B^{(\groupn)}\) band \cite{Krichever:1996qd}   simply as follows
\begin{equation}
\label{eq:generalWronskian}
        T_{a,s}=q_{(a)}^{[+s]}\wedge\, p_{(\groupn-a)}^{[-s]}\;,
\end{equation}
where we identify the exterior \(\groupn\)-form with a scalar,
\(e^1\wedge\dots\wedge e^\groupn\equiv 1\).
One can easily check that for arbitrary \(q\)'s and \(p\)'s, \(T_{a,s}\)  defined in this way  satisfy
indeed Hirota equation with  \(T_{a,s}= 0\) outside the band
\(B^{(\groupn)}\).
Notice however, that there is a certain freedom in choosing  different sets of \(q_{\emptyset}\), \(q_i\) and \(p_{\emptyset}\), \(p_i\)  for the same  solution of Hirota equation.  Indeed, the construction possesses the following \(\sl(\groupn)\) symmetry:
\begin{eqnarray}
\label{Hsym}
p_i\to {H_{i}}^j p_j\;\;,\;\;
q_i\to {H_{i}}^j q_j\;\;,\;\;
p_{\emptyset}\to p_\es\;\;,\;\;
q_{\emptyset}\to q_\es,\quad (\det H=1)
\end{eqnarray}
 which  leaves \(T_{a,s}\) invariant for an arbitrary non-degenerate \(i\)-periodic
 matrix \(H(u)\).
There are also two scalar symmetries:
\begin{subequations}
\label{eq:CFsym}
\begin{align}\label{eq:Csym}
& & & & & &
p_i&\to C^{\groupn-2}p_i\,,&
q_i&\to C^{\groupn-2}q_i\,,&
p_{\emptyset}&\to C^\groupn p_\es\,,&
q_{\emptyset}&\to C^\groupn q_\es\,;
& & & & & \\
\label{eq:Fsym}
& & & & & &
p_i&\to F\,p_i\,,&
q_i&\to F^{-1} q_i\,,&
p_{\emptyset}&\to F\,p_\es\,,&
q_{\emptyset}&\to F^{-1} q_\es& & & & &
\end{align}
\end{subequations}
with arbitrary \(i\)-periodic functions \(C\) and \(F\).

 Eqs.(\ref{Hsym}) and (\ref{eq:CFsym}) form the complete set of  transformations of the Wronskian solution which leave T's invariant. Indeed, \(q_1,q_2,\ldots q_\groupn\) should be \(\groupn\) independent solutions of the Baxter equation (see for instance \cite{Krichever:1996qd}). Any other solution of the Baxter equation should be a linear combination of these \(\groupn\) solutions with coefficients being  \(i\)-periodic functions, i.e. it should be related to those  \(\groupn\)   solutions by a combination of transformations (\ref{Hsym}) and (\ref{eq:CFsym}).  The same is true for \(p_1,p_2\ldots,p_\groupn\). We have also an additional freedom in rescaling \(p_{\es}\) and \(q_{\es}\)  which explains why there are two scalar symmetries (\ref{eq:CFsym}) and not one.
\subsection{Right band}
In this subsection, to demonstrate to the reader the method of Wronskians,   we rewrite the results of
\secref{sec:analyticity} for the right band in  Wronskian notations.
 Also we will  discuss rigorously
the analyticity properties of Q-functions of the right band
 previously assumed in \secref{sec:analyticity} as being just natural. All this will be very helpful when we generalize these methods, in the next couple of subsections and in
 \appref{app:wronskian},  to the considerably more complicated case of the upper band.
\label{sec:rightband}

 Using Eq.~(\ref{eq:WrSol}) we can write the T-functions in the right band  in terms of
\(6,\) so far  arbitrary functions  \(\hq _\emptyset,\;\hp _\emptyset,\;\hq _{1},\;\hq _{2},\;\hp _1,\;\hp _2\)\footnote{In the notations of  eq.(4.11) of \cite{Gromov:2010km}
we have \(
\hq _{\emptyset}=\Qs_{\emptyset},\;
\hq _1=\Qs_1,\;
\hq _2=\Qs_2,\;
\hp _{\emptyset}=\Qs_{\overline{12}},\;
\hp _1=\Qs_{\bar 2}/\CF,\;
\hp _2=\Qs_{\bar 1}/\CF\).}
\begin{equation}
\hat\wT_{0,s}=\hq _{\emptyset}^{[+s]}
\hp _{(2)}^{[-s]}\;\;,\;\;
\hat\wT_{1,s}=\hq ^{[+s]}\wedge
\hp ^{[-s]}\;\;,\;\;
\hat\wT_{2,s}=\hp _{\es}^{[-s]}\hq _{(2)}^{[+s]},
\end{equation}
where
\begin{equation}
\hq _{(2)}=\frac{\hq ^+\wedge \hq ^-}{\hq _\emptyset}\;\;,\;\;\hp _{(2)}=\frac{\hp ^+\wedge \hp ^-}{\hp _\emptyset}\;.
\end{equation}

\begin{figure}[t]
\centering
\includegraphics[width=6cm]{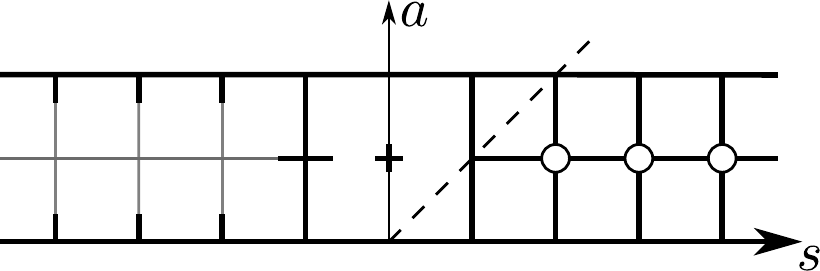}
\caption[The structure of \(\mathbb{Z}_4\) symmetric band
\(B^{(2)}\)]{\label{fig:2strip} 
\textbf{The structure of \(\mathbb{Z}_4\) symmetric band \(B^{(2)}\)}
obtained by analytic continuation from the right band of the Y-system. 
T- and Y-functions to the right of the dashed diagonal and T-functions at the dashed diagonal are the ones of the AdS/CFT \(\wT\)-hook and are given by the Wronskian \(2\times 2\) determinants.
T-functions to the left of the diagonal are the analytic continuation of these Wronskians.}
\end{figure}

This is the most general solution of Hirota equation in the infinite horizontal \(B^{(2)}\)   in \((a,s)\) plane, without any analyticity assumptions.   Let us show  how \(\mathbb Z_4\) symmetry constrains the form of this solution and its analyticity properties.
\begin{itemize}
\item From \(\hat \wT_{1,0}=0\), which is a consequence of \(\mathbb{Z}_4\) property \eqref{Z4right2},  we conclude that the forms \(\hp \) and \(\hq \) are  linearly dependent: \(\hp =\alpha\, \hq \), where \(\a\) is a function of the spectral parameter.
\item  \(\hat\wT_{1,1}=-\hat\wT_{1,-1}\) implies that \(\alpha^{+}=\alpha^{-}\), which means that \(\alpha\) is \(i\)-periodic.  It is therefore possible to absorb \(\a\) into \(\hat p\) and \(\hat q\) using (\ref{eq:Fsym}) with \(F=\sqrt{\a}\). Hence we put \(\a=1\) in what follows.
\item From \(\hat\wT_{0,s}=1\) we conclude that \(\hq _\es^{+}=\hq _\es^{-}\) and \(\hp _{\emptyset}=\hq _\es \hat \wT_{1,1}\).
\end{itemize}
From all these properties we get the solution which depends only on \(\hq _1\) and \(\hq _2\):
\begin{equation}\label{wronskiansimplified}
\hat\wT_{0,s}=1\;\;,\;\;
\hat\wT_{1,s}=\hq ^{[+s]}\wedge
\hq ^{[-s]}\;\;,\;\;
\hat\wT_{2,s}=\hat\wT_{1,1}^{[+s]}
\hat\wT_{1,1}^{[-s]}\;.
\end{equation}

The solution (\ref{wronskiansimplified})  literally coincides with  the parameterization of \(\su(2)\) XXX spin chain transfer matrices in terms of the Baxter Q-functions\footnote{up to relabeling \(\hat\wT_{a,s}\to T_{a,s-1}\). Of course, in our case Q-functions are not polynomials.} \cite{Krichever:1996qd} .
This implies that the following system of Baxter equations should be satisfied:
\begin{eqnarray}\label{Baxtergeneral}
\hq ^{[+2r-1]}\hat\wT_{1,1}=\hq ^+\hat\wT_{1,r}^{[+r-1]}-\hq ^-\hat\wT_{1,r-1}^{[+r]}\;,\\
\nonumber \hq ^{[-2r+1]}\hat\wT_{1,1}=\hq ^-\hat\wT_{1,r}^{[-r+1]}-\hq ^+\hat\wT_{1,r-1}^{[-r]}\;,
\end{eqnarray}
which is easy to check explicitly by substitution of \eqref{wronskiansimplified}.

The equations (\ref{Baxtergeneral})
are very useful for the analysis of analytic properties of Q-functions induced by
the analyticity of \(\wT\)-functions. Using them we can prove that the analyticity of \(\wT_{1,s}\) inside the strip \(\CA_s\)
and of \(\hat\wT_{1,1}\) everywhere except \(\hbZ_{\pm 1}\),
implies the existence of such a symmetry transformation \(H\) from \eq{Hsym} that \(\hq _1\) and \(\hq _2\)
are analytic everywhere except a single cut \(\hbZ_0\) on the real axis.
In particular, this shows that all \(\hat\wT_{1,s}\) have only two magic cuts.

Here is the idea of the proof. From the analyticity properties of T-functions and (\ref{Baxtergeneral}) it follows:
\begin{equation}\label{Baxterdisc}
\hat\wT_{1,1}^{[-3]}\disc \hq ^{[2r-4]}=\hat\wT_{1,r}^{[r-4]}\disc \hq ^{[-2]}
-\hat\wT_{1,r-1}^{[r-3]}\disc \hq ^{[-4]}\;,\ \ r>2.
\end{equation}
In other words, the discontinuities of \(\hq \)'s on the cuts \(\hbZ_{-2}\) and \(\hbZ_{-4}\) define all  discontinuities on \(\hbZ_{2r}\), \(r>0\).

We see that if a solution is regular at \(\hbZ_{-2}\) and \(\hbZ_{-4}\), it is then automatically regular in the upper
half-plane.  The conjugate of equation (\ref{Baxterdisc})
expressing the discontinuities of \(\hq \)'s on \(\hbZ_{-2r}\) in terms of the discontinuities on
\(\hbZ_{2}, \hbZ_{4}\)
implies the analyticity in the lower half-plane.
It is enough to find such a symmetry transformation (\ref{Hsym}) that
\begin{equation}\label{RHforH}
        \disc({H_i}^j)\hq _j^{[-2n]}+{H_i}^j\disc(\hq _j^{[-2n]})=0,\ \ n=1,2\;,
\end{equation}
which represents a Riemann-Hilbert problem.
Assuming that it has a solution we can prove the above-mentioned statement.
In order to argue that the solution exists let us recast Eq.~\eq{RHforH} into
a linear integral equation
\beq\la{int}
H_i^j=\CP_i^j+\frac1{2i}\coth(\pi (u-v))\hat * \left[H_i^k\disc(A_k^n)(A^{-1})_n^j\right],\;\;\;
A_i^n=\hq_i^{[-2n]}\;,
\eeq
where \(\CP_i^j\)  are some \(i\)-periodic functions without cuts.  It is convenient first to relax the  condition  \(\det H=1\). Then \eq{int} is
a linear equation for four independent matrix elements of \(H\). Suppose we found a solution of \eq{int}
which means that there is indeed a linear combinations \(\tilde q_i\) of the initial \(\hat q_i\) with periodic coefficients
without cuts \(\hbZ_{-2}\) and \(\hbZ_{-4}\). As it is easy to see, \(\tilde q_i\)  also satisfy \eq{Baxtergeneral}
and, as a consequence, \eq{Baxterdisc} which then implies that \(\tilde q_i\) could have only one single cut \(\hbZ_0\).
We should remember, however, that the expression for \(\wT_{1,s}\) in terms of \(\tilde q_i\) will also contain \(\det H\) in denominator.
This denominator is easy to get rid of since it is a periodic function and cannot contain any cuts. Indeed,
\beq
\det H=\frac{\hat \wT_{1,s}}{\tilde q^{[+s]}\wedge \tilde q^{[-s]}}\;\;,\;\;s\neq 0
\eeq
and so we can absorb \(\det H\) into \(\tilde q^{[+s]}\) keeping the nice analyticity properties of \(\tilde q^{[+s]}\).

 It is important to mention that the  explicitly found in \cite{Gromov:2010km} large volume asymptotics of Q-functions has indeed  only one cut.

We still have some freedom in the \(H\)-transformations for  \({H_i}^j\) being regular \(i\)-periodic functions. A part of this freedom can be used to make
\(\hq _1\) real and \(\hq _2\) purely imaginary. Both functions have only one cut
and thus can be very efficiently written in terms of a spectral representation (i.e.
from the discontinuities on the cuts).

We finish the discussion of the right band by relating the \(\hat q\)'s to the quantities introduced in \secref{sec:analyticity}.
One can easily identify \(\hq _1=(-i u+\hat G)\hat h\) and \(\hq _2=\hat h\), where \(\hat G\)  was defined in (\ref{Ghat}).

\subsection{Upper band}
\label{sec:upper-band}

\begin{figure}[t]
\centering
\includegraphics[width=3cm]{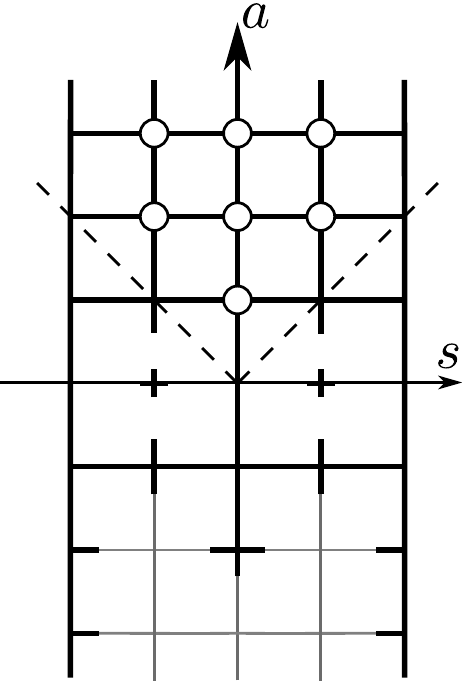}
\caption[The structure of the \(\mathbb{Z}_4\)
  symmetric band \(B^{(4)}\)]{\label{fig:4strip} \textbf{The structure
      of the \(\mathbb{Z}_4\) 
  symmetric band \(B^{(4)}\)} obtained by analytic continuation from the upper band of the
  Y-system.
 }
\end{figure}
For the upper band we can perform a similar procedure, but now in  the \(\bT\)-gauge. The following 10 functions can be used as a basis: \(\qs_{\emptyset}\), \(\ps_{\emptyset}\) and \(\qs_i\),\(\ps_i\), \(i=1,\dots,4\).\footnote{In the notations of \cite{Gromov:2010km} they are
\(    \qs_{\emptyset}=\Qs_{12},\ \ \ps_{\emptyset}=\Qs_{34},\ \ \qs_i=Q_{12\hat i},\ \ \ps_i=\Qs_{34\hat i}.
\)
}

The multi-indexed Q-functions \(\qs_{ij}\), \(\qs_{ijk}\) and \(\qs_{ijkl}\) can be determined from
the Pl\"ucker determinant formulae \eqref{eq:WrSol}  in  the vertical \(B^{(4)}\) band of the \(\mathbb{T}\)-hook:\footnote{There exists a Wronskian parameterization of the general solution in the full \(\wT\)-hook \cite{Gromov:2010km,Tsuboi:2011iz} and it  can be also represented in terms of exterior forms (we postpone it to future publications). But in this paper we prefer to build our physical solution from three \({\mathbb Z}_4\)-symmetric bands.}
\begin{equation}\label{wronskian4}
        \qs_{(2)}=\frac{\qs^+\wedge\qs^-}{\qs_\emptyset},\ \ \qs_{(3)}=\frac{\qs^{++}\wedge\qs\wedge\qs^{--}}{\qs_{\emptyset}^+\qs_{\emptyset}^-},\ \ \qs_{(4)}=\frac{\qs^{[+3]}\wedge\qs^+\wedge\qs^-\wedge\qs^{[-3]}}{\qs_{\emptyset}^{++}\qs_{\emptyset}\qs_{\emptyset}^{--}}\;.
\end{equation}
Wronskian solution of the vertical band is then written in an extremely compact form\footnote{In this section we systematically remove hats over \(\qs\)-s and \(\ps\)-s for convenience, however we consider them as functions with short cuts.}:
\begin{equation}\label{Wronskianupper}
        \hat\bT_{a,s}=\qs_{(2-s)}^{[+a]}\wedge\ps_{(2+s)}^{[-a]}\;.
\end{equation}

The Wronskian ansatz (\ref{Wronskianupper})
gives the formal  general solution of Hirota equation in the vertical \(B^{(4)}\) shown in \figref{fig:4strip}.
Further restrictions on the underlying \(\qs\)'s are needed to ensure that the \({\mathbb Z}_4\) properties are satisfied, as well as the other
analyticity properties listed in \secref{sec:prop}.
Moreover, for simplicity
we consider the LR-symmetric states only\footnote{This includes  in particular all \(sl(2)\) and \(su(2)\) states, and many more.}
and hence  an additional,  LR wing exchange symmetry is imposed:
\begin{equation}\label{LRsymmetry}
\hat\bT_{a,-s}=\hat\bT_{a,s}\;.
\end{equation}
In \appref{app:wronskian} we study all these conditions in detail. For the moment let us notice that from  \eqref{eq:Fnice},\eqref{wronskiansimplified} we have
\(\hat\bT_{a, 2}=\hat\bT_{2,a}=\hat\wT_{2,a}=\hat\wT_{1,1}^{[+a]}\hat\wT_{1,1}^{[-a]}\)  and
\(\hat\bT_{a,2}=\hat\bT_{a,-2}\). Therefore we
can set
\begin{equation}\label{q0=q4=p0=p4=1}
\qs_{\emptyset}=\ps_{(4)}=\ps_{\emptyset}=\qs_{(4)}=\hat\wT_{1,1}\;,
\end{equation}
by making  appropriate transformations \eq{eq:CFsym}.

After that we still have a residual \(\sl(4)\) symmetry \eq{Hsym}. As we prove in \appref{app:compbasis}, this symmetry can be partially used to choose \(\qs\)-s and \(\ps\)-s  satisfying the following relations:
\begin{align}
        \qs_{123}=\qs_1&=-\bar\ps_{134}=-\bar\ps_{3},&\qs_{124}=\qs_2&=\bar\ps_{234}=\bar\ps_{4},\nonumber\\
        \qs_{134}=\qs_{3}&=-\bar\ps_{123}=-\bar\ps_1, & \qs_{234}=\qs_4&=\bar\ps_{124}=\bar\ps_2.
 \label{QfunPfun}
 \end{align}
As a consequence of (\ref{QfunPfun}) we also get
\begin{equation}\label{q-barp}
        \qs_{12}=\bar\ps_{34},\ \ \qs_{13}=\bar\ps_{13},\ \ \qs_{14}=-\bar\ps_{23},\ \ \qs_{23}=-\bar\ps_{14},\ \ \qs_{24}=\bar\ps_{24},\ \ \qs_{34}=\bar\ps_{12}\,.
\end{equation}
With these identifications,  we obtain from \eqref{Wronskianupper} an  explicitly real and LR-symmetric parameterization \cite{Gromov:2010km}:
\begin{eqnarray}\label{T0pm1q}
        \hat\bT_{a,\pm1}&=&\qs_1^{[+a]}\bar\qs_2^{[-a]}+\qs_2^{[+a]}\bar\qs_1^{[-a]}+\qs_3^{[+a]}
        \bar\qs_4^{[-a]}+\qs_4^{[+a]}\bar\qs_{3}^{[-a]}\;,\\
                 \hat\bT_{a,0}&=&\qs_{12}^{[+a]}\bar\qs_{12}^{[-a]}+\qs_{34}^{[+a]}\bar\qs_{34}^{[-a]}-\qs_{14}^{[+a]}\bar\qs_{14}^{[-a]}-\qs_{23}^{[+a]}\bar\qs_{23}^{[-a]}-\qs_{13}^{[+a]}\bar\qs_{24}^{[-a]}-\qs_{24}^{[+a]}\bar\qs_{13}^{[-a]}.\nonumber
\end{eqnarray}

One may wonder whether the Wronskian solution (\ref{Wronskianupper}) possesses
 a finite cut structure for the Q-functions, as it was the case for the right band.
 We performed a detailed analysis of this question in \appref{app:wronskian} and
 came to the conclusion that unfortunately at least some of Q-functions
 should have  infinite number of cuts.
 However, we were able to show that there is  a choice of Q-functions such that \(\qs_{(k)}\) is analytic in the upper half-plane above \(Z_{-1+|k-2|}\) and \(\ps_{(k)}\) is analytic in the lower half-plane below \(Z_{1-|k-2|}\), where the Z-cuts are absent.
 This analyticity condition fixes a part of the \(\sl(4)\)
 symmetry.
 It can be shown that the transformations
 (\ref{Hsym},~\ref{eq:Csym},~\ref{eq:Fsym}) used to enforce
 \eqref{QfunPfun} do not spoil this analyticity condition, so that we
 can impose that \eqref{QfunPfun} holds for Q-functions such that
 \(\qs_{(k)}\)
 is analytic in the upper half-plane above \(Z_{-1+|k-2|}\).
In the rest of the paper we stick to this analytic choice of the Q-functions.

A  nice property of the relations \eqref{QfunPfun}, especially  important for the numerical applications, is that for the large volume \(L\)   the terms in the Wronskian formulas \eqref{T0pm1q} are well distinguished by their magnitude: the first two terms in \(\bT_{a,\pm 1}\) are of the order one whether as the other two are exponentially small w.r.t. the length \(L\). Similarly, the first term in \(\bT_{a,0}\) is of the order 1 whether  as the other are exponentially suppressed.

Let us show that it is enough to know \(\qs_1,\qs_2,\qs_{12}\)
to restore unambiguously all the \(\qs\)-functions of the upper band.  For that we use the Pl\"ucker relations which follow from (\ref{Wronskianupper}) and are explicitly given in  \cite{Kazakov:2007fy,Tsuboi:2009ud,Gromov:2010km}:
\begin{eqnarray}
\qs_{\es}\qs_{ij}&=&\qs_i^+\qs_j^--\qs_j^+\qs_i^-\;,\label{Plucker1}\\
\qs_{ijk}\qs_i&=&\qs_{ij}^+\qs_{ik}^--\qs_{ik}^+\qs_{ij}^-\;.\label{Plucker2}
\end{eqnarray}
 Reconstruction of the rest of \(\qs\) functions goes as follows:
\begin{itemize}
\item From (\ref{Plucker1}) with \(ij=12\)  one finds
\begin{equation}\label{q0plucker}
\qs_\emptyset=\frac{\qs_1^+\qs_2^--\qs_2^+\qs_1^-}{\qs_{12}}\,.\end{equation}
\item Relations (\ref{Plucker2}) for \(ijk=123\), \(124\), \(213\) and \(214\) can be explicitly written in view of (\ref{QfunPfun}) as
\begin{eqnarray}\label{q1q2Plucker}
        \qs_1^2&=&\qs_{12}^+\qs_{13}^--\qs_{13}^+\qs_{12}^-,\label{exp_plucker1}\\
        \qs_{1}\qs_2&=&\qs_{12}^+\qs_{14}^--\qs_{14}^+\qs_{12}^-=\qs_{12}^+\qs_{23}^--\qs_{23}^+\qs_{12}^-,\label{exp_plucker2}\\
        \qs_{2}^2&=&\qs_{12}^+\qs_{24}^--\qs_{24}^+\qs_{12}^-.\label{exp_plucker3}
\end{eqnarray}
We can unambiguously and explicitly define \(\qs_{13}\), \(\qs_{14}\), \(\qs_{23}\) and \(\qs_{24}\)    through \(\qs_1,\qs_2,\qs_{12}\) using their  regularity above \(Z_{-1}\), by inverting the linear difference operators in these Pl\"ucker relations. Note that \(\qs_{14}=\qs_{23}\).

\item Similarly, the relations (\ref{Plucker1}) for \(ij=13\) and \(14\) define \(\qs_3\) and \(\qs_4\).
\item Finally, the relation (\ref{Plucker2}) for \(ijk=134\) fixes \(\qs_{34}\).
\end{itemize}

Let us summarize the analyticity properties of the Q-functions introduced in the previous section:
 \(\qs_1,\qs_2\)  are regular above \(Z_0\) and \(\qs_{12}\) is regular above \(Z_{-1}\).
 A simple inspection of relations \eqref{Plucker1} and \eqref{Plucker2} shows that this automatically
 ensures the correct analyticity of all the rest of  \(\text{\qs}\)-functions, and,
 consequently, of all \(\bT\)-functions of the upper band:
 \(\qs_\emptyset\) is regular above \(Z_1\),  \(\qs_i\) are regular above \(Z_0\) and \(\qs_{ij}\) are regular above \(Z_{-1}\).

\section{Finite set of equations}
\label{sec:finlie}

In the previous section we managed to parameterize the T-functions of all three bands of the \(\mathbb{T}\)-hook by Wronskian determinants of the
Q-functions and establish their symmetry and  analyticity properties.
Here we will derive  yet missing equations of the system FiNLIE gluing
all three bands together into the single \(\wT\)-hook by means of a few transition functions and
constraining these functions by  their analyticity properties.
The method will not  appeal
to the TBA equations but rather will be based on
the properties listed in \secref{sec:prop}, from
which the TBA equations also follow (as is shown in \appref{sec:eqtoTBA}).
They look more fundamental and simple then the set of analyticity properties for the construction  of  TBA from the Y-system given in \cite{Cavaglia:2010nm} but their equivalence is demonstrated in  \appref{sec:eqtoTBA}.

\subsection{Density parameterization of the upper band}

In \secref{sec:upper-band} we explained how all the T-functions of the upper band can be written
in terms of only three independent functions: \(\qs_1,\qs_2\) and \(\qs_{12}\).
Let us now introduce a suitable parameterization for them:
\begin{equation}\label{Param}
\qs_1=\nU f^+f^-,\ \ \qs_2=\nU f^+f^- W,\ \ \qs_{12}=f^2 \VP\;,
\end{equation}
where \(\tilde Q=\prod_{j=1}^M(u-\tilde u_j)\) is simply a polynomial of degree \(M\) containing all zeros of \(\qs_{12}\)~\footnote{
 In the large volume limit \(\VP\) coincides with the Baxter polynomial
\(Q(u)=\prod_{j=1}^M(u-u_{j})\). Recall that \(\bT_{0,0}=\qs_{12}\bar \qs_{12}+{\rm subleading\,\, terms} \,\). The zeros \(\tilde u_j\) of \(q_{12}\) are chosen in such\ a way that \(\bT_{0,0}\) has zeros at the positions of  Bethe roots \(u_j\). In the large volume limit \(\tilde u_j=u_j\)
}
 and normalized so that \(\VP=u^M+\ldots\), whereas \(f,\nU,W\)
are nontrivial functions. We see that
this is indeed just a parameterization
which does not change the number of independent functions, and the roots of the polynomial \(\VP\) encode (but   are not equal to) the Bethe roots of an \(\sl(2)\)  state.
One can think about this parameterization as being a gauge transformation
 to a new gauge \(\sT_{a,s}=q_{(2-s)}^{[+a]}\wedge p_{(2+s)}^{[-a]}\) defined by
\begin{equation}
\qs_{\emptyset}=U^+U^-f^{++}f^{--}q_{\emptyset},\ \ \qs=Uf^+f^-q,\
\qs_{(2)}=f^2q_{(2)},\ \qs_{(3)}=\frac{f^+f^-}{U}q_{(3)},\
\qs_{(4)}=\frac{f^{++}f^{--}}{U^+U^-}q_{(4)}
\label{eq:qsq}
\end{equation}
so that:
\begin{equation}q_1=1\,,\qquad q_2=W\,,\qquad q_{12}=\VP\,,\qquad q_{123}=\nU^2\,.\end{equation}
The \({\sT} \)-functions and the \(\bT\)-functions are related as
follows\begin{eqnarray}\label{TtoT}
        \bT_{a,s}&=&{\sT}_{a,s} \,f^{[a+s]}f^{[a-s]}\bar f^{[-a-s]}\bar f^{[-a+s]}\left({\nU^{[+a]}\bar \nU^{[-a]}}\right)^{[s]_D}\;.
\end{eqnarray}
In particular, one can see that the LR symmetry of \(\bT_{a,s}\)  implies  that
\begin{equation}\label{TUUT}
\sT_{a,-1}=(\nU^{[+a]}\bar \nU^{[-a]})^2\sT_{a,1}\;.
\end{equation}
The Wronskian representation \eq{T0pm1q} becomes:
\begin{eqnarray}
\label{sTofq}
        \sT_{a,1}&=&\bar W^{[-a]}+W^{[+a]}+ q_3^{[+a]}\bar q_4^{[-a]}+ q_4^{[+a]}\bar q_{3}^{[-a]}\;,\\
                 \sT_{a,0}&=& q _{12}^{[+a]}\bar q _{12}^{[-a]}+ q _{34}^{[+a]}\bar q _{34}^{[-a]}-q _{14}^{[+a]}\bar q_{14}^{[-a]}-q _{23}^{[+a]}\bar q_{23}^{[-a]}-q_{13}^{[+a]}\bar q_{24}^{[-a]}- q_{24}^{[+a]}\bar q_{13}^{[-a]}\;.
\end{eqnarray}

Knowing the analytic structure of the functions \eqref{Param} we can parameterize them in terms of spectral densities with the support on \(\mathbb{R}\).
Not only it is conceptually important for the reformulation of the whole Y-system as a finite Riemann-Hilbert problem  but it is also very convenient   for the numerics.

 First, as we see from \eq{Param} \(W=\frac{\qs_1}{\qs_2}\)
and thus it  should be regular in the upper half-plane.
This allows us to introduce the spectral representation for \(W\).
Defining a real function
\beq
\tilde\rrho_2=W^{[+0]}+\bar W^{[-0]}\;,
\eeq
for the states with two symmetric magnons we can write
\beqa\label{Wspecold}
        W&=&-iu+{\cal K}*{\tilde\rrho_2}\;\;,\;\;\IM u>0,\\
        \bar W&=&+iu-{\cal K}*{\tilde\rrho_2}\;\;,\;\;\IM u<0.
\eeqa
For the states with more than two magnons the linear polynomial \(-iu\) in \(W\)
should be replaced by a polynomial of degree \(M-1\) (see \appref{app:asymptoticsolution} for more details).

This function \(W\) used for parameterization of the upper band is a direct analog of   the function \(\hat Q_1\) parameterizing the right band and defined before in eqs.\eqref{Q12b1b2},\eqref{Gdef} as
\begin{equation}
\hat Q_1=-iu+{\cal K}\cz\rrho\;.
\end{equation}
In the  large volume limit,
the densities \(\rrho\) and \(\tilde\rrho_2\) become semi-circle distributions
with a finite support on \(\hbZ_0\).
At a finite volume  \(\rrho\) still has the same finite support whereas
\(\tilde\rrho_2\) becomes non-zero everywhere on the real
axis. However,
if we define \(\rrho_2\equiv\tilde\rrho_2+q_3^{[+0]}\bar q_4^{[-0]}+q_4^{[+0]}\bar q_3^{[-0]}\) then the \(\mathbb Z_4\) constraint \(\hat\sT_{0,1}=0\) implies\footnote{
{One can define
    \(\rrho_2(u)\equiv\lim_{\epsilon\to+0}\hat\sT_{\epsilon,1}(u)\). For \(u\in\bZ_0\) the limit is zero since
    it is equal to \(\hat\sT_{0,1}\) which must vanish due to \({\mathbb Z}_4\). From \eqref{sTofq} we get
    \(\rrho_2=\tilde\rho_2+ q_3^{[+0]}\bar q_4^{[-0]}+
    q_4^{[+0]}\bar q_{3}^{[-0]}\).
}} that \(\rrho_2\) has a finite support, in complete analogy with \(\rrho\). In terms of this density \(W\) is parameterized as follows:
\begin{equation}\label{Wspec}
        W=-iu+{\cal K}\cz{\rrho_2}-\CC*(q_3^{[+0]}\bar q_4^{[-0]}+q_4^{[+0]}\bar q_3^{[-0]})\;,\;\;\IM u>0\;.
\end{equation}
Contribution from \(q_3\) and \(q_4\) vanishes for large volume and thus the parameterization \eq{Wspec}
is very handy for numerics.

Similar spectral representation can be written for \(f\) and \(U\) and thus we conclude that the spectral problem
reduces to a problem of finding a  few densities  as well as  a few additional parameters, such as the Bethe roots.
In the next subsection we show however that the function \(f\) can be explicitly excluded from the final FiNLIE.
The further restrictions on the densities are due to the symmetries, analyticity properties and the condition
that the Hirota equation is satisfied not only inside the bands but also in all nodes of the \(\mathbb{T}\)-hook.
In other words one should sew the three bands of the \(\mathbb{T}\)-hook together to close FiNLIE.

\subsection{Closing the system of FiNLIE }

In this subsection we show how the system of equations can be closed by sewing the \(\mathbb{T}\)-hook from three bands
and imposing the analyticity properties discussed earlier.

\subsubsection{Equation for \texorpdfstring{$f$}{f}}The regularity of the Q-functions in a half-plane allowed us to write a spectral representation of $W$
in terms of a single density \(\rrho_2\) \eq{Wspec}.
Here we will exploit the analyticity properties of \(\nU \) and \(f\), which follow from their relation
to the Q-functions \eq{Param}, to write the equations for  fermionic Y-functions \(Y_{1,\pm 1},Y_{2,\pm 2}\).
Namely, we will use the fact that the functions \(\nU, f^{-}, \hh\) are regular in the upper half-plane.
Moreover, for a sufficiently large \(L\) (or sufficiently small coupling constant)
these functions
do not have poles and zeroes in the upper half-plane
and behave as a power of \(u\) at \(u\to\infty\),
as one can see from the asymptotic limit
(see \appref{app:asymptoticsolution}).

Considering a function \(\left(\frac{f^-}{f^+}\right)^2\) we see from (\ref{TtoT}) that it can be written through T-functions of the upper band and the fermionic Y-functions as
follows:
\begin{equation}\label{Bdefin}
\bB\equiv\left(\frac{f^-}{f^+}\right)^2=\frac{\bT_{0,0}^-}{\bT_{1,0}}\frac{\sT_{1,0}}{\sT_{0,0}^-}=
\frac{1}{Y_{1,1}Y_{2,2}}\frac{\sT_{1,0}}{\sT_{0,0}^-}\,.
\end{equation}
Moreover, due to the analyticity properties of \(f\) this function is analytic in the upper half-plane and goes to \(1\) at infinity there.
Also we assume (and our numerics seems to confirm it)  that it has neither zeros nor poles there.
Thus one can  construct a spectral representation for \(\log \bB\) in the upper half-plane
\begin{equation}\label{BB}
\log\bB=\CC*\rrho_b\;\;,\;\;\IM u>0,
\end{equation}
from its real part on the real axis
\begin{equation}
\rrho_b(v)\equiv\log{{\Bup}(v+i0)}{\Bdown(v-i0)}
=\left\{
\begin{array}{lc}
\log\frac{\sT_{1,0}^2}{\sT_{0,0}^{+}\sT_{0,0}^{-}Y_{1,1}^2Y_{2,2}^2}\;\;&  ,\;\;|v|<2g\\
\log\frac{\sT_{1,0}^2}{\hat\sT_{0,0}^{+}\hat\sT_{0,0}^{-}}\;\;&,\;\;|v|>2g
\end{array}
\right.\;.
\end{equation}
Alternatively, one can also reconstruct \(\bf B\)
\begin{equation}\label{BBim}
\log\bB=\CC*\irho_b\;\;,\;\;\IM u>0
\end{equation}
from its imaginary part
\begin{equation}
\irho_b(v)\equiv\log{\frac{{\Bup}(v+i0)}{\Bdown(v-i0)}}
=\left\{
\begin{array}{lc}
\log\frac{\sT_{0,0}^+}{\sT_{0,0}^{-}}\;\;&  ,\;\;|v|<2g\\
\log\frac{\hat\sT_{0,0}^+}{\hat\sT_{0,0}^{-}}\left(\frac 1 {Y_{11}^{[+
0]}Y_{22}^{[+0]}}\right)^2\;\;&,\;\;|v|>2g
\end{array}
\right.\;.
\end{equation}
Notice that in both representations \eqref{BB},\eqref{BBim} the term \(\log{{\Bdown}(v-i0)}\) does not contribute in the r.h.s. We also used the fact that due to  \eqref{eq:Y11Y22} \(\log (Y_{1,1}Y_{2,2})\) is real between the branch points \(\pm 2g\)
and is purely imaginary outside that interval.

The former representation has an advantage with respect to the latter one
since
\(\rrho_b(v)\) tends to zero
faster than \(\irho_b\) when \(v\to\pm\infty\). This is because at
  large \(v\) \(\bB\) is asymptotically a phase\footnote{In \eqref{Bdefin}, the
    factor \(\sT_{1,0}\) is real, \(Y_{11}Y_{22}\) is a
  phase as soon as \(u>2g\)  and \(\sT_{1,0}/\sT_{0,0}^-\) approaches 1  at large values of \(u\).} as can be seen
  from \eqref{Bdefin}.

Since \(\Bup\) is a simple combination of \(f\)'s, one can determine \(f\) from the following  finite difference equation:
\begin{equation}
\log(f^-)^2-\log(f^+)^2=\log\bB=\CC*\rrho_b\;.
\label{fdef}
\end{equation}

Keeping in mind that \(f\)
should be regular in the upper half-plane we write the solution as follows
\begin{equation}
\label{eq:log-f2sim-1rrh}
\log f^2\simeq
\sum_{n=1}^\infty\CC^{[2n-1]}*\rrho_b\;.
\end{equation}
The sum in the r.h.s. is divergent, however the divergence is
just an infinite constant which can be regularized, as in \eqref{PsiK}, so that
\begin{equation}\label{eqf}
\log f^2=\Psi^+*\rrho_b\;,
\end{equation}
where the kernel \(\Psi\) is defined in \eq{PsiK}. Note that in principle
one can add a constant to the r.h.s. of the previous equation.
This however would change only the normalization of \(f\) and as a result
the normalization of \(\bT\). We fix the normalization by \eq{eqf}.

\subsubsection{Equations for \texorpdfstring{$Y_{1,1}, Y_{2,2}$ and $U$}{Y11, Y22 and U}}

\paragraph
{Equation for \(Y_{1,1}Y_{2,2}\):}
To derive an equation for
\(Y_{1,1}Y_{2,2}\) we consider instead
of \eqref{Bdefin}
\begin{equation}\label{B/sqrt}
\log \tilde\Bup(u)\equiv \frac{\log \Bup(u)}{\sqrt{4g^2-u^2}}\;.
\end{equation}
Again we will use the fact that due to  \eqref{eq:Y11Y22} \(\log (Y_{1,1}Y_{2,2})\) is real between the branch points \(\pm 2g\)
and is purely imaginary outside that interval. Together with the
square root the product \(Y_{1,1}Y_{2,2}\)
drops out from the imaginary part \(\Im(\log\tilde\bB)\) and therefore does
not appear in the r.h.s. of the spectral representation
\begin{eqnarray}\label{Btilde}
        \log\tilde{\Bup}&=&\CC*{\tilde\irho_b},\ \ {\rm Im}(u)>0\;,
\end{eqnarray}
where
\begin{equation}
\tilde\irho_b
=\log\frac{~\tilde\bB~}{\overline{\tilde\bB}}=\left\{
\begin{array}{ll}
\frac{1}{\sqrt{4g^2-v^2}}\log\frac{\sT_{0,0}^+}{\sT_{0,0}^{-}}\;\;&  ,\;\;|v|<2g\\
\frac{i}{\sqrt{v-2g}\sqrt{v+2g}}\log\frac{\sT_{1,0}^2}{\hat\sT_{0,0}^{-}\hat\sT_{0,0}^+}\;\;&,\;\;|v|>2g
\end{array}
\right.\;.
\end{equation}
 Inserting the definition of
 \(\tilde\Bup\)  (\ref{Bdefin},\ref{B/sqrt}) into the l.h.s of \eqref{Btilde} and shifting the contours of integration in the integrals involving \(\log\sT_{0,0}^\pm\)
we obtain an equation for  \(Y_{11}Y_{22}\)
expressing it through \(\sT_{1,0}\) and \(\sT_{0,0}\)  and the polynomial \(Q(u)=\prod_{j=1}^M(u-u_j)\) encoding the Bethe roots of the state:

\begin{equation}\label{Y11Y22throughT}
{\log {Y_{1,1}Y_{2,2}}}=\log{\frac{R^{(+)}B^{(-)}}{R^{(-)}B^{(+)}}}+
{\cal Z}_{+0}* \log\frac{\sT_{1,0}}{Q^+Q^-}-
{\cal Z}_1*\log\frac{\sT_{0,0}}{Q^2}\,,
\end{equation}
where the kernels are defined in
\secref{Notations}.
This equation can be also obtained from the TBA equations
as is shown in \appref{sec:tba-y11-y22}.
The derivation from TBA is an important check of our basic analyticity assumptions in \secref{sec:analyticity}.

\paragraph{Equation for \(Y_{1,1}/Y_{2,2}\):}
   Similarly to \(Y_{1,1}Y_{2,2}\), we can derive a separate equation  for the ratio
\(Y_{1,1}/Y_{2,2}\).
For that we construct a combination of T-functions which
contains only \(\nU \) and \(f\) and no conjugate quantities, namely
\begin{equation}
\label{ratioY11Y22throughT}
\frac{\bT_{1,0}(\bT_{1,1}^-)^2}{\bT_{0,0}^-(\bT_{2,1})^2}=
\left(\frac{\nU }{\nU ^{[2]}}
  \frac{f^{[1]}}{f^{[3]}}\right)^2\frac{\sT_{1,0}(\sT_{1,1}^-)^2}{\sT_{0,0}^-(\sT_{2,1})^2}\,.
\end{equation}
Next, writing the ratio of the Y-functions in terms of \(\bT\),
\begin{equation}
\frac{Y_{1,1}}{Y_{2,2}}=\left[\frac{\bT_{1,0}(\bT_{1,1}^-)^2}{\bT_{0,0}^-(\bT_{2,1})^2}\right]
\frac{(\bT_{1,2})^2}{(\bT_{1,1}^-)^2}\;,
\end{equation}
and using that, according to (\ref{eq:Fnice}) and \eqref{wT-->h*calT}, \(\bT_{1,1}^-/\bT_{1,2}=-\frac{\hat h}{\hat h^{[2]}}\cT_{1,1}^-/\cT_{1,2}\) we construct from here the quantity\begin{equation}\label{relatingBUC2}
   \bC=\left(\frac{\nU }{\nU ^{[2]}}
     \frac{f^{[1]}\hat h^{[2]}}{f^{[3]}\hat h}\right)^2=\frac{Y_{1,1}}{Y_{2,2}}\frac{\sT_{0,0}^-}{\sT_{1,0}}\left(\frac{\sT_{2,1}}{\CT_{1,2}}\frac{\cT_{1,1}^-}{\sT_{1,1}^-}\right)^2\;
\end{equation}
which is written completely in terms of   functions analytic in the upper half-plane and
  approaching \(1\) at infinity. Its spectral representation
is again straightforward:
\begin{eqnarray}\label{spectralC}
        \log\bC&=&\CC*\irho_{c}\;\;,\;\;{\rm Im}(u)>0\,
\end{eqnarray}
with the spectral density given by
\begin{equation}
    \irho_c=\log\frac{\bC(u+i0)}{\bar\bC(u-i0)}=\log\frac{\sT_{0,0}^-}{\sT_{0,0}^+}  \left(\frac{\sT_{1,1}^+}{\cT_{1,1}^+}\frac{\cT_{1,1}^-}{\sT_{1,1}^-}\right)^2\;.
\end{equation}
Here we use the reality of T-functions as well as the reality of \(\log Y_{1,1}/Y_{2,2}\)
so that the Y-functions again drop out from the density.
Furthermore, we can get rid of the shift in  arguments of the functions
in the definition of the density by shifting the integration contours in \eqref{spectralC}.
As before, one should be careful with the singularities at the positions of  Bethe roots. This gives
\begin{equation}
\log\frac{Y_{11}}{Y_{22}}=\log\frac{\sT_{1,0}}{Q^+Q^-}\left(\frac{\CT_{1,2}}{\sT_{2,1}}\right)^2-\slash\!\!\!\!\!\: {\cal
  K}_1*\log\frac{
  \sT_{0,0}}{Q^2}\left(\frac{\CT_{1,1}}{\sT_{1,1}}\right)^2.
\label{eq:TBAY11oY22}
\end{equation}
This equation is also equivalent to the corresponding combination of TBA equations for fermionic nodes
as it is shown in \appref{sec:tba-y11-y22}. Note however that, as we see from  \eqref{eq:Y11Y22},  the functions  \(Y_{1,1}\) and \(1/Y_{2,2}\) are not independent but rather they are the values of the same function on two consecutive Riemann sheets. So the ratio or the product of \eqref{Y11Y22throughT} and \eqref{eq:TBAY11oY22} are enough to fix any  of them (up to a sign).

\paragraph{Equation for \(\nU \):}

One can find \(\nU \) from the left equality in \eqref{relatingBUC2}.
Again we get a finite difference equation which can be solved
 up to a constant \(\Lambda\)
as follows:
\begin{equation}\label{UthroughTfh}
        \log \nU=\log\Lambda+
        \log
        \frac{\hh}{f^+}+\frac{1}{
          2
        }\Psi*\rrho_c\;.
\end{equation}
Similarly to \eqref{eqf}, \eqref{UthroughTfh}
  is not sensitive to the choice between
  \(\irho_c=\log\frac{\bC(u+i0)}{\bar\bC(u-i0)}\) and \linebreak
  \(\rrho_c=\log{\bC(u+i0)}{\bar\bC(u-i0)}\) which differ by a  function holomorphic in the lower half-plane.
  Numerically, it is more convenient to use \(\rrho_c\) in \eqref{UthroughTfh} because
it quickly decreases to   zero when \(u\to\infty\).
We fix the constant \(\Lambda\)  in  \eqref{eq:Normalization}.
The yet unknown function \(\hh\) is determined below.

\subsubsection{Equations for \texorpdfstring{$\rrho$ and $\rrho_2$}{rho and rho2}}
\label{sec:equations-rho-rho_2}
From \eq{eqr} and \eq{Ghat} we get an equation expressing the
resolvent \(\hat G\), or equivalently,  the density \(\rrho\) through
the Y-functions \(Y_{11}\) and \(Y_{22}\)
\begin{align}
\frac{1+1/Y_{2,2}}{1+Y_{1,1}}
&=\frac{(1+\sK^+_1\cz\rrho-\tfrac{1}{2}\rrho)(1+\sK^-_1\cz\rrho-\tfrac{1}{2}\rrho)}
{(1+\sK^+_1\cz\rrho+\tfrac{1}{2}\rrho)(1+\sK^-_1\cz\rrho+\tfrac{1}{2}\rrho)}\;\;,\;\;u\in \hbZ_0\;,
\label{rightmagic0}
\end{align}
where \(\sK^+_1\cz\rrho\) denotes the principal part of the convolution
along the  interval \([-2g,2g]\).
Expressing     \(\rrho\)  from the terms without convolution in the r.h.s. of  \eqref{rightmagic0}    we can easily determine
   \(\rrho\) numerically, by iterations, as a function of \(\frac{1+1/Y_{2,2}}{1+Y_{1,1}}\).

 To fix  \(\rrho_2\) we use a similar ratio of Y-functions which can be expressed only through the T-functions of the upper band,   namely
 \begin{align}
\label{eq:uppermagicratio0}
   \frac{1+Y_{2,2}}{1+1/Y_{1,1}}&=\frac{\sT_{2,2}^+\sT_{2,2}^-}{\sT_{3,2}\sT_{1,1}^+\sT_{1,1}^-}\;,
 \end{align}
 where each T-function can be expressed through q-functions according to \eqref{sTofq}\footnote{For instance, one gets \(
   \sT_{1,1}^+=1+\sK^+_1\cz(\rrho_2 -q_3^{[+0]}\bar
   q_4^{[-0]}-q_4^{[+0]}\bar q_3^{[-0]})+\tfrac{1}{2}\rrho_2 +
   (q_3^{[+2]}-\frac 1 2 q_3^{[+0]})\bar q_4^{[-0]}+(q_4^{[+2]}-\frac
   1 2 q_4^{[+0]})\bar q_3^{[-0]}\).}. The explicit expression for the r.h.s. of
 \eqref{eq:uppermagicratio0} is  similar to the r.h.s. of
 \eqref{rightmagic0}, up to the substitution \(\rrho\to\rrho_2\) and to a
 number of additional terms involving \(q_3\) and \(q_4\).
As a consequence, \(\rrho_2\) is expressed as a function of
\(\frac{1+Y_{2,2}}{1+1/Y_{1,1}}\) and \(q_3\),\(q_4\). Note that
\(q_3\) and \(q_4\) are relatively small and vanish in the large volume -- the fact which is important for our numerical iterative procedure described below.

\subsubsection{Equation for \texorpdfstring{$\hat h$}{h}}
\label{sec:eqnhh}
To obtain this last equation needed to complete our FiNLIE, we employ a couple of Hirota equations in the \(\wT\)-gauge:
\begin{eqnarray}
\wT_{2,2}^+\wT_{2,2}^-&=&\wT_{3,2}\wT_{1,2}+\wT_{2,1}\wT_{2,3}\,,\\
\wT_{1,1}^+\wT_{1,1}^-&=&\wT_{1,0}\wT_{1,2}+\wT_{2,1}\wT_{0,1}\,.
\end{eqnarray}
We notice that in this gauge \(\wT_{0,1}=1\), \(\hat\wT_{2,s}=\hat\wT_{1,1}^{[+s]} \hat\wT_{1,1}^{[-s]}\),
\(\wT_{3,2}=-{\cal F}^+\wT_{2,3}\), \(\wT_{1,0}=-Y_{1,1}Y_{2,2}{\cal F}^+\) and the magic and mirror functions are related for \(u\in\hbZ_0\) by \(\wT_{2,3}=\hat\wT_{2,3}\), \(\wT_{2,2}(u\pm\frac i2)=\hat\wT_{2,2}(u\pm\frac i2\mp i0)\),  \(\wT_{1,1}(u\pm\frac i2)=\hat\wT_{1,1}(u\pm\frac i2\mp i0)\). Using all these properties and excluding \(\wT_{2,1}\) from both  equations we get
\begin{equation}
\hat \wT_{1,1}^{[-1-0]}
\hat \wT_{1,1}^{[+1+0]}
+{\cal F}^+ \wT_{1,2}=\hat\wT_{1,1}^{[-1+0]}\hat\wT_{1,1}^{[+1-0]}+Y_{1,1}Y_{2,2}{\cal F}^+\wT_{1,2}\,,\ \ u\in\hbZ_0.
\end{equation}
Inserting the explicit  parameterization \eqref{MHir}
 \(\hat\wT_{1,s}=\hat h^{[+s]} \hat h^{[-s]}(s+\hat G^{[+s]}-\hat G^{[-s]})\)
we get the following equation on \(\hat h\):
\begin{equation}\label{discontinuitiyh}
\hat h(u+i0)\hat h(u-i0)=\frac{{\cal F}^+(1-Y_{1,1}Y_{2,2})}{\rrho},\ \ u\in \hbZ_0\,.
\end{equation}

We know that \(\hh\) has only one \(\hbZ_0\)-cut. Asymptotically, for large volume   \(\hh^2\) behaves at large \(u\) as \(u^{-L-2}\) (see \appref{app:asymptoticsolution}),
 and hence the monodromy of \(\hh^2\)
should be trivial for the  closed paths surrounding its single Z-cut.
This topological property is unlikely to be changed for a finite size
and therefore the large \(u\) behavior of \(\hh\) should
be the same as in the asymptotic limit.
Moreover, since in the asymptotic limit \(\hh\) has  no poles or zeroes
we conclude that at least for sufficiently large but finite \(L\) (or small but finite \(g\))
the discontinuity equation (\ref{discontinuitiyh}) uniquely fixes it as follows:
\begin{equation}\label{hhat_sol}
        \log \hh=-\frac{L+2}{2}\log{\hat x}+
        {\cal Z}\cz\log\left(\frac{\CF^{+}(1-Y_{1,1}Y_{2,2})}{\rrho}\right).
\end{equation}

We remind that the function \(\CF=\sqrt{\bT_{0,0}}\)
can be  written in terms of Q-functions. Alternatively we can use \eq{eq:discf} to write it in terms of the Y-functions \(Y_{1,1}, Y_{2,2}\)
which is a very convenient equation for numerics (see \appref{sec:finliedetails}).

This concludes the derivation of FiNLIE for the exact 
anomalous dimensions of the \(\sl(2)\) sector's
symmetric states in
AdS\(_5\)/CFT\(_4\). Our derivation uses only the properties of $\bT$-
and $\wT$-functions summarized in section \ref{sec:prop}, which can be
either  derived from TBA integral equations or postulated, from the explicit knowledge of large volume solution given in appendix \ref{app:asymptoticsolution} (weak coupling solution is enough as well), and an assumption, which can be verified in every explicit computation, that qualitative structure of poles and zeroes does not change if to compare with large volume expressions.

\section{List of FiNLIEs}
\label{sec:FiNLIEresume}

In this section we will collect together the results of  previous sections into the full list of FiNLIE.
To make the formulas a bit more compact we will employ here  exponential  notations for the convolutions. Namely, by definition, for any kernel \(\Xi\) and a function \(f\) we will define the exponential of convolution as
\begin{equation}
f^{*\Xi}\equiv\exp(\Xi*\log f).
\end{equation}

The equation for \(Y_{1,1}\) ( obtained as a sum of \eqref{Y11Y22throughT} and  \eqref{eq:TBAY11oY22}) now reads\footnote{The overall minus sign in  (\ref{Y11fin}) is not visible from  \(Y_{1,1}Y_{2,2}\) and \(Y_{1,1}/Y_{2,2}\). It is chosen so that \(Y_{1,1}\) and \(Y_{2,2}\) are positive on \([-2g,2g]\). The square root \(\sqrt{4g^2-v^2}\) in \(\CZ(u,v)\check*\) is evaluated slightly above real axis.  }
\begin{equation}\label{Y11fin}
Y_{1,1}=
-\sqrt{\frac{R^{(+)}}{R^{(-)}}\frac{B^{(-)}}{R^{(+)}}}\frac{\CT_{1,2}}{\sT_{2,1}}
\left(\frac{\sT_{1,0}}{Q^+Q^-}\right)^{1+{\check *}\CZ}\left(\frac{Q^2}{\sT_{0,0}}\right)^{*\frac 12(\CZ_1+\CK_1)}\left(\frac{\sT_{1,1}}{\cT_{1,1}}\right)^{*\slash\!\!\!\CK_1}\,.
\end{equation}
The equation for \(Y_{2,2}\) is essentially the same  since
\(Y_{2,2}\) is simply the analytic continuation of \(Y_{1,1}\) under the \({\bf Z}\)-cut: \(Y_{2,2}(u + i0)=1/{Y_{1,1}(u- i0)}\).
The \(\CT\)-functions in the r.h.s. of
\eqref{Y11fin} are given by
\begin{equation}
\label{eq:rho-to-CF}
      \hat\CT_{1,s}=s+{\cal K}_s{\hat*}\rrho
\end{equation}
and the equation for $\rho$ in terms of $Y_{1,1}$ and $Y_{2,2}$ is
\begin{align}
\frac{1+1/Y_{2,2}}{1+Y_{1,1}}
&=\frac{(1+\sK^+_1{\hat*}\rrho-\tfrac{1}{2}\rrho)(1+\sK^-_1{\hat*}\rrho-\tfrac{1}{2}\rrho)}
{(1+\sK^+_1{\hat*}\rrho+\tfrac{1}{2}\rrho)(1+\sK^-_1{\hat*}\rrho+\tfrac{1}{2}\rrho)}\,,
&u&\in[-2g,2g]\,.
\label{rightmagic}
\end{align}
The upper band \(\sT\)-functions are expressed through the  \(q\)-functions and  the LR wing exchange ``gauge'' function \(\nU(u)\) using \eqref{sTofq} and \eqref{TUUT}:
\begin{eqnarray}
        \sT_{a,+1}&=&q_1^{[+a]}\bar q_2^{[-a]}+q_2^{[+a]}\bar
        q_1^{[-a]}+ q_3^{[+a]}\bar q_4^{[-a]}+ q_4^{[+a]}\bar
        q_{3}^{[-a]}\;,
\label{Stofq}
\\
                 \sT_{a,0}&=& q _{12}^{[+a]}\bar q _{12}^{[-a]}+ q _{34}^{[+a]}\bar q _{34}^{[-a]}-q _{14}^{[+a]}\bar q_{14}^{[-a]}-q _{23}^{[+a]}\bar q_{23}^{[-a]}-q_{13}^{[+a]}\bar q_{24}^{[-a]}- q_{24}^{[+a]}\bar q_{13}^{[-a]},\\
\sT_{a,-1}&=&\left(\nU^{[+a]}\bar \nU^{[-a]}\right)^2\sT_{a,1}\;,
\label{sTofq2}
\end{eqnarray}with all the \(q\)-functions being parameterized through a base of 5 of them:

\begin{equation}\label{q's}
q_1=1,\qquad  q_2=P+{\cal K}*\tilde\rrho_2\;\;,\;\;\qquad  q_{12}=\tilde Q\,,\qquad q_{123}=U^2\,,\qquad q_{124}=U^2q_{2}\,,
\end{equation}
where for the Konishi state we introduce two polynomials 
\(\tilde Q=(u-\tilde u_1)(u+\overline{
\tilde u_1})\),\ \(P=-iu\).\footnote{
For the case of more than two symmetric roots we expect these polynomials to be
\(\tilde Q(u)=\prod_{j=1}^M(u-\tilde u_j)\), \(P(u)=-\frac{iu}{M-1}\prod_{j=1}^{M-2}(u-v_j)\)
} The value of \(\tilde u_1\) is determined by the condition that \(\sT_{1,0}^+\) should have zeroes at position of the Bethe roots.

The other \(q\)'s can be found through the set of Pl\"ucker relations \cite{Tsuboi:2009ud,Kazakov:2007fy}
\begin{eqnarray} q_{\es} q_{ij}&=& q_i^+ q_j^-- q_j^+ q_i^-,\label{Plucker1b}\\
 q_{ijk} q_i&=& q_{ij}^+ q_{ik}^-- q_{ik}^+ q_{ij}^-.\label{Plucker_final}
\end{eqnarray}

The density \(\tilde\rrho_2\) in \eq{q's} is represented as \(\tilde\rho_2=\rho_2-q_3^{[+0]}\bar q_4^{[-0]}-q_4^{[+0]}\bar q_3^{[-0]}\), where \(\rho_2\) has a finite support on \(\hbZ_0\) and the remainder is a small correction  fixed from the consistency with (\ref{Plucker1b}) and (\ref{Plucker_final}). \(\rho_2\) is then expressed
through \(Y_{1,1}\) and \(Y_{2,2}\)  as follows:
\begin{eqnarray}
\label{eq:uppermagicratio}
\frac{1+Y_{2,2}}{1+1/Y_{1,1}}&=&
\frac{\sT_{2,2}^+\sT_{2,2}^-\sT_{1,0}}{\sT_{1,1}^+\sT_{1,1}^-\sT_{3,2}}\,,\qquad u\in[-2g,2g]\,,
\end{eqnarray}
where \(\sT_{a,2}=q_\es^{[+a]}\bar q_\es^{[-a]}\).

The function \(\nU(u)\) is defined through  \eqref{UthroughTfh} with the spectral densities expressed through Y- and T-functions by means of \eqref{Bdefin},\eqref{eqf},\eqref{hhat_sol}:
\begin{equation}
        \nU^2=\frac{\L^2\sT_{00}^-}{\hat x^{L-2}Y_{1,1}Y_{2,2}\sT_{1,0}}
       \left(\frac{{1-Y_{1,1}Y_{2,2}}}{\rrho/{\cal F}^{+}}\right)^{\cz{2\CZ}}\left(\frac{\sT_{2,1}\CT_{1,1}^-}{\hat \sT_{1,1}^-\CT_{1,2}Y_{2,2}}\right)^{*2\Psi}\;\;,\qquad\IM u>0\label{eq:U}
\end{equation}
 where \(\Lambda\) is defined  by  relations \eqref{ImNorm},\eqref{eq:Normalization}.

The \(M\) Bethe roots are fixed by the exact Bethe equations which can be written as:
  \begin{equation}\label{BetheEQfin}
   \left(\frac{\hat x^-}{\hat x^+}\right)^{L+2}=-
\frac{Y_{2,2}^+}{Y_{2,2}^-}\frac{\CT_{1,2}^+}{\CT_{1,2}^-}\frac{\hat
  \CT_{1,1}^{[-2]}}{\hat \CT_{1,1}^{[+2]}} \left(
\frac{\rrho/{\cal F}^+}{{1-Y_{1,1}Y_{2,2}}}
\right)^{\cz 2 (\CZ^+-\CZ^-)}
\,\, ,\qquad \text{at}\,\,u=u_j\,.
\end{equation}

In \eqref{eq:U} and \eqref{BetheEQfin}, \({\cal F}\) is obtained from
\eqref{Fexplicit} as
\begin{align}
\label{eq:Fsummary}
  {\cal F}&=\Lambda_{\CF}
  \left(Y_{1,1}(v)Y_{2,2}(v)\right)^{\cZ \frac{\tanh\pi(u- v)+{\rm
        sign}(v)}{2i} }
\prod_{i=1}^M\sinh(\pi(u-u_i))\,,
\end{align}
where \(\L_\CF\) is just a constant expressed as \eqref{eq:LambdaF} in \appref{sec:fixing-norm-const}.

Finally, the energy of the state can be then found from the large \(u\)
asymptotics of the product of fermionic Y-functions:
\begin{equation} \log Y_{1,1}Y_{2,2}= \frac{iE}{u}+{\cal O}(1/u^{2})\,. \end{equation}

\section{Numerical implementation of FiNLIEs}
\label{sec:numer-impl-finl}
\begin{figure}[t]
\begin{center}
\begin{overpic}[scale=0.6]{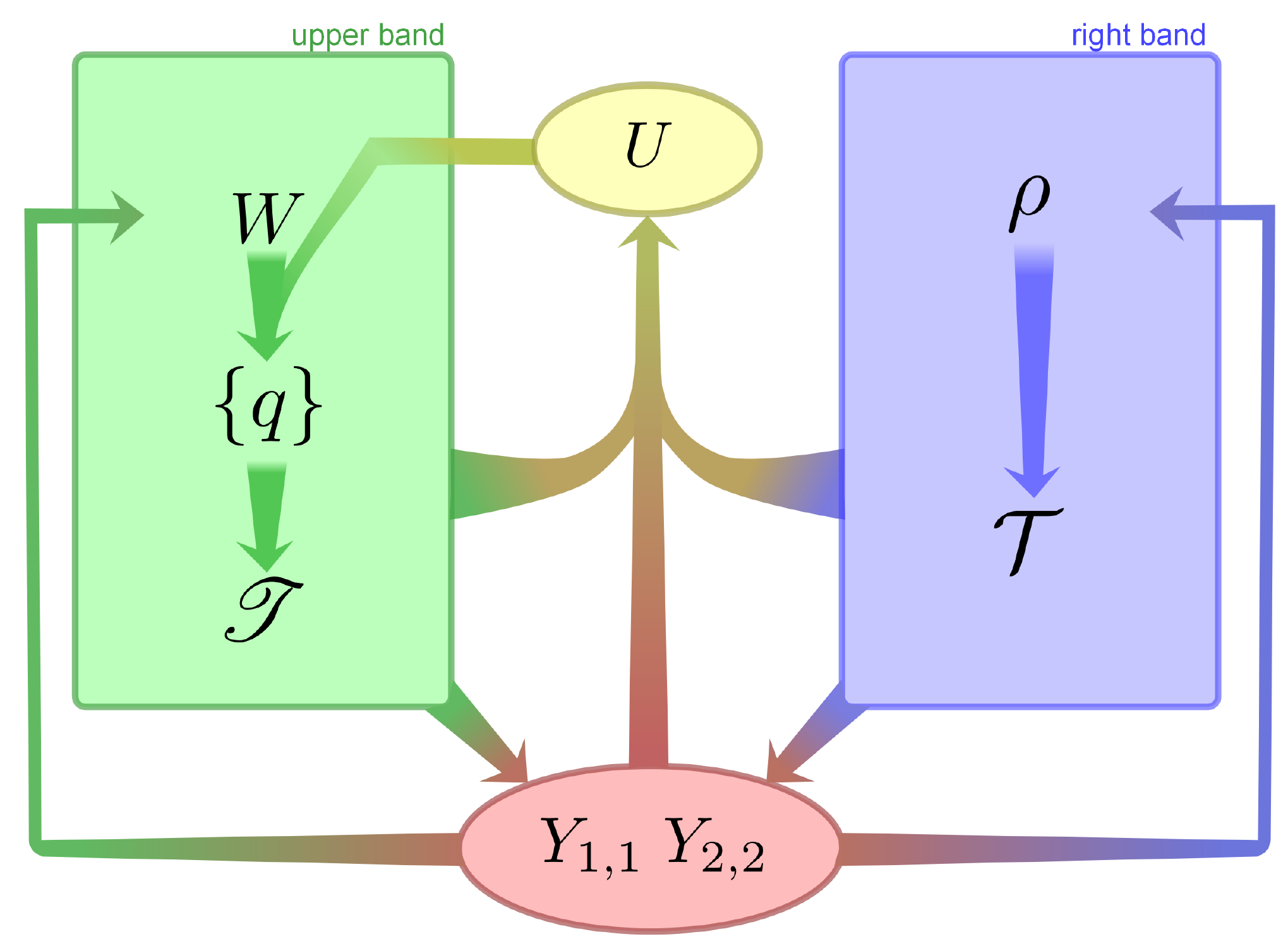}
    \put(-1,45){\textsf{\makebox[215mm][c]{
        \small \textsf{(\ref{eq:rho-to-CF})
        }
    }}}
    \put(15,45){\textsf{\makebox[215mm][c]{
        \small \textsf{(\ref{rightmagic})}
    }}}
    \put(-88,45){\textsf{\makebox[215mm][c]{
        \small \textsf{(\ref{eq:uppermagicratio})}
    }}}
    \put(-73,49){\textsf{\makebox[215mm][c]{
        \small \textsf{
          (\ref{Plucker1b},~\ref{Plucker_final})
        }
    }}}
    \put(-70.5,35){\textsf{\makebox[215mm][c]{
        \small \textsf{(\ref{sTofq2})}
    }}}
    \put(-41,50.5){\textsf{\makebox[215mm][c]{
        \small \textsf{(\ref{eq:U})}
    }}}
   \put(-47,18){\textsf{\makebox[215mm][c]{
        \small \textsf{(\ref{Y11fin})}
    }}}
   \put(-26,18){\textsf{\makebox[215mm][c]{
        \small \textsf{(\ref{Y11fin})}
    }}}

\end{overpic}
\end{center}
\caption{\textbf{Algorithm for our FiNLIE implementation.} \label{Alg}}
\end{figure}
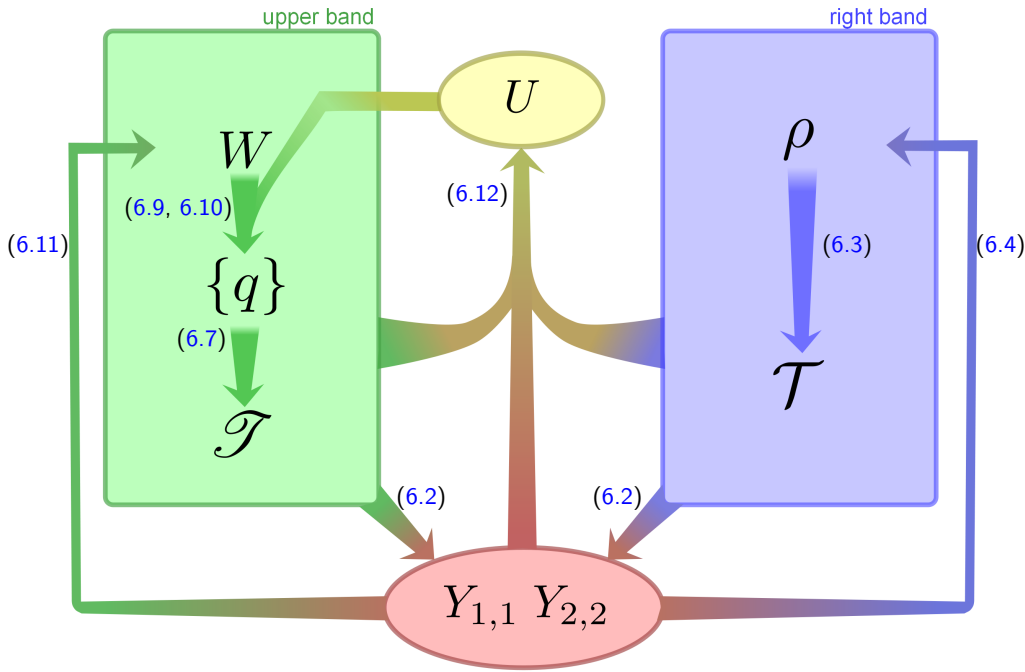

Using the system of FiNLIE  derived in the last two sections, the spectral problem can be solved iteratively for the densities of the parameterization 
(\ref{Ghat},\ref{Param}). We performed   the numerical computations for the Konishi operator characterized by two symmetric Bethe  roots \(u_1=-u_2\).
We expect the generalization to the other operators, with a larger number of roots, to be straightforward.

Let us denote as \(X=(\rrho,\nU, W,
\{\tilde u_j\},\{u_j\})\) the set of
parameters characterizing a solution which we want to find numerically.  It consists of two densities on a short  \(\hbZ\)-cut (the
density \(\rrho\) defined in \eqref{Gdef}
and the density \(\rrho_2\) which parameterizes \(W\) as in
\eqref{Wspec}), one function \(U\) (which can be parameterized from
knowing only its value on the real axis\footnote{The large-\(u\) behavior of
\(\nU\) is given by \(\nU\sim u^{(-L-\gamma)/2}\) (see
\eqref{eq:asUas} and the discussion at the end of
\appref{app:asymptoticsolution}),
allows us to write a Cauchy-kernel representation for it as follows
  \begin{align*}
\label{URhoU}
    \nU^2 =-\frac 1 {{\hat x}^{{L + \gamma} -1}} ~ \frac 1 {
\vphantom{{\hat x}^{\frac{L + \gamma} 2 -1}}
2 \pi i} \int_{-\infty}^{+\infty} \frac
    {\rhou(v)}{u-v}\mathrm{d}v\,,\qquad
{\rm where} \quad\rhou(u)=2\, \RE\left(\nU^2 \cdot {\hat x}^{
    {L + \gamma}
    -1} \right)
 \end{align*}
}), a set of     Bethe roots \(\{u_i\}\), and another set of roots \(\{\tilde u_i\}\) of the polynomial
\(\tilde Q\) from \eqref{Param} ( for the Konishi state \(i=1\) or 2, \(u_1=u_2\) and 
\(\tilde u_1=-\overline{\tilde u_2}\)).

The numerical solution of the FiNLIE system is achieved
using the
fixed point approach: first, the equations are put into a form \(X=F(X)\) which allows us to build a sequence of approximations \(X_{n+1}=F(X_n)\)  converging to the exact solution.
The solution is then reached by a large number of iterations, each of
them consisting of two main steps (see \figref{Alg}):
 \begin{itemize}
 \item On the first step, the T-functions and the fermionic Y-functions \(Y_{11},Y_{22}\) are
  computed for given  densities and Bethe roots:
  \begin{enumerate}
  \item The upper-band Q-functions are expressed in terms of \(\nU\),
    \(W\) and \(\tilde Q\) using the Pl\"ucker relations
    (\ref{Plucker1b},\ref{Plucker_final}).
It is  clear indeed that (\ref{Plucker1b},\ref{Plucker_final}) allow us
to write all the Q-functions in terms of the five functions given in
\eqref{q's}, (by essentially the same steps as at the end of
\secref{sec:upper-band}).
Numerically, the main tricky point is to invert the difference equations:
for instance \(q_{13}\) is obtained from
\begin{align}
  \nU^2\frac{{q_1}^2}{q_{12}^+q_{12}^-}&=\left(\frac{q_{13}}{q_{12}}\right)^--\left(\frac{q_{13}}{q_{12}}\right)^+\,.
\end{align}
The l.h.s. behaves as \(u^{-L-\gamma-2M}\) when \(u\gg 1\), hence \(\frac{q_{13}}{q_{12}} =  \sum_{k\geq 0}\left(\nU^2\frac{{q_1}^2}{q_{12}^+q_{12}^-}\right)^{[2k+1]}
\), where the sum converges very fast. It is  also possible
to rewrite this equation as an integral equation,
 similarly to \eqref{eq:log-f2sim-1rrh}.
At the end of this first step, once all these Q-functions are obtained,  one computes \(\sT\)'s
    using the equations (\ref{Stofq}-\ref{sTofq2});

  \item The right band \(\CT\)-functions are expressed in terms of \(\rrho\),
    using \eqref{eq:rho-to-CF};
  \item Then it is possible to compute \(Y_{1,1}\) and \(Y_{2,2}\) using
    equations (\ref{Y11Y22throughT},\ref{eq:TBAY11oY22}). To do this,
    we use the  \(\sT\)-functions, the \(\CT\)-functions, and the position of  Bethe
    roots.
\setcounter{savecounter}{\value{enumi}}
    \end{enumerate}
  \item On the second step, the set of parameters \(X\) is expressed from
    these T-functions and the fermionic Y-functions:
    \begin{enumerate}
      \setcounter{enumi}{\value{savecounter}}
    \item First, \({\tilde u_j}\) is obtained by the requirement that
      \(\sT_{1,0}(u_j+i/2)=0\);
    \item Then the density \(\rrho\) is extracted from \(Y_{11}\) and \(Y_{22}\)
      using {\eqref{rightmagic}}. In the same way \(W\) is extracted
      from \(Y_{11}\) and \(Y_{22}\) using
      (\ref{eq:uppermagicratio});
    \item The function \(\nU\) is then found from
      \eqref{eq:U}, with \(\CF\) fixed from \eqref{eq:Fsummary};
    \item \label{item:1}  Eq.~\eqref{BetheEQfin}
      is used to express
      the positions of the Bethe roots \(\{u_j\}\).
    \end{enumerate}
 \end{itemize}

Finally, for the initialization point \(X_0\)
one can either use the asymptotic solution, which
works perfectly for small \(g\)'s,
or extrapolate \(X\) from  smaller values of \(g\).

\paragraph{Numerical results:}
\label{sec:numeric-results}

Our, rather preliminary, numerical realization  of FiNLIE's gives for
the Konishi state the results written in table \ref{tab:KonishiNumericEnergy}.

\begin{table}
  \centering
  \begin{equation*}
  \begin{array}{|c|c|c|c|c|}
\hline
    g&\textrm{\(E_{\rm ABA}\)}&
\textrm{\(E_{\rm TBA}\) \cite{Gromov:2009zb}} & \textrm{\(E_{\rm FiNLIE}\)} &
\textrm{numerical error}\\
 \hline
0.25&4.6154 &4.6147&4.615(7) &10^{-3} \\
0.5&5.7362&5.7120&5.712(9)&10^{-3} \\
1&7.7285&7.6044&7.60(53)&10^{-3}\\
1.6&9.5749&9.3874&9.38(23)&5\cdot 10^{-3} \\
\hline
  \end{array}
\end{equation*}
\caption{Numerical energy of the Konishi state, for a few
 values of the coupling constant g. The outcome of our FiNLIE
 iterations is compared to the previous TBA-iterations [22] and to the
 prediction of the Asymptotic Bethe Ansatz (ABA).}
\label{tab:KonishiNumericEnergy}
\end{table}

These results are in perfect agreement with earlier numerical study, using TBA approach,
of the  Konishi state \cite{Gromov:2009zb} (confirmed by a  similar numerical
study of TBA  in \cite{Frolov:2010wt}): when \(g\leq 1\), the numerical
precision is essentially the same as in \cite{Gromov:2009zb}.

For a given state which we take as Konishi state here, our numerics
showed the convergence of the algorithm%
\footnote{For Konishi state, it is checked numerically at least in the range \(0\leq g\leq 2\).}
to the
exact solution of all the
above-written equations, and this solution coincides with the
previous numerical studies from the original TBA equations.
We notice however that numerically it is more efficient to use
the old exact Bethe equation derived in \cite{Gromov:2009tv} rather then \eqref{BetheEQfin}.

In \figref{fig:Yfuns}
we can see that the Y-functions \(Y_{a,0}\), \(Y_{1,1}\) and \(Y_{2,2}\)
obtained from FiNLIE iterations (dots) coincide with a good
precision with the data of TBA-iterations (solid curve). We can
also see a sensible deviation from the asymptotic limit
of these Y-functions (dashed curve).

 For Konishi state, the plots of various  densities resulting from the
 iteration procedure are presented in \figref{fig:FinDens}. The densities \(\rho\), \(\rho_2\) and \(\rho_U\) (black curve) are
compared to their asymptotic values
\eqref{rhoforKonishi},~\eqref{eq:rho2konishi},~\eqref{largeuhh}
(dashed gray
curve). We see that at \(g=1.6\), the densities are already quite
different from their asymptotic value. In particular,
\figref{fig:dt2}
shows that the density \(\rho_2\) sensibly deviates from its asymptotic
semi-circle shape.

In conclusion, our preliminary numerical results confirm the validity of our FiNLIE system. We plan
to improve the precision and speed of our numerical procedure in the near future.

{
\begin{figure}
  \centering
  \subfloat[Middle nodes Y-functions (\(Y_{a,0}\))]
{\label{fig:Ymid}\includegraphics[width=0.45 \textwidth]{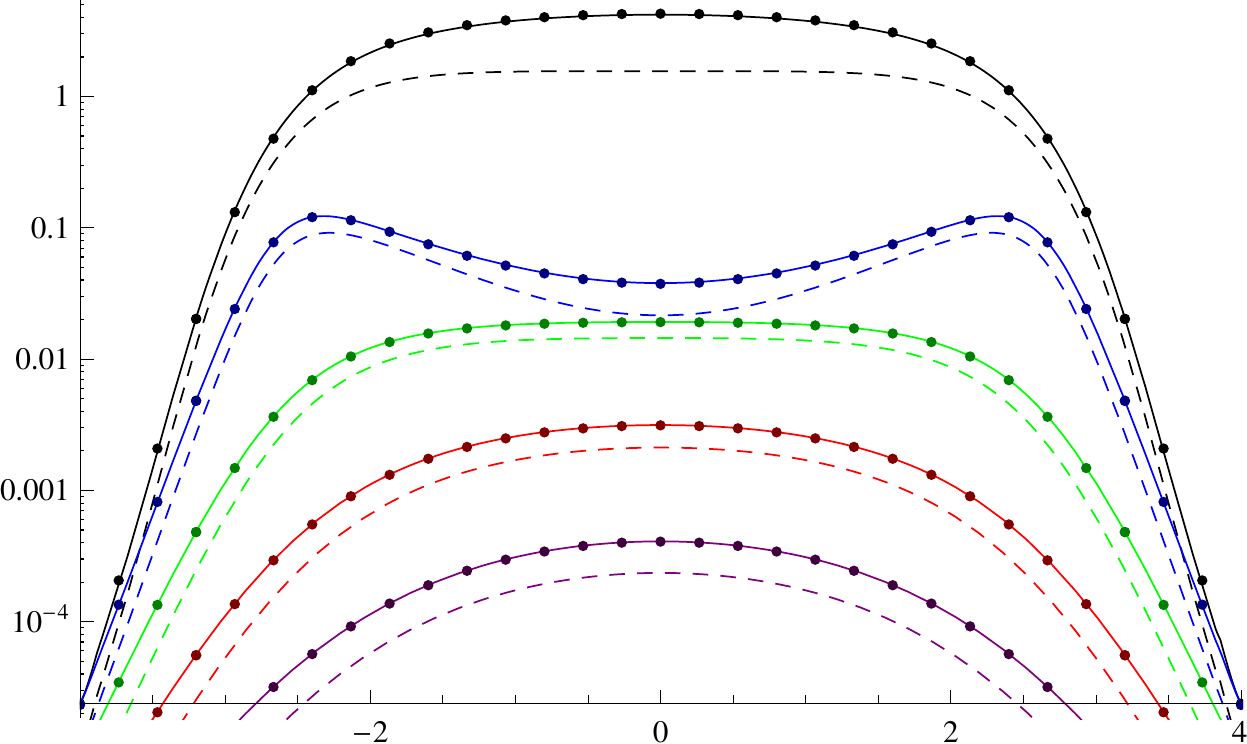}} \qquad
  \subfloat[Fermionic Y-functions (\(Y_{1,1}\) and \(1/Y_{2,2}\))]
{\label{fig:Yf}\includegraphics[width=0.45 \textwidth]{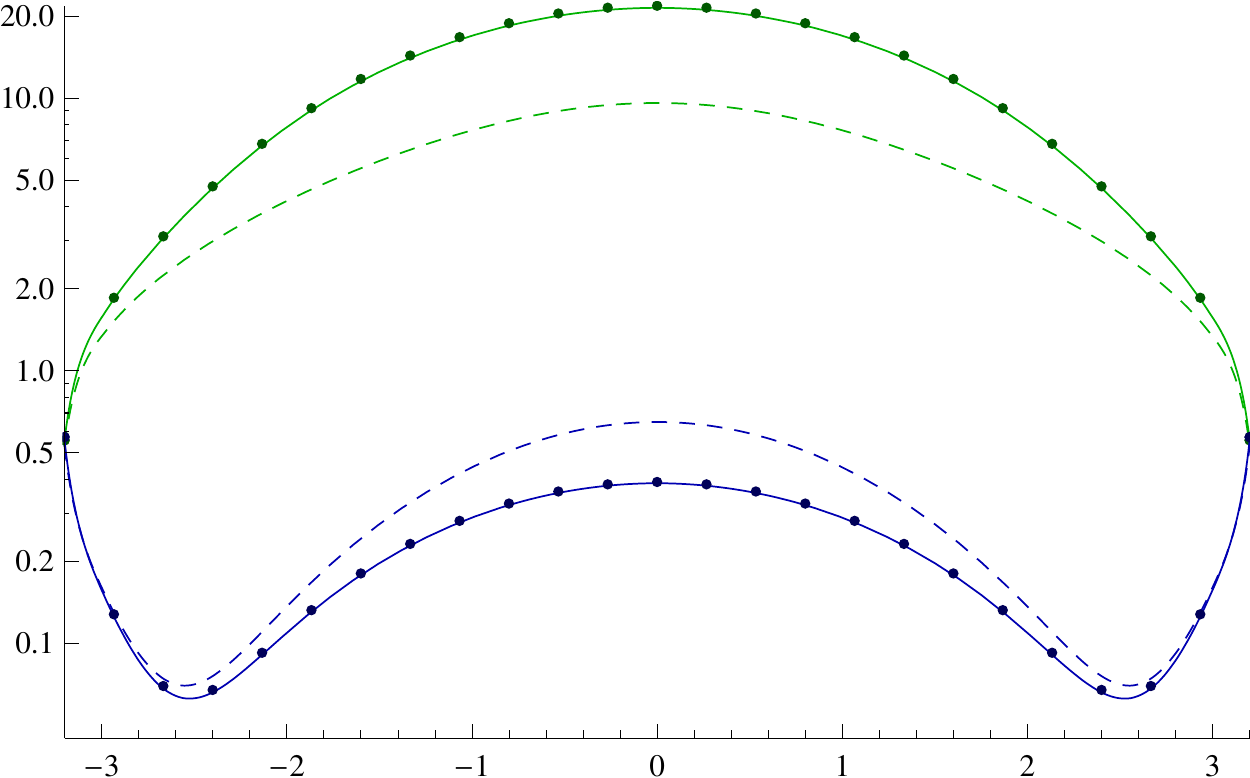}}
  \caption[Numerical Y-functions for Konishi state]{\textbf{Numerical Y-functions for Konishi state at g=1.6 :}
 The Y-functions obtained from FiNLIE  iterations (dots) are compared
 to the outcome of TBA-iterations  \cite{Gromov:2009zb}
 (solid lines) and to the ABA expression (dashed lines).
 The figure~\ref{fig:Ymid} shows \(Y_{1,0}\) (black), \(Y_{2,0}\) (blue), \(Y_{3,0}\)
 (green), \(Y_{4,0}\) (red) and \(Y_{5,0}\) (violet), while the
 figure~\ref{fig:Yf} shows \(Y_{1,1}\) (green) and its continuation
 \(1/Y_{2,2}\) (blue).}
  \label{fig:Yfuns}
\end{figure}

\begin{figure}
  \centering
  \subfloat[Density \(\rrho\) when \(g=1.6\)]
{\label{fig:dr2}\includegraphics[scale=0.35]{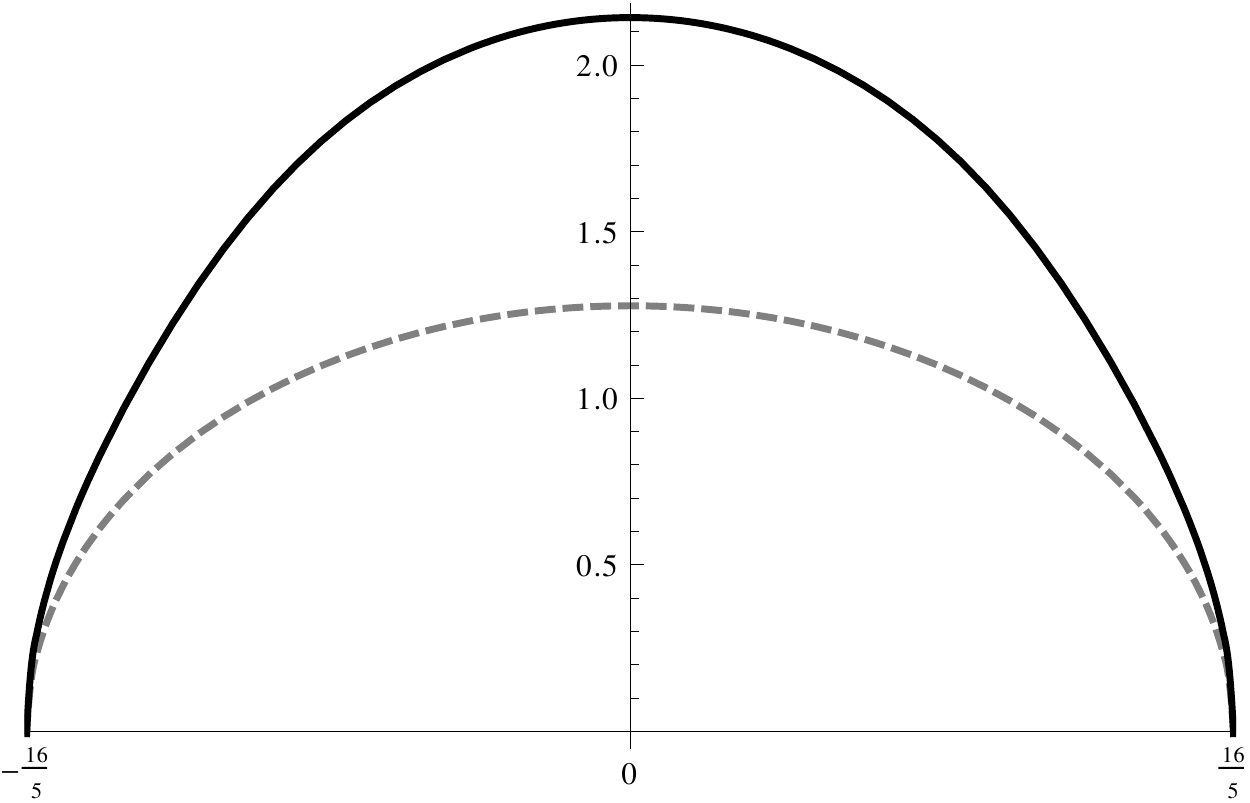}} \qquad
  \subfloat[Minus density \(\rrho_2\) when \(g=1.6\)]
{\label{fig:dt2}\includegraphics[scale=0.35]{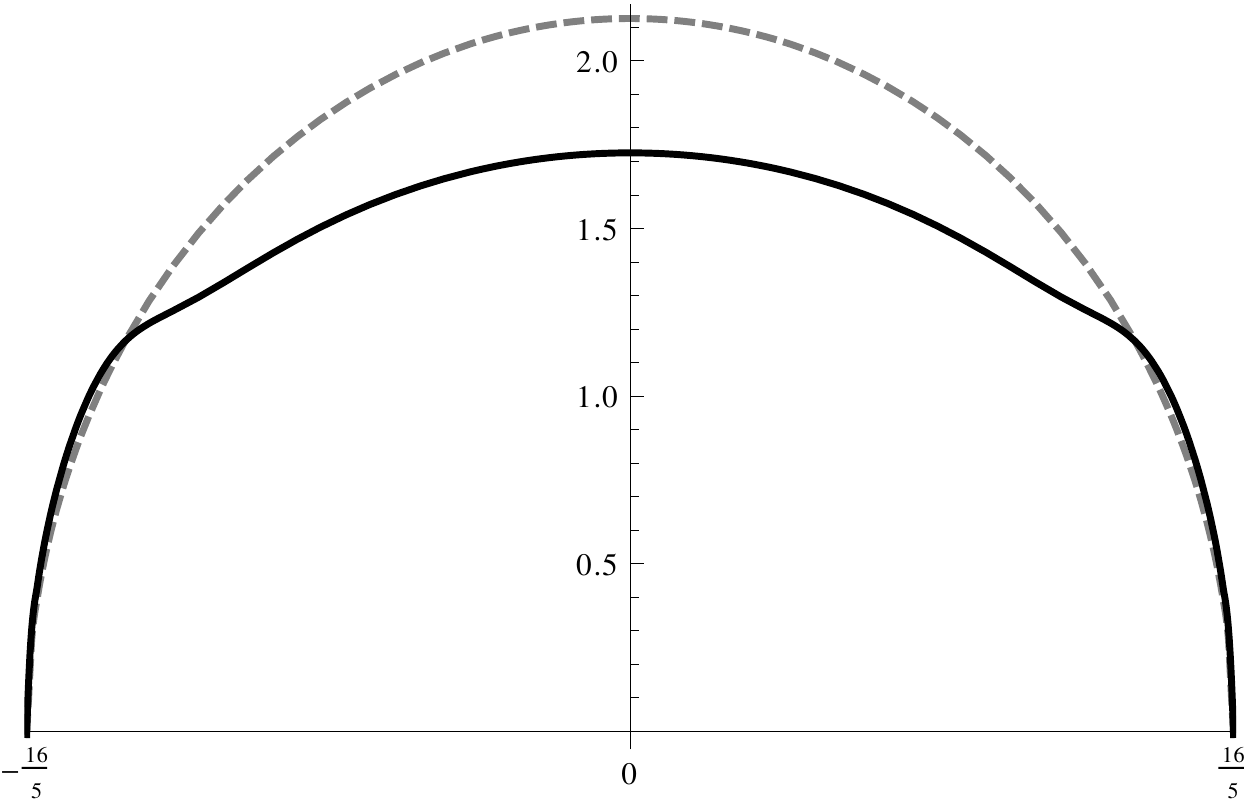}} \qquad
  \subfloat[Density \(\rhou\)  when \(g=1.6\)]
{\label{fig:du2}\includegraphics[scale=0.35]{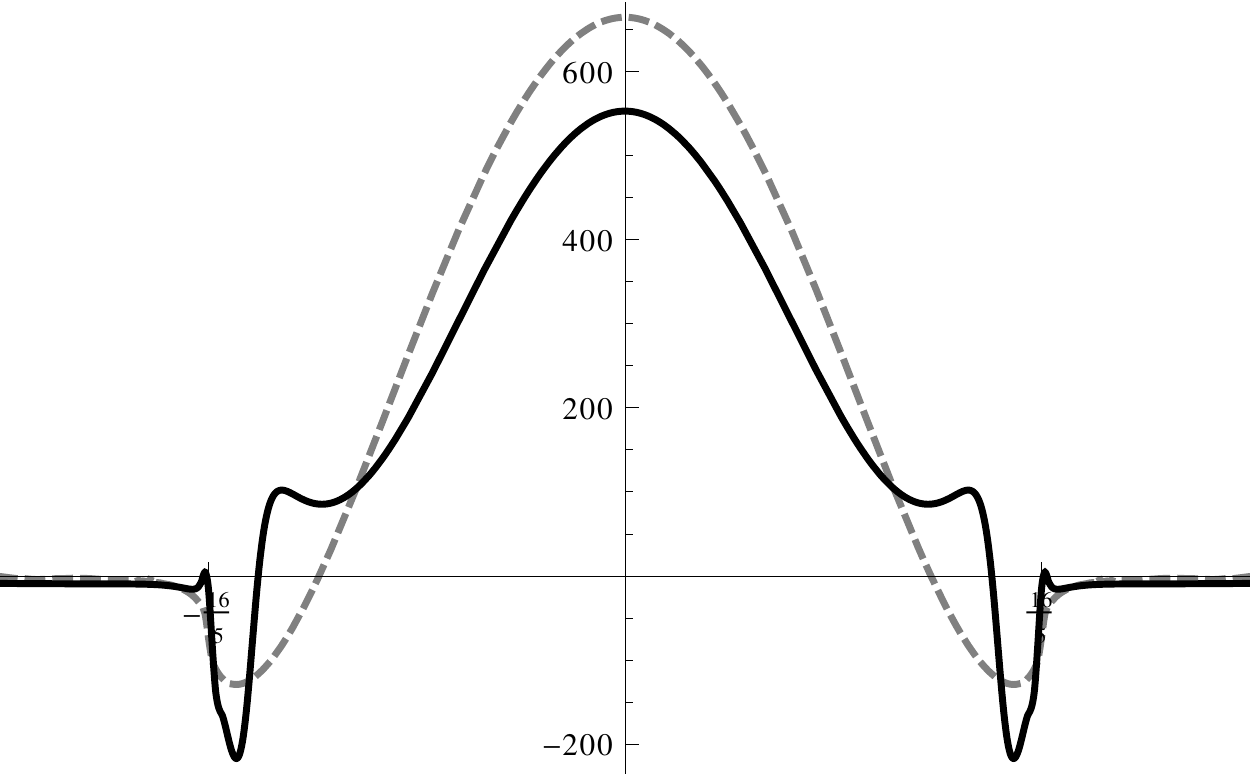}}\\
  \caption[Numerical densities for Konishi state]{\textbf{Numerical densities obtained for Konishi state by our FiNLIE
      algorithm:} In figure (a) the density \(\rho\) describing
    the right band is plotted (black curve) and compared to its asymptotic expression
    (\ref{rhoforKonishi}, dashed gray curve). The same comparison is presented
  in figure (b) for \(-\rho_2\) and in figure (c) for
  \(\rho_U\).}
  \label{fig:FinDens}
\end{figure}
}

\section{Conclusions}

In this paper we have shown that using the integrable Hirota dynamics, together with a few relatively simple and natural  assumptions about the analyticity properties of T-functions (parameterizing the Y-functions)  we can
transform the  AdS\(_5\)/CFT\(_4\)   Y-system  into a finite set of non-linear integral equations (FiNLIE).

Our FiNLIE  remotely resembles, at least  in spirit,   the famous  Destri-de Vega equations known in the literature for some other  2D QFT's. But the method of derivation,
first proposed in the context of the \(SU(N)\) principal chiral field (PCF) in \cite{Gromov:2008gj}, is conceptually very much different from the original approach of Destri and de Vega\footnote{though the resulting equations can be identical in some simplest cases, like Sine-Gordon or Gross-Neveu models, see \cite{Gromov:2008gj}.}. The relation between two approaches still needs to be clarified, though some interesting results on this way are obtained for the principal chiral field model
in \cite{Caetano2010}. It is relatively easy to formally solve the integrable Hirota equations in \(\wT\)-hook in terms of a few ``boundary'' functions,  which is done by means of the Wronskian determinant formulas \cite{Gromov:2010km,Krichever:1996qd}. However, it is much more difficult to establish the right analyticity properties of those functions: asymptotic conditions, analyticity strips, analytic continuation through Z-cuts present in all T-functions, etc., and thus fix the full physical solution.

 As for the case of the PCF, we observe that the analytic properties of T-functions can be reasonably simple (i.e. exhibiting well established analyticity strips) only in a well chosen gauge which is different for right, left and upper bands of the \(\wT\)-hook.

Our fundamental observation, mentioned already in \cite{Gromov:2010km}
but clearly understood and employed only in the current paper, is the fact
that the \(\mathbb Z_4\) symmetry of the string coset model, explicit
in the classical system, can be promoted to the quantum
level as a symmetry under the  analytic continuation of the
T-functions w.r.t. their representational indices \((a,s)\).  For a given T-function, this is
only possible  on a certain sheet of the  Riemann surface having only short Z-cuts. We call it the {\it magic} sheet. The
  \(\mathbb Z_4\) symmetry, the quantum analogue of unimodularity (quantum determinant\(\,=1\)), and a few other natural assumptions allow us to fix the analytic properties of each T-function in the appropriate gauge, as well as the transitional gauge functions relating T's for three different bands of the \(\wT\)-hook. Only then, knowing the whole structure of all T-functions, we have enough of analyticity input  not only to parameterize all T-functions and the gauge transition functions but also to constrain them by additional Riemann-Hilbert-type equations, thus closing the whole system of FiNLIE.

Our system of FiNLIE (summarized in \secref{sec:FiNLIEresume})  shows that we achieved our conceptual goal of getting a finite system of equations out of a  set of analyticity properties of T-functions. Probably the actual realization is still  perfectible and  we only begin to see the hidden ``simplicity'' of the whole problem. It is even conceivable that there exists a ``quantum spectral curve'' of an infinite genus, with  infinite number of sheets connected by Z-cuts, uniformizing the analytic structure of all basic functions (presumably Q-functions) of the model at once.

We also demonstrated here that our FiNLIE is a useful tool for the computation of the spectrum. Our first, preliminary numerical implementation of the FiNLIE (which, with a certain effort, can be certainly  improved in efficiency and precision) allows  to check that, within  reasonable error
margins, our numerics reproduces the known results  \cite{Gromov:2009zb}, thus perfectly confirming the correctness of our
FiNLIE.  We also  proved here the equivalence of our FiNLIE to the TBA equations of \cite{Bombardelli:2009ns,Gromov:2009bc,Arutyunov:2009ur,Gromov:2009zb} analytically (the paper \cite{Cavaglia:2010nm} was especially useful for that).

Our method rendering a finite system of equations for the AdS/CFT spectrum  can be certainly generalized to
any  state of the model but
 the details for the other states  still have to be worked out.  We also hope that our FiNLIE will allow in the nearest future an efficient  way for constructing the systematic weak coupling expansion.
  It will be more difficult to do the same for the strong coupling, thus making it possible an efficient higher loop calculation in the string sigma model, but our FiNLIE could hopefully provide us with some clues also for that.
 It is especially important in view of recent analytic results
 obtained by a very different method at strong coupling \cite{Gromov:2011de,Roiban:2011fe,Vallilo2011,Beccaria:2011uz,Gromov:2011bz}. Indeed  FiNLIE, unlike the TBA, provides an extensive knowledge about the analytic properties of underlying functions on the whole magic sheet.
 It should be also easier now to attack another important limit of the theory --- the BFKL approximation.
 Hopefully FiNLIE will allow  to re-derive the known leading  \cite{Kotikov2003} and next-to-leading  \cite{Balitsky2010} BFKL approximation for N=4 SYM and  to attempt a systematic study of this expansion in N=4 SYM.
 Another important quantity which would be interesting to compute is the ``slope function'', the exact form of which was conjectured recently in \cite{Basso:2011rs}.

Finally, we hope that our result will bring us closer to the understanding of the mystery of the AdS/CFT correspondence. At least in technical terms, due to FiNLIE the gap between two sides of duality seems  to be much narrower now than before. But an important ingredient -- a derivation of the  Y-system and of FiNLIE from the first principles, i.e. directly from both the string sigma-model  and  the    N=4 SYM theory  -- is still missing.

\section*{Acknowledgments}
The work  of   VK was partly supported by  the grant RFBI 11-02-01220. The work  of VK was also partly supported by  the ANR grants StrongInt  (BLANC-SIMI-4-2011)  and  GranMA (BLANC-08-1-313695). The work of VK and SL was also partially supported by the ESF grant HOLOGRAV-09-RNP-092. The work of DV was partly supported by the US
Department of Energy under contracts DE-FG02-201390ER40577. We  thank the Perimeter institute (Waterloo, Canada), and V.K. also thanks KITP (USA), for the kind hospitality on the last stage of  this project.  We  thank
D. Serban, P.Vieira,  Z.Tsuboi and K.Zarembo for useful comments and discussions.

\newpage
\appendix

\section{\texorpdfstring{$\mathbb Z_4$}{Z4} symmetry and strong coupling limit}
\label{app:QCanal}
In this appendix we remind the classical \(\mathbb Z_4\)
symmetry of the string coset at strong coupling and motivate its quantum generalization
given in the main text.

We know from the group theoretical analysis of finite gap
solutions for the superstring on  \linebreak \(\frac{PSU(2,2|4)}{Sp(2,2)\times Sp(4)}\)
coset \cite{Gromov:2009tq,Gromov:2010vb} that in the classical  limit our T-functions become characters,  explicitly
written in \cite{Gromov:2010km} (see eqs. (4.12-21) there),  in the highest weight unitary irreps
\(a^s\) of \(PSU(2,2|4)\):
\begin{equation}\label{eq:T-Omega}
T_{a,s}=\tr_{a,s}\Omega(u)\,,
\end{equation}
where \(\Omega(u) \) is the classical monodromy matrix with the eigenvalues \((\mu_1,\mu_2,\mu_3,\mu_4|
\lambda_{1},\lambda_{2},\lambda_{3},\lambda_{4})   \).
Note that this identification was made   in \cite{Gromov:2010km}  in the
mirror sheet and the expression \eqref{eq:T-Omega} should be considered in the
mirror kinematics. The classical monodromy matrix as a function of spectral parameter \(u\)
has only one large cut \((-\infty,-2g]\cup[+2g,+\infty)\) with an essential singularity at the branch points
\(u=\pm 2g\).

 The \(\mathbb Z_4 \)-symmetry of the coset sigma model \cite{Bena:2003wd} imposes the following relations among the
 eigenvalues \cite{Beisert:2005bm}:
\begin{align}\label{eq:Z4x}
\mu_1(u)&= 1/\mu_2(u^\gamma),& \mu_3(u)&= 1/\mu_4(u^\gamma),&
 \lambda_{1}(u)&= 1/\lambda_2(u^\gamma),& \lambda_3(u)&= 1/\lambda_4(u^\gamma)\;.
  \end{align}
Here we denote by \(f(u^\gamma)\) the result of the analytic continuation of a function \(f\) following
the full circle around the branch point \(u=2g\) along a path avoiding square root cuts of a finite gap solution which are present in the eigenvalues but are absent in the monodromy matrix\footnote{these cuts may be interpreted as a condensate of Bethe roots.}.

For classical T-functions,  or characters,
 given by the formula (2.19) of \cite{Gromov:2010vb}    the property (\ref{eq:Z4x}) implies
  the following symmetry
  \begin{equation}\label{eq:TaT-a}
  \left\{
    \begin{array}{rlrl}
T_{a,s}(u)&=&(-1)^s T_{a,-\widehat {s}}(u^\gamma),    \qquad\qquad \textrm{if}\,\,
  |s|\geq a \\
T_{a,s}(u)&=&(-1)^a T_{-\widehat{a}, s}(u^\gamma),    \qquad\qquad  \textrm{if}\,\,
  a\geq |s|
    \end{array}
  \right.\,,
\end{equation}
where the functions \(T_{a,- \widehat s}\) represent the analytic
continuations of the functions \(T_{a, s}\) with respect to the argument
\(s\) from the values \(s>a\).  \(T_{a,- \widehat s}\)
 should not be
confused with   \(T_{a,-s}\)  entering the Hirota
equation \eqref{eq:Hirota} on the T-hook.
\(T_{-\widehat a,s}(u)\)  is also simply the
analytic continuation of \(T_{a,s}(u)\) from \(a>|s|\) to negative \(a\).

The classical limit has a rather degenerate analytic structure w.r.t.  the full quantum case, with the eigenvalues
(and hence the T-functions) having only two branch points at \(\pm 2 g\). In
the full quantum case,
  we have, a priori, an
infinite system of branch points, at    \(
u=\pm 2g \pm i(\frac{\left|a-\left|s\right|\right|+1}{2}+ n)
,\quad n=0,1,2,3,\dots\).
At strong coupling \(g\to \infty\)
there is no way to distinguish between the branch points with different \(n\)'s. Hence, the quantum version of (\ref{eq:TaT-a}) is potentially ambiguous. However, discussed in this paper analytic properties of Q-functions
give a natural suggestion how the proper quantum version should look like.

Let us consider the right-band T-functions and use the Wronskian parameterization for them, e.g. \(T_{1,s}=q_1^{[+s]}p_2^{[-s]}-q_2^{[+s]}p_1^{[-s]}\). The transformation \(T_{a,s}\to T_{a,-\widehat s}\) simply becomes the shift of the spectral parameter\footnote{Note that for \(q\)-functions such a shift is not negligible, even in the character limit.} \(u\) in the mirror kinematics: \(q_i^{[+s]}\to q_i^{[-s]}\), \(p_i^{[-s]}\to p_i^{[+s]}\). Now we consider only \(q\)-s and note that combination of the shift with analytical continuation \(u^\gamma\) clockwise is nothing but the shift \(q_i^{[+s]}\to q_i^{[-s]}\) avoiding the short cut \([-2g,2g]\), see \figref{fig:appA}. The same observation holds for \(p\)-s if the continuation \(u^\gamma\) is made counterclockwise\footnote{There is no apparent contradiction in using the opposite directions of \(\gamma\) for  continuation of \(q\) and \(p\) and the same direction for  continuation in  (\ref{eq:TaT-a}). The reason is that \(\gamma\) defines a monodromy of order \(2\) as it follows from (\ref{eq:Z4x}).}. Therefore (\ref{eq:TaT-a}) for the horizontal band is reformulated as:
\begin{figure}
  \centering
  \includegraphics[scale=0.7]{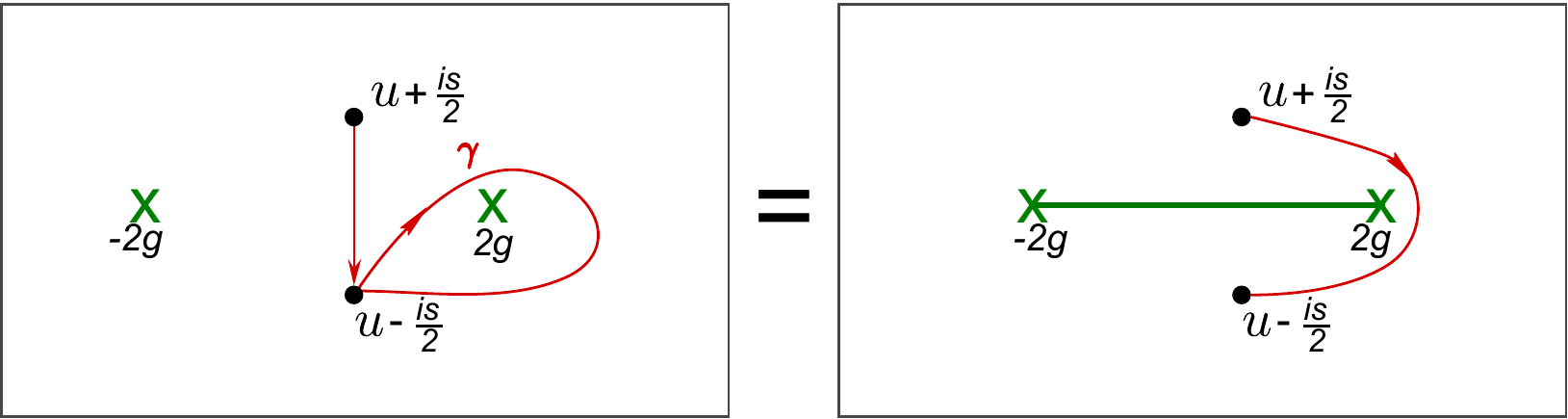}
  \caption{\label{fig:appA}\textbf{Equivalent representations of transformation of \(q\)-functions following from (\ref{eq:TaT-a})}}
\end{figure}
\be\label{appA4}
        T_{a,s}(u)=(-1)^a\hat T_{a,-s}(u),
\ee
where \(\hat T_{a,-s}\) is defined by the Wronskian ansatz for the right band considered at \(-s\) and with all shifts of \(q\)- and \(p\)-functions avoiding the short cut.

Generically, Q-functions on the quantum level have infinite number of cuts, which makes the above
arguments very ambiguous. However, in the \(\wT\)-gauge we know that the Q-functions can be chosen to have only one short cut.
This makes the \(\wT\)-gauge (and any other gauge with only two short cuts)
more suitable for the generalization of  (\ref{appA4}) to the quantum case. That is probably why one gets  \eq{Z4right2}.
A similar analysis for the upper band is more tricky.
However, in \appref{sec:zfode} we show that upper-band  \(\mathbb Z_4\) symmetry can be derived from the right-band \(\mathbb Z_4\) symmetry.

The outcome of our observations is that the quantum \(\mathbb Z_4\) symmetry
\eq{Z4up2} and \eq{Z4right2} is the most natural generalization of the   classical  \(\mathbb Z_4\) symmetry  of the string sigma model on
\(\frac{PSU(2,2|4)}{SO(1,5)\times S)(6)}\) coset. However, we should note that
it still remains to be proven that the generalization we propose
actually reduces to the classical \({\mathbb Z}_4\)
symmetry in the strong coupling limit.
One should carefully analyze the possibility of interchanging the order of making
analytical continuation and taking the strong coupling limit. Also note that in the quantum
version we use two different gauges for the upper and the right band
to formulate the \({\mathbb Z}_4\) invariance,
whereas in the classical case they should coincide up to a constant for \(|\Re(u)|<2g\). This effect is accounted by the \(i\)-periodic function \(\CF\) relating two gauges which in the scaling regime used for the classical approximation is indeed approximated by a constant for  \(|\Re(u)|<2g\).

\section{Large volume asymptotics}
\label{app:asymptoticsolution}
The asymptotic large \(L\) solution of the Y-system is used in this work as a starting point for the iterative numerical solution of FiNLIE.
The positions of cuts/poles/zeroes and the large \(u\)\ expansion for the asymptotic solution  are explicitly known. We use this information to
deduce the qualitative structure of cuts/poles/zeroes and to derive the large \(u\) expansion of the exact solution.

We consider only the \(\mathfrak{sl}(2)\) sector, hence all the Bethe roots are expected to be real. We also study here for simplicity only symmetric configurations of these roots, i.e \(u_k=-u_{M-k}\) with an even number of  roots \(M\). Some of the formulae written below are simplified by using these assumptions. In particular, the vanishing of the total momentum is already satisfied due to the relation\footnote{In the generic case the momentum  is given by (\ref{E:def}).} \(\prod_k\frac{\xp_k}{\xm_k}=1\).

It is useful to define the following functions
\begin{eqnarray}
\XpY=\frac {(-1)^{M/2}}{2}(B^{(+)}R^{(-)}+B^{(-)}R^{(+)})\;\;,\;\;
\XmY=\frac{(-1)^{M/2}}{2}(B^{(+)}R^{(-)}-B^{(-)}R^{(+)})\;.
\end{eqnarray}
Note that \(P=u^M+\ldots\)   is a polynomial in \(u\),
and \(S\) is a polynomial times pure square root.

\subsection{Right band}\label{app:as}
An expression for the large volume limit of T-functions was presented in \cite{Gromov:2009tv}. On can check that up to a gauge transformation this
expression for the case of the right band is equivalent to:
\begin{equation}\label{Tcalas}
\hat\CT_{1,s}= C_\CT\left(s+\frac 12\frac{1+\hat\Phi^{[+s]}}{1-\hat\Phi^{[+s]}}-\frac 12\frac{1+\hat\Phi^{[-s]}}{1-\hat\Phi^{[-s]}}\right),\ \ C_{\CT}=\frac{ \gas}{\gas+2},
\end{equation}
where \(\gamma_{as}\) is the asymptotic value of the anomalous dimension and   \(\hat\Phi=(-1)^{M/2}\frac{\hat B^{(+)}}{\hat B^{(-)}}\) is defined as a function with short \(Z\)-cut.
The normalization constant \(C_{\cT}\) is chosen so that \(\hat\CT_{1,s}\to s\), \(u\to\infty\).
For  configurations of the Bethe roots without poles in the analyticity strip of \(\hat\CT_{1,s}\)
one can write a spectral representation for \(\hat\CT_{1,s}\):
\be
\hat\CT_{1,s}=s+\CK_s\,\hat *\,\rho\,.
\ee
 The density \(\rrho\) is given by:
\begin{equation}\label{rhoasymptotic}
\rrho= C_{\CT}\frac{\hat\Phi^{[+0]}-\hat\Phi^{[-0]}}{(1-\hat\Phi^{[+0]})(1-\hat\Phi^{[-0]})}=C_{\CT}\frac{2\XmY}{Q^++Q^--2\XpY}\,,\ \ u\in\hbZ_0\,.
\end{equation}
The above expressions give indeed the large volume limit in the \(\CT\)-gauge introduced in \secref{sec:analyticity}.
One can see that the denominator in the rightmost expression of (\ref{rhoasymptotic}) is a polynomial of degree \(M-2\).
For the case of two symmetric magnons (\(M=2\))  \eq{rhoasymptotic} simplifies to
\begin{equation}\label{rhoforKonishi}
\rrho=4\frac{\sqrt{4g^2-u^2}}{\gas+4}=\frac{x-1/x}{x_1^+-x_1^-}\;
\end{equation}
and \eq{Tcalas} simplifies to
\begin{equation}\label{TcalM2}
        \hat\CT_{1,s} = s - \frac{1/\hat x^{[s]}-1/\hat x^{[-s]}}{x_1^+-x_1^-}\;,
\end{equation}
where we choose \(u_1=-u_2>0\).

\subsection{Upper band}
For two symmetric magnons there is an expression for the T-functions in the  \(\sT\)-gauge which is very similar to
\eq{TcalM2}:
\begin{eqnarray}
        \hat{\sT}_{a,1}(u) = a - \frac{1/\hat x^{[a]}-1/\hat x^{[-a]}}{1/x_1^+-1/x_1^-}\;.
\end{eqnarray}
In this form it is easy to see that \(\sT_{1,1}(u_j)=0\).
For general \(M\) the asymptotic expressions for the \(\sT\)-gauge can be read off from the Wronskian parameterization in \cite{Gromov:2010km}. The explicit asymptotic values of our basis of Q-functions look as follows:
\begin{equation}\label{asymptoticupper}
       q_1=1\;\;,\;\;  \hat W=q_2= C_{\sT}\left({\bf \CM}-\hat\XmY\right)\;\;,\;\;  q_{12}=\tilde Q= Q\;.
\end{equation}
Here \(C_{\sT}=\frac{4}{\gas(\gas+2)}\) and \({\bf \CM}\) is a polynomial solution of the following equation:
\begin{equation}\label{Mequation}
        {\bf \CM}^+-{\bf \CM}^-=2Q-\XpY^+-\XpY^-\;.
\end{equation}
The additive constant  cannot  be fixed from (\ref{Mequation}) but it is irrelevant
due to the H-symmetry (\ref{Hsym}). A convenient choice for this constant is given below.

In the leading large volume approximation the T-functions in the \(\sT\)-gauge are given by
\begin{eqnarray}
\hat\sT_{a,1}= \hat W^{[+a]}-\hat W^{[-a]}\;\;,
\;\;\hat\sT_{a,0}=Q^{[+a]}Q^{[-a]}\;\;,\;\;
\hat\sT_{a,2}=\frac{\hat\sT_{1,1}^{[+a]}\hat\sT_{1,1}^{[-a]}}{Q^{[+a]}Q^{[-a]}}\;.
\label{asymsT}
\end{eqnarray}
One can see that
\begin{equation}\label{zer}
\hat\sT_{1,1}(u_j)=0\;,
\end{equation}
 so the poles in the denominator of \(\hat\sT_{a,2}\) cancel. The normalization constant \(C_\sT\) introduced in (\ref{asymptoticupper}) is chosen so that
\(\sT_{a,1}=a\, u^{M-2}+\dots\) for large \(u\).

Since \(W\) has no poles and for large \(u\) it behaves as \(\frac{a u^{M-1}}{i(M-1)}\) one can construct the following spectral representation:
\begin{equation}
\hat W={P}_{M-1}+\CC\cz\rrho_2\;\;,\;\;\rrho_2=-2C_{\sT}\XmY\,,
\end{equation}
where the zeros of the polynomial \(P_{M-1}=-\frac{i}{M-1}u\prod_{j=1}^{M-2}(u-v_j)\)
are fixed by  condition \eq{zer}. Note that we used a freedom in the choice of the additive constant of \({\bf M}\) to constrain the form of \(P_{M-1}\).
For \(M=2\) we get simply
\begin{equation}
\hat W=-iu-\frac{1/\hat
  x}{1/x_1^+-1/x_1^-}\;\;,\;\;\rrho_2=\frac{x-1/x}{1/x_1^+-1/x_1^-}\;.
\label{eq:rho2konishi}
\end{equation}

\paragraph{Expression for \(\nU\):}
Comparing
\begin{equation}
Y_{a,0}=\frac{\hat\sT_{a,1}^2 (U^{[+a]}\bar U^{[-a]})^2}{
\hat\sT_{a+1,0}\hat\sT_{a-1,0}}\;\;,\;\;\IM u> a/2
\end{equation}
to the asymptotic expression for \(Y_{a,0}\) in \cite{Gromov:2009tv} we find
\begin{eqnarray}\label{nU}
\nU(u)&=&\frac{e^{+i\Lambda_U}}{\sqrt{C_{\sT}}}B^{(-)}(u) \frac{V(u)}{x^{L/2}(u)}\;\;,\;\;\IM u>0\;,\\
\bar\nU(u)&=&\frac{e^{-i\Lambda_U}}{\sqrt{C_{\sT}}} R^{(+)}(u)\frac{x^{L/2}(u)}{V(u)}\;\;,\;\;\IM u<0\;,
\end{eqnarray}
where \(\Lambda_U\) is a real number and \(V\) is a solution to the equation
\begin{equation}
\frac{V^+}{V^-}=\frac{B^{(+)+}}{B^{(-)+}}\prod_{k=1}^M\sigma_{\rm BES}(u,u_k)=\left(\frac{B^{(-)+}}{B^{(-)-}}\prod_{k=1}^{M}\sigma_{\rm mirr}(u,u_k)\right)^{-1}\;.
\end{equation}
The function \(\s_{\rm mirr}\) first time appeared in
\cite{Arutyunov:2009kf,Gromov:2009bc} and  was later identified in
\cite{Volin:2009uv,Volin:2010cq} with solution of the mirror crossing equation. It
has a particularly nice behavior under fusion: \(\s(u,v)^{[a]_\D}\) as
a function of \(u\) has only two cuts, \(\check Z_a\) and \(\check
Z_{-a}\), on the mirror sheet. As a consequence, \(V\) has only one
cut, \(\check Z_0\), on the mirror sheet.
From integral representations for \(\s_{mirr}\) given in \cite{Gromov:2009bc} or \cite{Volin:2010cq} one derives the following explicit expression for \(V\):
\begin{eqnarray*}
        V(u)&=&\prod_{k=1}^M\exp\left(\sum_{i=0}^2\chi_i\left(u,u_{k}+\frac i2\right)-\chi_i\left(u,u_{k}-\frac i2\right)\right)\!,
\end{eqnarray*}
\begin{eqnarray*}
\chi_0(u,v)&=&-\frac {ig}{y}\log x,\\
\chi_1(u,v)&=&-\left(\int_{-\infty}^{-1}+\int_1^{\infty}\right)\frac{dz}{2\pi i}\left(\frac{1}{z-x}-\frac 1{z-\frac 1x}\right)\log\left(1-\frac 1{z\,y}\right),\\
\chi_2(u,v)&=&\int_{2g}^\infty\frac{d\omega}{2\pi}\frac{x-\frac 1x}{x_\o-\frac 1{x_\o}}\int_{-2g}^{2g}dq\frac 1{e^{2\pi(q+\o)}-1}\left(\frac 1{\o-u}\log\left(\frac{1+\frac{x_q}{y}}{1+\frac 1{x_q\,y}}\right)+\frac 1{\o+ u}\log\left(\frac{1-\frac{1}{x_q\,y}}{1-\frac {x_q}{y}}\right)\right)\,,
\end{eqnarray*}
where \(x=x(u),y=\hat x(v),x_\o=\hat x(\o), x_q=x(q).\)

 Note that \(\log V\) is a purely imaginary function in the mirror kinematics (which is consistent with
\eq{nU}).
The large \(u\) behavior of \(V\) is defined solely by \(\chi_0\). One gets \(\log V\simeq -\frac{\gamma}{2}\log u\) which implies
\begin{equation}
\log U\simeq -\frac{L+\gamma}{2}\log u\;\;,\;\;u\to \infty+i 0\;.
\label{eq:asUas}
\end{equation}

\paragraph{Asymptotics of \(\hat h\):}
The large volume expression for \(\hat h\) is rather complicated.
However, we would need only the large \(u\) asymptotics of this expression. From the condition \(\bT_{a,2}=\wT_{2,a}\) one can relate \(\hat h\) to the other, already known functions:
\begin{equation}
\hat h^+
\hat h^-{\hat\cT}_{1,1}=e^{i C}f^{++}f^{--} U^+ U^-\frac{{\hat\sT}_{1,1}}{Q}\,,
\end{equation}
where \(C\) is some real number.
One can see from (\ref{appBas}) and (\ref{fdef})  that \(f\sim u^{\gamma/2}\) and thus from the previous relation one has
\begin{equation}\label{largeuhh}
\hh^2\simeq \frac{f^2 \nU^2(W^+-W^-)}{Q}\sim \frac{u^{\gamma}u^{-\gamma-L} u^{M-2}}{u^M}\sim u^{-L-2}
\end{equation}
or
\begin{equation}
\hat h\sim u^{-\frac{L+2}{2}}\;.\label{largeuh}
\end{equation}
We expect that in a finite volume this asymptotic behavior remains unchanged, see discussion after (\ref{discontinuitiyh}).

\paragraph{Some useful expressions:}
It is known that the product of fermionic Y-functions is very simple
\begin{equation}\label{asymYfer}
Y_{1,1}
Y_{2,2}=\frac{B^{(-)}R^{(+)}}{B^{(+)}R^{(-)}}\;,
\end{equation}
so that from \eq{Bdefin} one can easily determine the large volume expression for \(\Bup\)\,:
\begin{equation}
{\bf B}
=\left(\frac{B^{(+)}}{B^{(-)}}\right)^2\;.
\label{appBas}
\end{equation}
The function \(\cal F\) can be determined by its discontinuities and zeroes. By applying arguments of \appref{sec:alteqnonCF} to the large volume expression (\ref{asymYfer}) one gets:
\begin{equation}
{\cal F}^2=\Lambda_F\prod_{a=-\infty}^\infty \frac{\exp\left(\frac{2E}{|2a-1|}\right)}{Y^{[2a-1]}_{1,1}Y^{[2a-1]}_{2,2}}\;.
\end{equation}
The exponential numerical factor in the numerator is needed to ensure the convergence and \(\Lambda_F\) is a normalization constant.

\paragraph{Large \(u\) behavior of exact quantities:}
\label{sec:large-u-behavior}
 Above we derived the large \(u\) behavior of different quantities in
 the large volume approximation. The result however remains the same
 at finite volume provided that we use the exact value for the energy
 (\ref{E:def})
and anomalous dimension. This was already shown for \(Y_{1,1}Y_{2,2}\)
(see Eq.~(\ref{Ylauexpan})) and for \(\hat h\). Finite volume
corrections  also do not change  the leading large \(u\) term of
\(\sT_{a,s\geq 0}\) and \(\cT_{a,s}\). Indeed, the magnitude of
corrections in these quantities is defined by  \(U^\a\bar U^{-\a}\) for
some positive integer \(\a\) which is smaller than \((\hat x^{[+\a]}\hat
x^{[-\a]})^{-L/2}\).  From the definition (\ref{Bdefin}) for \(\bf B\)
and \(f\) we can now conclude that \(f\sim u^{\gamma/2}\) and
\(B\sim1-\frac{i\gamma}{u}\) at any volume. Finally, we can invert
relation (\ref{largeuhh}) to prove that (\ref{eq:asUas}) holds at any
volume.

\section{Equivalence of FiNLIE to the TBA equations}
\label{sec:eqtoTBA}
The finite volume AdS/CFT spectrum problem  was previously analyzed in the literature using an infinite set of the TBA equations.
 These equations were proposed in \cite{Bombardelli:2009ns,Gromov:2009bc,Arutyunov:2009ur}, following the discovery of the Y-system \cite{Gromov:2009tv}, and  passed a number of important checks
\cite{Bajnok:2008bm,Gromov:2009zb,Gromov:2010vb,Gromov:2009tq,Arutyunov:2010gb,Balog:2010xa,Frolov:2010wt,Gromov:2011de,Roiban:2011fe,Vallilo2011}.
It is important to verify that  FiNLIE  is equivalent to the TBA equations and this is the goal of this appendix.
The TBA equations can be also viewed as a departing point to derive the properties of the transfer matrices which we
summarized in \secref{sec:prop}. Alternatively, these properties were considered there as the  initial assumptions for the derivation of FiNLIE.

The TBA equations and the notations we use are the ones from \cite{Gromov:2009bc,Gromov:2009zb}.
The papers \cite{Bombardelli:2009ns,Arutyunov:2009ur} contain an equivalent set of TBA equations
in different notations, but without the driving terms for the excited states.
\subsection{TBA  for \texorpdfstring{$Y_{1,s}$ and $Y_{a,1}$}{Y1s and Ya1}}
\label{app:stringTBA}
We start our analysis by the TBA equations for the right band:
\begin{equation}\label{TBAright}
        \log(Y_{1,n+1})=\sum_{m=1}^{\infty}\CC_{n,m}*\log(1+1/{Y_{1,m+1}})+\CK_n\hat*\log\frac{1+Y_{1,1}}{1+1/Y_{2,2}}\;,\ \ n\geq 1\,,
\end{equation}
\begin{equation}
  \CK_{n,m}=\left(\D^{+2}-\D^{-2}\right)[n]_\D [m]_\D {\cal K}\;\;,\;\;{\D=e^{\frac i2\partial_u}}\;.
\end{equation}
The fusion operator \([s]_\D\) is defined in \secref{Notations}.

Now one replaces the Y-functions with a combination of the
T-functions using
\(1+1/Y_{1,m}=\frac{\CT_{1,m}^+\CT_{1,m}^-}{\CT_{1,m+1}\CT_{1,m-1}}\). This
leads to the  chain cancellations  of an infinite number of terms in the sums leaving us only with  a finite number of terms in the r.h.s of (\ref{TBAright}).
We call this procedure ``telescoping''\footnote{The same kind of simplification
  was also used in \cite{Balog:2011nm,Balog:2011cx} under the name of ``chaining relations''.}. Indeed, after the substitution of T-functions and the shifts of the contours of integration in the terms containing \(\log\CT_{1,m}^{\pm}\) one gets:
\begin{eqnarray}
    \CC_{n,m}*\log(1+1/{Y_{1,m+1}})=&&\nonumber\\
  ({\D}^{+2}-{\D}^{-2})[n]_{\D}\left(\sum_{m=1}^\infty (\D+\D^{-1})[m]_\D\right.&-&\left.\sum_{m=0}^\infty [m\!+\!1]_\D-\sum_{m=2}^\infty [m\!-\!1]_\D \right){\cal K}*\log{\CT_{1,m}}\,.\\
\end{eqnarray}
Since \(({\rm D}+{\rm D}^{-1})[m]_{\rm D}-[m\!+1]_{\rm D}-[m-1]_{\rm D}=0\), all the terms in the infinite sums above cancel out except the boundary ones. We therefore get:
\begin{eqnarray}\label{telescopingreslut}
   \sum_{m=1}^\infty\CC_{n,m}*\log(1+1/{Y_{1,m+1}})&=&\log(\CT_{1,n+2}\CT_{1,n})-(\CC_{n+1}+\CC_{n-1})*\log(\CT_{1,1})\nonumber\\
   &=&\log(\CT_{1,n+2}\CT_{1,n})-\CC_{n}*\log(\CT_{1,1}^{[+1-0]}\CT_{1,1}^{[-1+0]}).
\end{eqnarray}
The formal manipulations that we made are valid only if the
integration and summation can be interchanged and only if  the
contours of integration do not hit  poles or branch points when being
shifted. The \(\CT\)-gauge given by (\ref{eq:rho-to-CF}) enjoys these
nice properties and that is why we used it for the derivation of (\ref{telescopingreslut}). Note that the r.h.s of (\ref{telescopingreslut}) is not invariant under an arbitrary gauge transformation while the l.h.s is; that is so because the telescoping procedure can be performed only in a specially chosen gauge.

One can  substitute \(Y_{1,n+1}=\frac{\CT_{1,n+2}\CT_{1,n}}{\CT_{2,n+1}}\) at  the l.h.s of the TBA equation (\ref{TBAright}). The numerator is cancelled against \(\log(\CT_{1,n+2}\CT_{1,n})\) term in the telescoped expression (\ref{telescopingreslut}).  Therefore this TBA equation is reduced to
\begin{eqnarray}\label{apptelescopedTBAright}
        -\log(\CT_{2,n+1})&=&-\CC_{n}*\log(\CT_{1,1}^{[+1-0]}\CT_{1,1}^{[-1+0]})+\CC_n\cz\log\frac{1+Y_{11}}{1+1/Y_{22}}\nonumber\\
        &=&-\CC_{n}*\log(\CT_{1,1}^{[+1-0]}\CT_{1,1}^{[-1+0]})+\CC_n\cz\log\frac{\CT_{2,3}\CT_{1,1}^{+}\CT_{1,1}^{-}}{\CT_{2,2}^{-}\CT_{2,2}^{+}}\nonumber\\
        &=&-\CC_{n}{\check *}\log(\hat\CT_{1,1}^{+}\hat \CT_{1,1}^{-})+\CC_n\cz\log\frac{\CT_{2,3}}{\CT_{2,2}^{-}\CT_{2,2}^{+}}\,.
\end{eqnarray}

Using the representation (\ref{TGrepr}) for \(\CT_{2,n+1}\) we get:
\begin{eqnarray}
\log(\CT_{2,n+1})&=&\log(1+G^{[n+2]}-G^{[n]})+\log(1+\bar G^{[-n-2]}-\bar G^{[-n]})\nonumber\\
&=&\CC_n*\log(1+G^{[2]}-G^{[+0]})(1+\bar G^{[-2]}-\bar G^{[-0]})\equiv \CC_n*\log\frac{\CT_{2,2}^{[+1-0]}\CT_{2,2}^{[-1+0]}}{\CT_{2,3}}\label{introducingkn}.
\end{eqnarray}
Note that the expression in the argument of the \(\log\)
is equal to \(\hat \CT_{2,1}\) for \(u>2g\) and
 \(\frac{\CT_{2,2}^+\CT_{2,2}^-}{\CT_{2,3}}\) for \(u<2g\) so that (\ref{introducingkn}) reduces to:
\begin{equation}\label{finaltelescopingresult}
        \CC_n\check *\log\frac{\hat\CT_{2,1}}{\hat\CT_{1,1}^{+}\hat\CT_{1,1}^{-}}=0\;.
\end{equation}
This equation should be valid everywhere on the real axis and for any positive integer \(n\), which is only possible if
\begin{equation}
{\hat\CT_{1,1}^{+}\hat\CT_{1,1}^{-}}={\hat\CT_{2,1}}.
\end{equation}
From the Hirota equation for \(\hat\CT_{1,1}\) we have  \(\hat\CT_{1,1}^+\hat\CT_{1,1}^-=\hat\CT_{2,1}+\hat\CT_{1,0}\hat\CT_{1,2}\), and consequently\footnote{The property (\ref{resultmagic1}) can be equivalently reformulated in terms of Y-functions as \(\hat Y_{1,2}^+\hat Y_{1,2}^-=(1+Y_{1,3})\). The latter       equation was first time derived from the TBA system in \cite{Arutyunov:2009ur}, however the magic properties of the \(\cT\)-gauge following from (\ref{resultmagic1}) were not discovered there.
}
\begin{equation}\label{resultmagic1}
\hat\CT_{1,0}=G^{[+0]}+\bar G^{[-0]}=0.
\end{equation}
 We thus derived from the TBA equations the statement (\ref{QbarQright}).

 The  derivation presented here can be reversed. Hence the set of TBA equations for \(Y_{1,s\geq 2}\) is   equivalent to the condition \(\hat\CT_{1,0}=0\), with
       an additional assumption that \(\CT_{1,s}\in\CA_s\) and with
a large volume asymptotics \(\CT_{1,s}\sim s+\CO(1/u)\).
The condition of the absence of poles and zeroes in the ``telescoping strip"\footnote{This is the strip in which the contours are shifted.} \(-1/2<{\rm Im}(u)<1/2\) is also needed,
however it is relevant only because we consider here the \(\mathfrak{sl}(2)\) sector.
In  case of a general state  zeroes may be present in these T-functions and they should lead to the appearance of additional driving terms in the TBA equation (\ref{TBAright}).
\ \\

\paragraph{Equations for ``pyramids'':} Let us now consider the TBA equations for \(\{a,1\}\) nodes of the \(\wT\)-hook:
\begin{eqnarray}\label{appTBAupright}
        \log(Y_{n+1,1})&=&[n\!-\!1]_\D \log\frac{B^{(+)}}{B^{(-)}}+[n\!+\!1]_\D\log\frac{R^{(+)}}{R^{(-)}}+{\cal M}_{n+1,m}*\log(1+Y_{m,0})\nonumber\\
&&-\CC_{n,m}*\log(1+{Y_{1,m+1}})-\CC_n\cz\log\left(\frac{1+Y_{1,1}}{1+\frac 1{Y_{2,2}}}\right),\ \ n\geq1.
\end{eqnarray}
Here and below the summation sign over the doubly repeated \(m\) from \(1\) to \(\infty\) is systematically omitted.

We want to perform the telescoping procedure similar to the one for the right band. For that it is natural to use the \(\sT\)-gauge which has the simplest possible large \(u\) behavior.
One should be especially careful with the terms containing \(Y_{m,0}\) in the r.h.s. which have a complicated kernel\footnote{we use the notations
from \cite{Gromov:2009tv} }. Moreover the term with fermions does not
have a form, which is easily expressible in terms of the T-functions of the upper band.
Both  complications can be overcome at once by subtracting from (\ref{appTBAupright}) the TBA equation for \(Y_{1,1}\)
 after applying to it  \(\CC_n\ *\hspace{-1em}\circlearrowright\):
\begin{equation*}
  \CC_n\ *\hspace{-1em}\circlearrowright \left(      \log Y_{1,1}=\log\left[-\frac{R^{(+)}}{R^{(-)}}\right]+\CR^{(0m)}*\log(1+Y_{m,0})-\CC_m*\log\frac{1+Y_{m+1,1}}{1+1/Y_{1,m+1}}\right).
\end{equation*}
 The result of subtraction reads:
 \begin{gather}
        \log Y_{n+1,1}=\log\frac{Q^{[+n+1]}}{Q^{[-n+1]}}+\CC_{n,m-1}^{\neq}\log(1+Y_{m,0})-\CC_{n,m}*\log(1+Y_{m+1,1})\nonumber
        +\CC_n\cz\log\left(\frac{1+Y_{2,2}}{1+\frac 1{Y_{1,1}}}\right),\nonumber\\
{\rm where}\,\,\,\CC_{n,m-1}^{\neq}=(\D^{}-\D^{-1})[n]_\D[m]_\D{\cal K}\;.
\end{gather}

Now one can perform the same steps as for the right band. The difference is that \(\sT_{a,s}\) functions do contain zeroes inside the telescoping strip. However, the only fate of the residues coming from these zeroes is to cancel the driving term \(\log\frac{Q^{[n+1]}}{Q^{[n-1]}}\). Note that the driving term is definitely such only for Konishi-like states   at a  sufficiently small coupling. In general this term may be more complicated
 \cite{Gromov:2009zb,Arutyunov:2009ax}.
In any case this term is  constructed precisely in the way to cancel residues coming from the contour shifts. Therefore the result of  telescoping procedure is not sensible to the structure of the driving term
and in this sense it is  more universal than the
TBA equations. In complete analogy with the right band, the final equation we get for the upper band is:
\begin{equation}\label{resultTBAupperright}
        \hat \sT_{0,1}=0.
\end{equation}
Again, the equivalence works in both directions: one can derive the set of TBA equations for \(Y_{a,1}\),   \(a>1\) from (\ref{resultTBAupperright}). The required assumption,
in addition to \eq{resultTBAupperright},
is the existence of the \(\sT\)-gauge with \(\sT_{a,s}\in\CA_{a-|s|+1}\) and a polynomial behavior at infinity.

\subsection{TBA for \texorpdfstring{$Y_{1,1}$ and $Y_{2,2}$}{Y11 and Y22}}
\label{sec:tba-y11-y22}
One can make the telescoping procedure  for the TBA equations for \(Y_{1,1}Y_{2,2}\) and \(Y_{1,1}/Y_{2,2}\) and get precisely the integral equations (\ref{Y11Y22throughT}) and  (\ref{ratioY11Y22throughT}). We leave this exercise to a curious reader, while here we will use the results of \cite{Cavaglia:2010nm} for the derivation of  equivalence between (\ref{Y11Y22throughT}), (\ref{ratioY11Y22throughT}) and the TBA equations. In \cite{Cavaglia:2010nm} it was shown that the fermionic vacuum TBA equations are a consequences of the discontinuity conditions: TBA for \(Y_{1,1}Y_{2,2}\) is equivalent to the discontinuity condition (1.7) of \cite{Cavaglia:2010nm} and the TBA for \(Y_{1,1}/Y_{2,2}\) is equivalent to the discontinuity condition (F.5) in that paper\footnote{Though (F.5) is claimed in \cite{Cavaglia:2010nm} to be a consequence of other discontinuity relations and Y-system equations, we found that the derivation of this claim contains a logical gap, at least in the way it is presented. Therefore we consider (F.5)  as an independent requirement on the discontinuities of the Y-functions. }.
The discontinuity conditions are proved to be correct also when the excited states are considered \cite{Balog:2011nm}. We will show now that these discontinuity conditions are equivalent   to the analyticity of \(\bB\) and \(\bC\) functions,  in the upper half-plane --- the essential property needed for the derivation of  (\ref{Y11Y22throughT}) and (\ref{ratioY11Y22throughT}), respectively.

\paragraph{TBA for the product of \(Y_{1,1}\) and \(Y_{2,2}\):} The condition (1.7) of \cite{Cavaglia:2010nm} is written in our notations as
\begin{equation}\label{discY11Y22our}
        \disc \left((\log Y_{1,1}Y_{2,2})^{[+2n]}\right)=-\sum_{a=1}^{n}\disc\left(\log(1+Y_{a,0}^{[2n-a]})\right)\;\;,\;\;n\geq 1\,.
\end{equation}
We express \(1+Y_{a,0}\) through the T-functions using \(1+Y_{a,0}=\frac{\sT_{a,0}^+\sT_{a,0}^-}{\sT_{a+1,0}\sT_{a-1,0}}\). Again, massive cancellations occur in the sum and the final result is:
\begin{equation}\label{discY11Y22simp}
\disc \left((\log Y_{1,1}Y_{2,2})^{[+2n]}\right)=-\disc\left(\log\frac{\sT_{1,0}^{[2n]}\sT_{n,0}^{[n-1]}}{\sT_{0,0}^{[2n-1]}\sT_{n+1,0}^{[n]}}\right).
\end{equation}
moreover since \(\sT_{n,0}^{[n-1]}\) and \(\sT_{n+1,0}^{[n]}\)
does not have a cut on the real axis
we conclude that \eq{discY11Y22simp} is precisely the condition
of analyticity of \(\Bup\), defined in \eq{Bdefin}, in the upper half-plane.
As we know, the analyticity of \(\Bup\) allows the construction of the gauge transformation \(f\) via \(\eq{eqf}\),
which then implies existence of \(\bT\) as we show below.
Let us consider then the LR-symmetric gauge \(T_{a,s}=\sqrt{\sT_{a,s}\sT_{a,-s}}\) and perform the following gauge transformation:

\begin{equation}\label{C15}
    \bT_{a,s}=f^{[a+s]}f^{[a-s]}\bar f^{[-a+s]}\bar f^{[-a-s]}T_{a,s}\;.
\end{equation}
Since \(f\) by construction is regular above \(Z_{-1}\), \(\bT\)-functions  have the same analyticity strips as \(\sT\)-functions.
Moreover from \eq{C15} one can show that
\beq\la{Bblack1}
\frac{\bT_{3,2}\bT_{0,1}}{\bT_{2,3}\bT_{0,0}^-}=1\;.
\eeq
Next, since \(\bT\)-functions are real by construction, the complex conjugate  of (\ref{Bblack1}) reads
\(\frac{\bT_{3,2}\bT_{0,1}}{\bT_{2,3}\bT_{0,0}^+}=1\). This immediately  implies  \(\bT_{0,0}^+=\bT_{0,0}^-\) and \(\bT_{3,2}=\bT_{2,3}\),  and thus the  \(\bT\)-gauge is
indeed the gauge introduced in
\secref{sec:analyticity} (\(\mathbb Z_4\)-symmetry in this gauge is shown below).
We therefore arrive at the initial assumptions that we used in the main text for derivation of regularity of \(\bB\) in the upper half-plane.

\paragraph{TBA for the ratio of \(Y_{1,1}\) and \(Y_{2,2}\):}
The TBA equation for the ratio can be reformulated as a discontinuity equation (F.5) of \cite{Cavaglia:2010nm}.
In our notations it reads as
\begin{equation*}
   \disc\left(2\sum_{a=2}^\infty\log\left(\frac{1+Y_{a,1}}{1+1/Y_{1,a}}\right)^{[2n+1-a]}-\sum_{a=1}^\infty\log\left(1+Y_{a,0}\right)^{[2n-a]}
-\log\left(\frac{Y_{2,2}}{Y_{1,1}}\right)^{[2n]}
   \right)=0\,.
\end{equation*}
After the substitutions \(1+Y_{a,s}=\frac{\sT_{a,s}^+\sT_{a,s}^-}{\sT_{a+1,s}\sT_{a-1,s}}\), \(1+1/Y_{1,a}=\frac{\cT_{1,a}^+\cT_{1,a}^-}{\cT_{1,a+1}\cT_{1,a-1}}\) and the telescoping of the sums we get
\begin{equation}
        \disc\left(2\log\frac{\sT_{2,1}}{\cT_{1,2}}\frac{\cT_{1,1}^-}{\sT_{1,1}^-}-\log\frac{\sT_{1,0}}{\sT_{0,0}^-}-\log\left(\frac{Y_{2,2}}{Y_{1,1}}\right)\right)^{[2n]}=0\,.
\end{equation}
It easy to see that this equation is equivalent to the condition of regularity of \(\bC\) (\ref{relatingBUC2}) in the upper half-plane.

Again, the only assumption we used is the existence of a \(\CT\)-gauge with \(\CT_{a,s}\in\CA_{s-a}\), \(s>a\). Let us show now that the \(\wT\)-gauge defined by (\ref{eq:Fnice}) has the same analyticity strips as the \(\sT\)-gauge. For that it is enough to show that the function \(h\) relating \(\wT\)- and \(\CT\)- gauges by
\begin{equation}
\label{D18eqn}
        \wT_{a,s}=(h^{[+s]}\bar h^{[-s]})^{[a]_{\rm D}}\cT_{a,s}
\end{equation}
is analytic in the upper half-plane. To show this, we note that from analyticity of \(\bC\) in the upper half-plane
and (\ref{relatingBUC2}) it follows that \(h^{++}/h\) is  analytic in the upper half-plane. Therefore discontinuities $\delta=\Delta(\log h^{[2n]})$, if non-zero, are the same for arbitrary $n\geq1$. Let us introduce an $i$-periodic function $\CP$ with $\Delta(\log \CP)=\delta$. Then $h' \equiv h/\CP$ is analytic in the upper half-plane. From $(h^{[+s]}\bar h^{[-s]})^{[2]_{\rm D}}=\wT_{2,s}/\cT_{2,s}=\bT_{2,s}/\cT_{2,s}\in\CA_{s-1}$ we conclude that $2\Delta(\log\CP\bar\CP)=0$ and hence we can choose $\CP$ to be such that $\CP\bar\CP=1$. Therefore \(h^{[+s]}\bar h^{[-s]}=h'^{[+s]}\bar h'^{[-s]}\), i.e. we can always adjust $h$, by multiplying by irrelevant $i$-periodic function, to be analytic in the upper half-plane.

\subsection{TBA for \texorpdfstring{$Y_{a,0}$}{Ya0}}
It is again useful to start from the discontinuity condition (1.6) of \cite{Cavaglia:2010nm,Balog:2011nm} instead of the equivalent  TBA equation for \(Y_{a,0}\).  This condition reads in our notations as follows:
\begin{eqnarray*}
\disc(\disc(\log Y_{1,0}^+)^{[2n]})&=&2\disc\!\!\left(\log\left(1+{Y_{1,1}^{[2n]}}\right)+\sum_{m=1}^{\infty}\log(1+{Y_{m+1,1}^{[2n-m]}})\right)
-\log{Y_{1,1}}{Y_{2,2}},
\end{eqnarray*}
\(n\geq1\).
Let us first compute \(\disc\!\left(\log Y_{1,0}^+\right)\). We use \(Y_{1,0}=\frac{\bT_{1,1}^2}{\bT_{0,0}\bT_{2,0}}=\frac{\wT_{1,1}^2}{\bT_{2,0}}\) and \(\bT_{2,0}\in\CA_3\). Then \(\disc\left(\log Y_{1,0}^+\right)=2\disc\left(\log \wT_{1,1}^+\right)=2\log{\wT_{1,1}^+}/{\tilde\wT_{1,1}^+}\), where \(\tilde \wT_{1,1}\) is defined as follows: It coincides with \(\wT_{1,1}\) below \(i/2\) and has a short \(\hbZ_1\) cut;  all other cuts of \(\tilde \wT_{1,1}\) above \(Z_1\) are of  the \(\bZ\)-type (long).

After telescoping we get
\begin{equation}\label{discmiddleround2}
        \disc\log\left(\frac{\wT_{1,1}^+}{\tilde \wT_{1,1}^+}\frac{\bT_{0,1}}{\bT_{1,1}^+}\right)^{[2n]}=-\log Y_{1,1}Y_{2,2}\,.
\end{equation}
Now we use \(\bT_{0,1}/\bT_{1,1}^+=\cF^+/\wT^+_{1,1}\). Since the analyticity strips for \(\bT\)-gauge and the periodicity of \(\bT_{0,0}\equiv\CF^2\) were already established (from TBA for \(Y_{1,1}Y_{2,2}\)) one can use (\ref{eq:discf}), which is derived
indeed from the periodicity of \(\bT_{0,0}\) and the analyticity of \(\bT_{1,0}\). It states: \(\disc\CF^{[2n+1]}=-\log Y_{1,1}Y_{2,2}\).
Therefore (\ref{discmiddleround2}) simplifies  to
\begin{equation}
        \disc\log\tilde\wT^{[2n+1]}_{1,1}=0,\ \ \ n=\overline{1,\infty}. \end{equation}
This means that \(\hat\wT_{1,1}\) has only two short cuts \(\hbZ_{\pm 1}\), and is regular otherwise. This condition is also in our list in  \secref{Notations}.
In the previous subsection we showed that
\(\hat h\) is analytic in the upper half-plane.
Since \(\hat\wT_{1,1}=\hat h^+\bar{\hat h}^{-}\hat\cT_{1,1}\)
we see that  \(\hat h\) has only one \(\hat Z_0\)-cut and is regular elsewhere in the complex plane.

Let us now show that \(\hat h\) can be chosen to be a real function.
For that we use equation (\ref{discontinuitiyh}). It was derived for real \(\hat h\) but this assumption can be relaxed
in the derivation:
\begin{equation*}
       \hat h^{[+0]}\bar \hh^{[-0]}=\frac{\CF^+(1-Y_{1,1}Y_{2,2})}{\rrho}\,.
\label{discontinuitiyhhbar}
\end{equation*}
The r.h.s of this equation is regular on the real axis. Indeed,
\(\rrho=\hat G^{[+0]}-\hat G^{[-0]}\), hence it behaves as a pure square root on the real axis.  Since \(Y_{1,1}^{[+0]}Y_{2,2}^{[+0]}=1/(Y_{1,1}^{[-0]}Y_{2,2}^{[-0]})\) and \(\CF^{[1+0]}/\CF^{[1-0]}=1/(Y_{1,1}^{[+0]}Y_{2,2}^{[+0]})\), the numerator of the r.h.s  also behaves  as a pure square root on the real axis.
Therefore we conclude that \(\hat h^{[+0]}\bar \hh^{[-0]}=\hat h^{[-0]}\bar \hh^{[+0]}\). Hence \(\bar\hh/\hh\) is a function  analytic everywhere. Its  large \(u\) behavior should be at most polynomial as it follows from the asymptotic solution, see (\ref{largeuhh}). For the asymptotic solution one can check   the absence of poles/zeroes in \(\bar\hh/\hh\) and we assume that this is also the case  for the exact \(\hh\). Therefore \(\bar\hh/\hh\) should be a constant. It is always possible to redefine the normalization of the \(\wT\)-gauge so that this constant is equal to one. Hence  \(\hh\)
can be indeed chosen as a real function.

\subsection{Deriving \texorpdfstring{${\mathbb Z}_4$}{Z4} invariance}
\label{sec:zfode}

\paragraph{\(\protect\mathbb Z_4\) invariance of the \(\wT\)-gauge:}
In \secref{sec:genesis} we show that the condition \(\hat\cT_{1,0}=0\)
implies the \(\mathbb Z_4\) symmetry of \(\hat\cT_{a,s}\) and only two magic Z-cuts for \(\hat\cT_{1,s}\). As a consequence, \(\hat\cT_{2,s}\) is given by \(\hat\cT_{2,s}=\hat\cT_{1,1}^{[+s]}\hat\cT_{1,1}^{[-s]}\) and hence it has only four magic Z-cuts.
Using the reality of \(\hh\) we see that Since \(\wT\)- and \(\cT\)- gauges are related by
(\ref{wT-->h*calT}),
the  \(\mathbb Z_4\) symmetry of the \(\wT\)-gauge follows from the \(\mathbb Z_4\) symmetry of the \(\cT\)-gauge. Since \(\hat h\) has only one cut, the magic cut structure of the \(\cT\)-gauge is also preserved in the \(\wT\)-gauge.
\paragraph{\(\protect\mathbb Z_4\) invariance of the  \(\bT\)-gauge:}  The simplest is to derive \(\mathbb Z_4\) symmetry on the border. Since \(\bT_{a,2}=\bT_{2,a}=\wT_{2,a}\) and \(\wT\)-gauge is \(\mathbb Z_4\)-symmetric, we get:
\begin{equation}\label{zmagic1}
        \hat\bT_{a,2}=\hat\bT_{-a,2}\;.
\end{equation}
Moreover the TBA equations \(Y_{1,a}\) imply (see (\ref{resultTBAupperright}))
\begin{equation}\label{zmagic2}
\hat\bT_{0,1}=0\;.
\end{equation}
Then, the followings equalities
\begin{equation}
\frac{\hat\bT_{1,1}^2}{\hat\bT_{0,0}}\frac{1}{\hat\bT_{0,2}}={\hat\wT_{1,1}^2}\frac 1{\hat\wT_{2,0}}=1
\end{equation}
and Hirota equation in the magic at the node \((0,1)\): \(\hat\bT_{1,1}\hat\bT_{-1,1}+\hat\bT_{0,0}\hat\bT_{0,2}=0\) imply that
\begin{equation}\label{zmagic3}
        \hat\bT_{1,1}=-\hat\bT_{-1,1}.
\end{equation}
Finally, consider \(u\) slightly above real axis with \(\Re(u)>2g\), so that \(\hat\CF^-=\CF^-=\CF^+\). Then:
\begin{equation}\label{eqqeqqq}
        \frac{\bT_{1,0}^2}{\hat\bT_{0,0}^+\hat\bT_{0,0}^-}=\left(\frac{\bT_{0,1}Y_{1,1}Y_{2,2}}{\hat\cF^+\hat\cF^-}\right)^2=\left(\frac{(\cF^-)^2}{\hat\cF^+\hat\cF^-}\frac {\hat\CF^+}{\CF^+}\right)^2=1.
\end{equation}
From the Hirota equation in the magic \(\hat\bT_{0,0}^+\hat\bT_{0,0}^-=\hat\bT_{1,0}\hat\bT_{-1,0}\) and (\ref{eqqeqqq}) it follows that
\begin{equation}\label{zmagic4}
        \hat\bT_{1,0}=\hat\bT_{-1,0}.
\end{equation}

\paragraph{Relation between \(Z_4\) symmetries in the right and upper bands:}
Let us show that the property \(\hat\bT_{0,1}=0\) can be obtained from the property \(\hat\wT_{1,0}=0\)
instead of the TBA equations for \(Y_{1,a}\).
To see this  let us use that \(\bT_{1,1}=\hat\wT_{1,1}\CF\). Then:
\begin{equation}\label{preT010}
   \hat\bT_{1,1}^+\hat\bT_{1,1}^-=\hat\wT_{1,1}^+\hat\wT_{1,1}^-\frac{\hat\cF^{+}}{\CF^{+}}(\cF^+\hat \cF^-).
\end{equation}
Let us consider this equation for  \(\Re(u)>2g\) and slightly above the real axis. Now let us note that
\(\CF^+=\hat\CF^-=\sqrt{\bT_{0,0}^+}=\sqrt{\bT_{0,1}}\) and \(\log\hat\cF^+/\cF^+=-\disc(\CF^+)=\log Y_{1,1}Y_{2,2}=\bT_{1,0}/\bT_{0,1}\).
Since the \(\wT\)-gauge is \(\mathbb Z_4\)-symmetric, \(\hat\wT_{1,1}^+\hat\wT_{1,1}^-=\hat\wT_{2,1}\).
On the other hand, from (\ref{typical}), (\ref{eq:Fnice}) and  Hirota equations in the magic one gets
\(\hat\wT_{2,1}=\hat\bT_{1,2}\).  Plugging these relations into (\ref{preT010}) we get
\begin{equation}
\hat\bT_{1,1}^+\hat\bT_{1,1}^-=\hat\bT_{1,2}\hat\bT_{1,0}.
\end{equation}
Since the full Hirota equation for \(\hat\bT_{1,1}\) in the magic is \(\hat\bT_{1,1}^+\hat\bT_{1,1}^-=\hat\bT_{1,2}\hat\bT_{1,0}+\hat\bT_{2,1}\hat\bT_{0,1}\) and \(\hat\bT_{2,1}\neq0\) we conclude that
\begin{equation}\label{zmagic2b}
\hat\bT_{0,1}=0\;.
\end{equation}

Therefore we can derive equations (\ref{zmagic1}),(\ref{zmagic2}),(\ref{zmagic3}) from the \(\mathbb Z_4\) symmetry of the \(\wT\)-gauge.
Note that this derivation is reversible and we can use equations (\ref{zmagic1}),(\ref{zmagic2}),(\ref{zmagic3}) as an input to derive  the \(\mathbb Z_4\) symmetry of the \(\wT\)-gauge. Finally, (\ref{zmagic1}),(\ref{zmagic2}),(\ref{zmagic3}),(\ref{zmagic4}) is the  only input needed for the derivation of the formulae (\ref{repbt})
from the next appendix that explicitly show that the generic  form of the \(\mathbb Z_4\) symmetry (\ref{Z4up}) holds.

\section{Details of the Wronskian solution for the upper band}
\label{app:wronskian}
In this appendix, the Wronskian formalism of \secref{sec:wronskian-solution-} is applied to derive
  the analyticity properties in the upper band.
  We also derive here the relation  between various \(\ps\)- and \(\qs\)- functions and find a basis of these functions  where all the \(\qs\)-functions are analytic in the upper half-plane. We also show that, unfortunately, there is no basis in which all \(\qs\)-functions of the upper band  have a finite number of \(\hbZ\)-cuts.

\subsection{\texorpdfstring{$1/2$}{1/2}-Baxter equations}
The simplest \(\mathbb Z_4\) condition \(\hat\bT_{0,1}=0\) implies that
the columns in the corresponding determinant \eq{Wronskianupper} are linearly dependent:
 \begin{equation}\label{q-lin-p}\qs=\a\,\ps^{++}+\beta\,\ps+\gamma\,\ps^{--}\,,\end{equation} where \(\a,\b,\gamma\) are some functions of the spectral parameter.
 We can easily determine these coefficients by computing \(\hat\bT_{\pm1,1}\) and \(\hat\bT_{2,-1}\).
 For example\footnote{We remind here that, as discussed in \secref{sec:upper-band}, the relation \(\qs_\es = \qs_{(4)}=
     \ps_\es = \ps_{(4)} = \hat\wT_{1,1}\) holds.
}
\begin{eqnarray}
        \hat\bT_{+1,1}&=&\qs^{+}\wedge \ps_{(3)}^-=\a^+\frac{\ps^{[+3]}\wedge\ps^{+}\wedge\ps^-\wedge\ps^{[-3]}}{\ps_{\emptyset}\ps_{\emptyset}^{--}}={\a^+}{\hat\bT_{1,2}^+}\,,\nonumber\\
        \hat\bT_{-1,1}&=&\qs^{-}\wedge \ps_{(3)}^+=\gamma^-\frac{\ps^{[-3]}\wedge\ps^{[+3]}\wedge\ps^+\wedge\ps^{-}}{\ps_{\emptyset}^{++}\ps_{\emptyset}}=-{\gamma^-}\hat\bT_{1,2}^-\,,\\
    \nonumber \hat\bT_{2,-1}&=&\qs_{(3)}^{++}\wedge\ps^{--}=
    -\qs_{(3)}^{++}\wedge\left(
    \frac{\a}{\gamma}\ps^{++}+
    \frac{\beta}{\gamma}\ps
    \right)=-\frac\a\gamma\hat\bT_{0,-1}^{++}-\frac\beta\gamma\hat\bT_{1,-1}^+.
\end{eqnarray}
Therefore we get  \(\a=\hat\bT_{1,1}^-/\hat\bT_{1,2}\), \(\gamma=\hat\bT_{1,1}^+/\hat\bT_{1,2}\), \(\beta=-{\hat\bT_{2,1}}/{\hat\bT_{1,2}}\).
Analogously, from \(\hat\bT_{0,-1}=0\) a linear relation follows
\begin{equation}\ps=\a\,\qs^{++}+\beta\,\qs+\gamma\,\qs^{--}\end{equation}  with the same \(\a,\b,\gamma\) due to the LR symmetry, and we get the following pair  of ``1/2-Baxter'' equations:

\begin{equation}
\label{halfBaxter}
\begin{array}{ccc}
        \hat\bT_{1,2}\,\qs+\hat\bT_{2,1}\ps&=&\hat\bT_{1,1}^-\ps^{++}+\hat\bT_{1,1}^+\ps^{--}\,,\\
        \hat\bT_{1,2}\,\ps+\hat\bT_{2,1}\qs&=&\hat\bT_{1,1}^-\qs^{++}+\hat\bT_{1,1}^+\qs^{--}\,.
\end{array}
\end{equation}

Excluding \(\ps\) (resp \(\qs\))
from these two equations 
we get a finite difference equation of the 4th order  on
\(\ps\) (resp \(\qs\)) only,  which  has  a form of the Baxter
equation for the \(\su(4)\) spin chain.
In what follows, we will make some important observations based on (\ref{halfBaxter}).

\subsection{Periodic 2-forms}
The sum and the difference of the \(1/2\)-Baxter equations read
\begin{equation}
\label{halfBaxterexplicit}
\begin{array}{ccc}
(\hat\bT_{2,1}+\hat\bT_{1,2})(\qs+\ps)&=&\hat\bT_{1,1}^-(\qs+\ps)^{++}+\hat\bT_{1,1}^+(\qs+\ps)^{--}\,,\\
(\hat\bT_{2,1}-\hat\bT_{1,2})(\qs-\ps)&=&\hat\bT_{1,1}^-(\qs-\ps)^{++}+\hat\bT_{1,1}^+(\qs-\ps)^{--}\,.
\end{array}
\end{equation}
These are the second order difference equations and thus the Wronskians
\(\tau_\pm=(\qs\pm\ps)^+\wedge(\qs\pm\ps)^-/{\hat\bT_{1,1}}\) should be \(i\)-periodic functions (playing the same  role as constants for the differential equations).
To see this one can for instance multiply the first equation in \eq{halfBaxterexplicit} by \((\qs+\ps)\wedge\).
The l.h.s. gives zero whereas the r.h.s. gives precisely the condition of \(i\)-periodicity of \(\tau_+\).
It is convenient to introduce instead of \(\tau_{\pm}\) their linear
combinations \(\omega=\frac{\tau_++\tau_-}2\) and
  \(\chi=\frac{\tau_+-\tau_-}2\)~:
\begin{equation}\label{omega-chi}
        \o=\frac 1{\hat\bT_{1,1}}(\qs^+\wedge\qs^-+\ps^+\wedge\ps^-)\;\;,\;\;
        \chi=\frac 1{\hat\bT_{1,1}}(\qs^+\wedge \ps^-+\ps^+\wedge \qs^-)\;.
\end{equation}
These periodic 2-forms allow to relate \(\ps_{(3)},\;\qs_{(3)}\) and \(\qs,\;\ps\):
\begin{subequations}
\la{pqrelationsboth}
\begin{eqnarray}\label{pqrelations}
\qs\wedge\o^-=+\ps_{(3)}\;\;&,&\;\;\ps\wedge\o^-=+\qs_{(3)}\\
\qs\wedge\chi^-=-\qs_{(3)}\;\;&,&\;\;\ps\wedge\chi^-=-\ps_{(3)}\;.
\end{eqnarray}
\end{subequations}
The derivation is straightforward. For example, using \eqref{q-lin-p}:
\begin{eqnarray*}\label{p3toq1}
\qs
\wedge\o^-&=&
\frac{\qs\wedge\ps\wedge\ps^{--}}{\hat\bT_{1,1}^-}
=
\frac{\ps^{++}\wedge\ps\wedge\ps^{--}}{\hat\bT_{1,2}}
=
  +\ps_{(3)}\;,\\
 \qs\wedge\chi^-&=&
 \frac{\qs\wedge\ps\wedge\qs^{--}}{\hat\bT_{1,1}^-}
 =
 \frac{\qs\wedge\qs^{++}\wedge\qs^{--}}{\hat\bT_{1,2}}
 =
  -\qs_{(3)}\;.
\end{eqnarray*}

\subsection{Expressions for \texorpdfstring{$\bT$}{T}}
Using the relations from the previous section we can exclude \(\ps\) from the expressions for T-functions to get:
\begin{subequations}
\label{repbt}
\begin{eqnarray}
\label{repbt2}
        \hat\bT_{a,2}&=&\qs_\es^{[+a]}\qs_\es^{[-a]}\;,\\
\label{repbt1}
        \hat\bT_{a,1}&=&\qs^{[+a]}\wedge\qs^{[-a]}\wedge\o^{[a-1]}\;,\\
\label{repbt0}
        \hat\bT_{a,0}&=&-\qs_{(2)}^{[+a]}\wedge\qs_{(2)}^{[-a]}+\DF^{[+a]}\DF^{[-a]}.
\end{eqnarray}
\end{subequations}
Here is the derivation of the last, the most complicated
formula. First, we 
notice that the \(\mathbb Z_4\)
symmetric Hirota equation
\(\hat\bT_{0,1}^+\hat\bT_{0,1}^-=\hat\bT_{0,0}\hat\bT_{0,2}-\hat\bT_{1,1}^2=0\) allows to substitute   \(\hat\bT_{1,1}\)
with \(\hat
\qs_0 \sqrt{\hat\bT_{0,0}}= \CF ~\qs_0\). This allows to rewrite the
first equality in \eqref{omega-chi} as
\begin{equation}\label{omegaq2p2}
        \o=\frac 1\CF(\qs_{(2)}+\ps_{(2)}).
\end{equation}
Explicit calculation then gives:
\begin{eqnarray}
\hat\bT_{a,0}=\qs_{(2)}^{[+a]}\wedge\ps_{(2)}^{[-a]}&=&-\qs_{(2)}^{[+a]}\wedge\qs_{(2)}^{[-a]}+\CF^{[-a]}\qs_{(2)}^{[+a]}\wedge\o^{[\pm a]}=-\qs_{(2)}^{[+a]}\wedge\qs_{(2)}^{[-a]}+\frac{\CF^{[-a]}}{\CF^{[+a]}}\qs_{(2)}^{[+a]}\wedge\ps_{(2)}^{[+a]}\nonumber\\
&=&-\qs_{(2)}^{[+a]}\wedge\qs_{(2)}^{[-a]}+\frac{\CF^{[-a]}}{\CF^{[+a]}}\hat\bT_{0,0}^{[+a]}=-\qs_{(2)}^{[+a]}\wedge\qs_{(2)}^{[-a]}+{\CF^{[-a]}}{\CF^{[+a]}}\,.
\end{eqnarray}

The representations (\ref{repbt2}),  (\ref{repbt1}) and (\ref{repbt0}) explicitly show that the expressions for  \(\hat\bT_{a,s}\) in terms of the Wronskian ansatz (\ref{Wronskianupper}) explicitly satisfy the general \(\mathbb Z_4\) property \(\hat\bT_{-a,s}=(-1)^s\hat\bT_{a,s}\)  announced in
\secref{sec:wronskian-solution-}.

\subsection{Relation between \texorpdfstring{$\ps$ and $\qs$}{p and q}}
\label{subsec:V^2=1}
Both \(\qs_i\)  and \(\ps_i\) satisfy the same \(4^{th}\) order finite difference equation and thus each \(\ps\) should be a linear combination  of \(\qs\)'s with  \(i\)-periodic coefficients. It is easy to see from \eq{pqrelations}
that in fact these coefficients are:
\begin{equation}\label{pq}
\ps_i = {V_{i}}^j \qs_j\;\;,\;\;{V_i}^j=- \omega_{ik}\chi^{kj}\;,
\end{equation}
where \(\chi^{ij}\equiv\frac 12\e^{ijkl}\chi_{kl}\). For that we use that the
Pfaffian of \(\chi_{ij}\) is \(-1\):
\begin{equation}
\chi\wedge\chi=-2\frac{\wT_{1,1}^2}{\bT_{1,1}^2}\qs_{(2)}\wedge\ps_{(2)}=-2\frac{\wT_{1,1}^2}{\bT_{1,1}^2}\bT_{0,0}=-2\,,
\end{equation} where we used \eqref{q0=q4=p0=p4=1}.
Similarly, the Pfaffian of \(\o_{ij}\) can be shown to be \(1\)
and so
\begin{equation}
        \o^{ij}\equiv\frac 12\e^{ijkl}\o_{kl}=-(\o^{-1})^{ij},\ \ \chi^{ij}\equiv\frac 12\e^{ijkl}\chi_{kl}=(\chi^{-1})^{ij}.
\end{equation}
As a consequence \(V^2=1\) which implies that the inverse relation has precisely the same form \(\qs_i={V_i}^j\ps_j\)
and thus it ensures  the LR symmetry. In addition one can check that \(\Tr\,V=0\). Hence the eigenvalues of \(V\) are  \((+1,+1,-1,-1)\).

\subsection{\label{app:darbouxbasis}Darboux basis}
Let us now use the freedom \eq{Hsym} in the choice of Q-functions
to simplify the relation among \(\ps,\;\qs,\ps_{(3)}\) and \(\qs_{(3)}\) (\ref{pqrelationsboth},\ref{pq}).
In order to preserve \(\qs_{\emptyset}=\qs_{\bar\emptyset}=\ps_{\emptyset}=\ps_{\bar\emptyset}=\wT_{1,1}\)
we restrict ourselves to the transformations with \(\det H=1\).
Under the transformation \eq{Hsym} \(\o_{ij}\) and \(\chi_{ij}\) transform covariantly while \(V_i^{\ j}\) transforms by the adjoint action:
\begin{equation}
        \o\to H\o H^{\rm T},\ \ \chi\to H\chi H^{\rm T},\ \ V\to HV H^{-1}.
\end{equation}
Note that \(V\) by itself is an \(H\)-transformation. In addition, since \(V\) just interchanges \(\ps\) and \(\qs\), it leaves \(\o\) and \(\chi\) invariant.

 Since \(\o_{ij}\) is a skew-symmetric matrix, it is always possible to choose an \(H\)-transformation that brings \(\o_{ij}\) to the Darboux form, i.e. to the matrix with \(\o_{12}=-\o_{21}=\o_{34}=-\o_{43}=1\) and with all other matrix elements being zero.
The remaining freedom in \(H\)-transformation is the \(\mathfrak{sp}(4)\) subgroup, and \(V\) is its particular element. The action of \(H\) on \(V\) generates an adjoint orbit which always passes through a Cartan element,  hence \(V\) can be brought to the
form \({\rm diag}(1,1,-1,-1)\). Therefore it is possible to find a Q-basis in which:
\begin{equation}
        \o=\left(\begin{array}{cccc}
0 & 1 & 0 & 0 \\
-1 & 0 & 0 & 0 \\
0 & 0 & 0 & 1 \\
0 & 0 & -1 & 0 \\
\end{array}\right),\ \ \chi=\left(\begin{array}{cccc}
0 & 1 & 0 & 0 \\
-1 & 0 & 0 & 0 \\
0 & 0 & 0 & -1 \\
0 & 0 & 1 & 0 \\
\end{array}\right).
\end{equation}

In this canonical basis we get simple relations\footnote{these relations are not the same  as in  \eqref{QfunPfun}.}
\begin{eqnarray}
\ps_{123}&=&\qs_3=-\ps_3=-\qs_{123},\nonumber\\
\ps_{124}&=&\qs_4=-\ps_4=-\qs_{124},\nonumber\\
\ps_{234}&=&\qs_2=+\ps_2=+\qs_{234},\nonumber\\
\ps_{134}&=&\qs_1=+\ps_1=+\qs_{134}.
\end{eqnarray}
Notice that  we still have a residual symmetry \(\mathfrak{sp}(2)\times\mathfrak{sp}(2)\) preserving both \(\omega\) and \(\chi\).

The Darboux basis is a direct analog of the 1-cut basis for the horizontal \(\su(2)\)-band of the section \ref{sec:rightband}. It is tempting to ask whether it is  possible also to find a basis of \(\qs\)-functions all having a finite number of cuts. Unfortunately, as we shell show later this appears to be not possible.

\subsection{Existence of a basis  regular in a half-plane}

In this subsection we generalize the analyticity arguments, given for the right band in the main text,
to the upper band. The situation in the upper band is a little bit more complicated and in particular
we will see in the subsequent section that it is not possible to choose a basis with just a single cut. However, we will prove that there exists a basis in which \(\qs_{(k)}\)-forms are regular everywhere above \(\hat Z_{|k-2|-1}\) cut and \(\ps_{(k)}\)-forms are regular everywhere below \(\hat Z_{-|k-2|+1}\) cut.

We start by writing an analog of \eq{Baxtergeneral}. The idea of derivation is  simple:
we want to write \({\qs}_i^{[2a-1]}\) in terms of \(\bT\) functions with a shift \(\sim a\) and of the other \({\qs}_i\) or \({\ps}_i\)
functions with shifts independent of \(a\). This is straightforward to achieve using \eqref{Wronskianupper} which we copy here for convenience:
\beq
\bT_{a,1}=\bT_{a,-1}={\qs}_{(3)}^{[+a]}\wedge
{\ps}^{[-a]}={\qs}_{}^{[+a]}\wedge
{\ps}_{(3)}^{[-a]}\;.
\eeq
Indeed, we notice that   expressions for \({\bT}_{a-2,1}^{[a+1]},\;{\bT}_{a-1,1}^{[a]},\;{\bT}_{a,1}^{[a-1]},\;{\bT}_{a+1,1}^{[a-2]}\)
are four independent linear functions of four components of \({\qs}_{(3)}^{[2a-1]}\)
with coefficients containing only \(a\)-independent shifts. Therefore \({\qs}_{(3)}\)
can be written as  certain linear combinations of these functions. Explicitly:
\beqa\la{njashka}
\nonumber
({\ps}^{[+3]}\wedge {\ps}^{[+1]}\wedge {\ps}^{[-1]}\wedge {\ps}^{[-3]})
{\qs}_{(3)}^{[2a-1]} &=&
-{\ps}^{[+1]}\wedge {\ps}^{[-1]}\wedge {\ps}^{[-3]}\bT_{a-2,1}^{[a+1]}\\
\nonumber&&+{\ps}^{[+3]}\wedge {\ps}^{[-1]}\wedge {\ps}^{[-3]}\bT_{a-1,1}^{[a]}\\
\nonumber&&-{\ps}^{[+3]}\wedge {\ps}^{[+1]}\wedge {\ps}^{[-3]}\bT_{a,1}^{[a-1]}\\
&&+{\ps}^{[+3]}\wedge {\ps}^{[+1]}\wedge {\ps}^{[-1]}\bT_{a+1,1}^{[a-2]}\;.
\eeqa
The coefficients in front of T-functions is very easy to check by acting with \(\wedge {\ps}^{[n]},\;n=-3,-1,1,3\)
on both sides of the equality.

Now we make the following observation. Let us first shift the argument in  both
sides of \eq{njashka} by \(-5i/2\) so that \({\qs}_{(3)}^{[2a-6]}\) is expressed in terms of
\({\bT}_{a-2,1}^{[a-4]},\;{\bT}_{a-1,1}^{[a-5]},\;{\bT}_{a,1}^{[a-6]},\;{\bT}_{a+1,1}^{[a-7]}\)
with coefficients depending on \({\ps}^{[-2]},\;{\ps}^{[-4]},\;{\ps}^{[-6]},\;{\ps}^{[-8]}\).
Notice that all these T-functions are shifted such that they are still regular on the real axis for \(a\geq 4\). Thus if we find a basis where \({\ps}\) does not have cuts \(\hbZ_{-2},\;\hbZ_{-4},\;\hbZ_{-6},\;\hbZ_{-8}\)
then \({\qs}_{(3)}\) will be immediately regular in the whole upper half-plane.

Note that we can reverse all shifts and interchange \({\qs}_{(3)}\) with \(*{\ps}\)~\footnote{the \(\star\) is the Hodge dualization.} in \eq{njashka}. From that equation we
can say that in the basis where \(\qs\) does not have the cuts \(\hbZ_{+2},\;\hbZ_{+4},\;\hbZ_{+6},\;\hbZ_{+8}\),
\({\ps}\) must be regular everywhere in the lower half-plane.

Using the argument from the main text we can argue that the cuts
\(\hbZ_{2},\;\hbZ_{4},\;\hbZ_{6},\;\hbZ_{8}\) in \({\qs}_{(3)}\) can be always canceled by an appropriate H-transformation.
Then in this basis we see that \({\qs}_{(3)}\) will be in fact free of all cuts
above the real axis and \({\ps}\) will have no cuts below the real axis.
Note that in this basis \({\ps}_{(3)}\), as a certain combination of \({\ps}^{[-2]},\;{\ps},\;{\ps}^{[2]}\),
should be regular at least starting from \(-i\) and below.

To complete our proof let us make two further observations.
First, due to the LR symmetry \(\bT_{a,1}=\bT_{a,-1}\)
we can also  make a replacement \({\qs}\to *{\qs}_{(3)}\) and
\({\ps}\to *{\ps}_{(3)}\)
in \eq{njashka}.
Second, we notice that the regularity argument also works
when in \({\ps}_{(3)}\) there are no cuts \(\hbZ_{-4},\;\hbZ_{-6},\;\hbZ_{-8},\;\hbZ_{-10}\) instead. For that
one should shift \eq{njashka} by \(-6i/2\) and use it for \(a\geq 5\).
Thus we conclude that in the same basis \({\qs}\) and \({\qs}_{(3)}\) are regular in the upper half-plane and
\({\ps}\) and \({\ps}_{(3)}\) are regular in the lower half-plane.
It remains to show that \({\qs}_{(2)}\) and \({\ps}_{(2)}\)
are also regular in their half-planes. It is immediately obvious that, say,
\({\qs}_{(2)}\) is regular starting from \(\hbZ_1\). In fact we can prove that
the cut \(\hbZ_1\) is absent and the function is regular everywhere
above \(\hbZ_{-1}\).

Indeed, let us rewrite the Pl\"ucker relation (\ref{Plucker2}) in the following form:
\begin{equation*}
        \frac{\qs_{kij}\qs_k}{\qs_{ki}^+\qs_{kj}^+}=\frac{\qs_{kj}^-}{\qs_{kj}^+}-\frac{\qs_{ki}^-}{\qs_{ki}^+}\,.
\end{equation*}
The l.h.s of this relation is regular everywhere above the real axis.
In particular, the discontinuity at \(\hbZ_2\) should be trivial, which gives for the r.h.s.:
\begin{equation}\label{discequality}
    o\equiv\frac{\disc(\qs_{kj}^{[+1]})}{\qs_{kj}^{[+3]}}=\frac{\disc(\qs_{kr}^{[+1]})}{\qs_{kr}^{[+3]}},\ \forall {k,j,r}.
\end{equation}
We see that \(o\) is some universal function which should be the same for any pair of indexes,
since \(\qs_{ij}\) is an antisymmetric tensor.
Consider now
\begin{eqnarray}
         \bT_{2,0}^{-}&=&\qs_{(2)}^{+}\wedge\ps_{(2)}^{[-3]}=\frac 14\e^{ijkl}\qs_{ij}^{[+1]}\ps_{kl}^{[-3]}.
\end{eqnarray}
Since \(\disc(\bT_{2,0}^-)=0\), one gets
\be
0=\frac 14\e^{ijkl}\disc(\qs_{ij}^{[+1]})\ps_{kl}^{[-3]}=o\frac 14\e^{ijkl}\qs_{ij}^{[+1]}\ps_{kl}^{[-3]}=o\bT_{3,0}\,.
\ee
We see therefore that \(o=0\) and hence \(\disc(\qs_{ij}^+)=0\), i.e. we proved
that \(\qs_{(2)}\) is regular above \(\hbZ_{-1}\)-cut. The same logic
 applies to \(\ps_{(2)}\) which is thus regular below the \(\hbZ_1\)-cut.
This accomplishes the proof about the possibility to construct a basis  which is regular in a half-plane announced in the beginning of this subsection.

\subsection{No-go theorem:  There is no \texorpdfstring{$\qs$}{q}-basis with a finite number of short cuts}

{\it Proof:} Suppose  there exists such a basis. Then by  considerations from the previous subsection \(\qs_{(2)}\) should be analytic in the upper half-plane up to \(\hbZ_{-1}\) and in the lower half-plane up to
\(\hbZ_{1}\), i.e. it should be analytic in the whole plane. The same is true for \(\ps_{(2)}\). Therefore \(\bT_{a,0}\) in \eqref{Wronskianupper} and hence \(Y_{a,0}=\frac{\bT_{a,0}^+\bT_{a,0}^-}{\bT_{a+1,0}\bT_{a-1,0}}-1\) should not have any cuts at all which leads to an obvious contradiction with the basic properties of the Y-system, starting from the large \(L\) asymptotics.

This no-go theorem shows that the Darboux basis cannot be in this respect as nice as the 1-cut basis of the right(left) band. It even cannot be  a half-plane regular one. Indeed, in the Darboux basis \(\ps_{12}=\qs_{34}\) and hence, due to \eqref{omegaq2p2} \(\DF=\DF \o_{12}=\qs_{(2)}+\ps_{(2)}=\qs_{12}+\qs_{34}\). \(\CF\) definitely has a
\(\hbZ_{1}\) cut, hence \(\qs_{(2)}\) cannot be regular at \(\hbZ_1\). However, the existence of a periodic form \(\omega\), with the periodic structure of  short Z-cuts, may potentially allow to deform the Darboux basis in a way to  parameterize the upper band in terms of densities with the finite \([-2g,2g]\) support only.

\subsection{\label{app:compbasis}Basis used for computations}
In this subsection we show existence of the basis which is used for explicit computations\footnote{Note that this not a Darboux basis of \appref{app:darbouxbasis}. In particular, structure of \(\o\) is complicated.}. This basis is regular in a half-plane and posses particularly chosen relations (\ref{QfunPfun}) among \(\ps\)-s and \(\qs\)-s.

In addition to LR and \(\mathbb Z_4\) symmetries which manifest themselves in existence of respectively 2-forms \(\chi\) and \(\omega\) there is also a reality property of T-functions which we will now encode into the properties of   an  \(i\)-periodic \(4\times 4\) matrix \(S\) defined by:\be
        \qs_k=i{S_k}^j\bar \ps_j.
\ee
The latter relation should exist since complex conjugated quantities \(\bar \ps\) and \(\bar\qs\) satisfy the same Baxter equations (\ref{halfBaxter}) as \(\ps\) and \(\qs\).
It also follows from (\ref{halfBaxter}) that
\be
        \ps_k=i{S_k}^j\bar \qs_j.
\ee

\(S\) transforms under the \(H\)-transformation
\eq{Hsym} as \(S\to HS\bar H^{-1}\) and enjoys the following properties:\begin{subequations}
\be
      \det\,S&=&1,\ \ \\
\la{SSSS}      {\bar S}\,S&=&1,\ \\
\la{SSSS2}      S\bar\chi S^{T}&=&\chi\;.
\ee
\end{subequations}
These properties follow correspondingly from (\ref{q0=q4=p0=p4=1}), reality of T-functions,  and
definition (\ref{omega-chi}) of \(\chi\).

Relations  (\ref{QfunPfun}) mean a particular choice of \(\chi\) and \(S\):
\begin{equation}
\chi_0\equiv\left(\begin{array}{cccc}
0 & 0 & 0 & 1 \\
0 & 0 & -1 & 0 \\
0 & 1 & 0 & 0 \\
-1 & 0 & 0 & 0 \\
\end{array}\right),\ \ S_0\equiv \frac1i\left(\begin{array}{cccc}
0 & 0 & -1 & 0 \\
0 & 0 & 0 & 1 \\
-1 & 0 & 0 & 0 \\
0 & 1 & 0 & 0 \\
\end{array}\right).
\end{equation}
Let us now show that this particular choice can be always done by using the \(H\)-symmetry
\eq{Hsym}.  First, we bring \(\chi\) to the form \(\chi_0\). After that we only can use
the H-transformations restricted by:
\beq
\chi_0=H\chi_0 H^T\;.
\eeq
It is easy to construct such transformation explicitly:\footnote{
Since \(S\) is an element of \(SP(4)\) (see \eq{SSSS2}) it can be written as \(e^{iR}\)
where \(R\) is an element of the algebra \({\frak{sp}}(4)\) and
 \(R=\bar R\) due to \eq{SSSS}.
 We define \(\sqrt{S}=e^{i R/2}\). One can see that \(\sqrt{S}\overline{\sqrt{S}}=1\) and also \eq{SSSS2} holds for \(\sqrt{S}\).
In the same way we define \(\sqrt{S_0}\).}
\beq
H={\sqrt{S_0}}\overline{\sqrt{S}}\;.
\eeq
Since \(S\) transforms as \(S\to H S \bar H^{-1}\) we get
\beq
S\to \left({\sqrt{S_0}}\overline{\sqrt{S}}\right)S\left(\overline{\sqrt{S}}{\sqrt{S_0}}\right)=S_0\;.
\eeq

Note that once \(\qs\)'s are chosen to be regular in the upper half-plane
(and \(\ps\)'s are chosen to be regular in the lower half-plane), \(\chi\) and \(S\) become also regular. Hence all the transformations discussed in the previous paragraph are done by the regular \(H\)-matrices and thus do not spoil analyticity properties of the \(\qs\)-basis.

\section{Further details about FiNLIE equations}
\label{sec:finliedetails}
\subsection{Uniqueness of the physical gauge}
\label{sec:uniqueness}

We are going to show that conditions (\ref{propbT}) completely constrain the \(\bT\)-gauge, up to an overall constant.
First, let us note that the gauge transformations that preserve the LR symmetry \(\bT_{a,-s}=\bT_{a,s}\), reality \(\bT_{a,s}=\overline \bT_{a,s}\), and analyticity \(\bT_{a,s}\in\CA_{1+|a-s|}\) are of the following form:
\begin{equation}
        \bT_{a,s}\to g^{[a+s]} g^{[a-s]}\bar{g}^{[-a-s]}\bar{g}^{[-a+s]}\bT_{a,s},
\end{equation}
where the function \(g\) is analytic in the upper half-plane above \(Z_{-1}\) and \(\bar{g}\) is the complex conjugate of \(g\).

To preserve the \(\bT_{2,3}=\bT_{3,2}\) condition \(g/\bar{g}\) should be a periodic function. To preserve the \(\bT_{0,0}^+=\bT_{0,0}^-\) condition \(g\bar{g}\) should be also a periodic function. Hence \(g=\sqrt{g\bar g\times g/\bar g}\) is periodic by itself. Then, since \(g\) is analytic in the upper half-plane and periodic, it is analytic everywhere. Such periodic analytic function should not have poles, due to the condition of  absence of singularities, and no zeroes, due to the condition of a minimal possible number of zeroes. Therefore this gauge transformation can be only a constant.

\subsection{Reduction of  Bethe equations to  computable quantities}
\label{app:BetheRoots}

We need to express (\ref{BetheExact}) in terms of quantities that can be explicitly computed.
We start from the  fact that \(Y_{1,0}=\frac{\bT_{1,1}^2}{\bT_{2,0}\bT_{0,0}}=\frac{\wT_{1,1}^2}{\bT_{2,0}}\). Since \(\bT_{2,0}\in\CA_2\), its analytic continuation along the contour \(\gamma\) defined in \figref{fig:contourbethe} is trivial. Therefore (\ref{BetheExact}) can be equivalently written as:
\begin{equation}\label{bethefirstround}
        \frac{(\wT_{1,1}^{\gamma}(u_j))^2}{\bT_{2,0}(u_j)}=\frac{(\wT_{1,1}^{\gamma_+}(u_j))^2}{\bT_{2,0}(u_j)}=-1.
\end{equation}
It is allowed to replace \(\gamma\) with \(\gamma^+\) because the branch points are of the square root type and analytic continuation along \(\gamma\) and \(\gamma^+\) leads to the same results.
To find explicitly the analytically  continued function \(\wT_{1,1}^{\gamma_+}\) one uses
\(\wT_{1,1}=\hat h^+\hat h^-\CT_{1,1}\) and the density representation (\ref{eq:rho-to-CF})  \(\CT_{1,1}=1+\CC_1*\rrho\). For the latter the continuation picks a pole in the Cauchy kernel leading to
\begin{equation}\label{Bbbb1}
\cT_{1,1}^{\gamma_+}=\CT_{1,1}-\rrho^-.
\end{equation}

Analytic continuation of \(\hat h\) can be  found from the discontinuity property (\ref{discontinuitiyh}) which gives
\begin{equation}\label{Bbbb2}
  (\hat h^-)^{\gamma_+}=\frac 1{\hat h^-}\frac{\CF(1-Y_{1,1}^-Y_{2,2}^-)}{\rrho^-}=-\frac 1{\hat h^-}\frac{\CF Y_{1,1}^-Y_{2,2}^-}{\rrho^-}\,,\qquad u=u_j.
\end{equation}
At the last step we used that \(Y_{2,2}^-\) has a pole at the Bethe root. Note that \(\CF\) has a zero at the same position, therefore the whole expression is regular.

 The function \(\rrho^-\) can be found from equation~(\ref{rightmagic}) shifted by \(- i/2\). Again, at the Bethe root \(Y_{2,2}^-\) is singular, therefore  we can simplify the l.h.s of (\ref{rightmagic}) just to \((1+Y_{1,1}^-)^{-1}\). Explicitly, we get after some algebra:
\begin{equation}\label{Bbbb3}
\rrho^-=\frac{\CT_{1,1}\hat\CT_{1,1}^{[-2]}Y_{1,1}^{-}}{\CT_{1,2}^{-}+\hat\CT_{1,1}^{[-2]}Y_{1,1}^-},\qquad u=u_j\,.
\end{equation}
By combining (\ref{Bbbb1}),(\ref{Bbbb2}) and (\ref{Bbbb3}) we derive the final answer for \(\wT_{1,1}\):
\begin{equation}\label{subT11hup}
\wT_{1,1}^{\gamma_+}=-\frac{\hat h^+}{\hat h^-}\CF Y_{2,2}^-\frac{\CT_{1,2}^-}{\hat\CT_{1,1}^{[-2]}}\,,\qquad u=u_j.
\end{equation}

Let us now check  the following property:
\begin{equation}\label{bethereality}
        \frac{\wT_{1,1}^{\gamma_+}(u_j)\wT_{1,1}^{\gamma_-}(u_j)}{\bT_{2,0}(u_j)}=1.
\end{equation}
Applying Hirota equations on the magic sheet and using the relation between \(\bT\)- and \(\wT\)-gauges one can get:
\begin{equation}
\bT_{2,0}=\hat\CF^{++}\hat\CF^{--}-\wT_{1,1}^2.
\end{equation}
The first term on the r.h.s.  can be further transformed,  using the \(i\)-periodicity of \(\CF\) in the mirror, the relation (\ref{eq:discf}), and the Y-system equation (\ref{eq:Y-system}) for \(Y_{1,1}\), to:\begin{equation}\label{FppFmm}
\hat\CF^{++}\hat\CF^{--}=\CF^2(Y_{1,1}Y_{2,2})^+(Y_{1,1}Y_{2,2})^-=\CF^2Y_{2,2}^+Y_{2,2}^-(1+Y_{1,0})(1+Y_{1,2}),\qquad u=u_j\,.
\end{equation}
Note that at the last step we used that \(1+1/Y_{2,1}=1\) at the Bethe root.

On the other hand,  it  follows from (\ref{subT11hup}) and its complex conjugate:
\begin{equation}\label{wTupwTdn}
        \wT_{1,1}^{\gamma_+}\wT_{1,1}^{\gamma_-}=\CF^2Y_{2,2}^+Y_{2,2}^-\frac{\CT_{1,2}^-\CT_{1,2}^+}{\CT_{2,2}}=\CF^2Y_{2,2}^+Y_{2,2}^-(1+Y_{1,2})\;.
\end{equation}
The difference of (\ref{FppFmm}) and  (\ref{wTupwTdn}) gives:
\begin{equation*}
   \CF^2Y_{2,2}^+Y_{2,2}^-Y_{1,0}(1+Y_{1,2})=\frac{\hat\CF^{++}\hat\CF^{--}}{1+\frac 1{Y_{1,0}}}=\frac{\hat\CF^{++}\hat\CF^{--}}{\bT_{1,0}^+\bT_{1,0}^-}\bT_{1,1}^2=\frac{\hat\CF^{++}\hat\CF^{--}}{\hat\CF^{++}\CF^2\,\hat\CF^{--}}\bT_{1,1}^2=\wT_{1,1}^2,
\end{equation*}
which indeed assures (\ref{bethereality}).

The property (\ref{bethereality}) in particular implies that if
(\ref{bethefirstround}) holds then the conjugated equation \linebreak \({{(\hat\wT_{1,1}^{\gamma_-}(u_j))^2}=-{\bT_{2,0}(u_j)}}\) also holds, i.e. the Bethe roots always come in the conjugated pairs.
For the \(\sl(2)\) sector all the Bethe roots are expected to be real and the eq.(\ref{bethereality})  follows from (\ref{bethefirstround}) up to a sign. We hence checked what sign should be chosen.

The ratio of  \eqref{bethefirstround} and   (\ref{bethereality}) leads to:
\begin{equation}
         \frac{\hat\wT_{1,1}^{\gamma_+}(u_j)}{\hat\wT_{1,1}^{\gamma_-}(u_j)}=-1,
\end{equation}
which  after substitution (\ref{subT11hup}) gives
\begin{equation}
   \left(\frac{\hat h^+}{\hat
       h^-}\right)^2=-\frac{Y_{2,2}^+}{Y_{2,2}^-}\frac{\CT_{1,2}^+}{\CT_{1,2}^-}\frac{\hat\CT_{1,1}^{[-2]}}{\hat\CT_{1,1}^{[+2]}}.
\label{eq:BetheAn}
\end{equation}
This is the equation that we are using to determine the positions of the Bethe roots.

One can show from (\ref{relatingBUC2}) and the explicit definitions for
\(\bB\)  that (\ref{eq:BetheAn}) is equivalent to
\begin{equation}\label{eqnpreasymBethe}
      -\left(\frac{\hat \nU^-}{\hat \nU^+}\right)^2=\left(\frac{\sT_{2,1}^-}{\hat\sT_{1,1}^{[-2]}}\right)^2\frac{\hat\sT_{0,0}^{[-2]}}{\sT_{2,0}}\,.
\end{equation}
Then, by using explicit asymptotic expressions from \appref{app:asymptoticsolution} it is easy to see that (\ref{eqnpreasymBethe}) reduces to
\begin{equation}
        \left(\frac{\hat x^+}{\hat x^-}\right)^L=-\frac{Q^{--}}{Q^{++}}\left(\frac{\hat B^{(+)+}}{\hat B^{(-)-}}\right)^2\s_{BES}^2
\end{equation}
in the large volume limit. This is nothing but  the asymptotic Bethe  Ansatz equations for the \(sl(2)\) sector, as it should be.

\subsection{\label{sec:alteqnonCF}An alternative equation on \texorpdfstring{$\CF$}{F}}

The \(i\)-periodic function \(\CF=\sqrt{\bT_{0,0}}\) plays the key role in our constructions
since it is the function that relates the right band with the upper band. In this section we derive a simple expression for
\(\CF\) in terms  of the product \(Y_{1,1}Y_{2,2}\) which allows for a better understanding of the properties of \(\cF\).
For that we depart from  (\ref{eq:discf}) which can be considered as the Riemann-Hilbert equation on \(\CF\).  It is not  hard to solve this equation. One of the possible solutions is\footnote{Recall that \(\cF\) is \(i\)-periodic and hence it has an infinite set of long \(\check Z\)-cuts.}
\begin{equation}\label{F0int}
\CF_0(u)=\exp\left[
\int_{{\bZ}_0}\frac {dv}{2i}\left(\tanh\pi(u- v)+{\rm sign}(v)
  \rule{0pt}{2.3ex} \right)\log \left(Y_{1,1}(v)Y_{2,2}(v)
  \rule{0pt}{2.3ex} \right)
\right]\;,
\end{equation}
where  \({\rm sign}(v)\) is added for the convergence of the integral.
In general one can multiply \(\CF_0(u)\) by an arbitrary periodic function analytic in the whole complex plane.
The remaining function can be fixed from the knowledge that \(\CF\) has simple zeros at the positions of  Bethe roots:
\begin{equation}\label{Fexplicit}
        \CF(u)=\CF_0(u) \Lambda_\CF\prod_{i=1}^M\sinh(\pi(u-u_i))\;,
\end{equation}
where \(\L_\CF\) is a constant.

One should be careful with the \(2\pi i {\mathbb Z}\) ambiguity of \(\log(Y_{1,1}Y_{2,2})\) in the definition of \(\CF_0\).
This ambiguity is however uniquely fixed by the requirement that \(\log\CF_0\) should not have any logarithmic poles
at \(u=\tfrac{i}{2}\pm 2g\). This is only possible if the integrand is zero at \(v=\pm 2g\) which uniquely fixes the branch.

Let us now analyze  the large \(u\) behavior of \(\CF_0(u)\).
Expansion (\ref{Ylauexpan})  implies that \(\log(Y_{1,1}Y_{2,2})\to2\pi i m_\pm\)
at \(u\to \pm \infty+ i0\).
In order to determine the integers \(m_\pm\) we analytically continue the logarithm from \(u=2 g\), where we already fixed
\(\log(Y_{1,1}Y_{2,2})=0\), to \(\pm\infty+i0\) along the real axis.
At  weak coupling the Y-functions are known explicitly, see (\ref{asymYfer}), and we get:
\begin{equation}
Y_{1,1}Y_{2,2}\simeq\prod_{j=1}^M \frac{u-u_j+i/2}{u-u_j-i/2}
\end{equation}
from where we find \(m_--m_+=M\).
By the continuity argument, since \(m_\pm\) are integers, this relation should hold  for a finite coupling\footnote{unless there is a critical point at some value of the coupling - the possibility which we ignore since it was not realized so far for explicit examples.}. For the case of symmetric roots \(m_-=-m_+=M/2\).

One can check that
the exponential factor from \(\CF_0(u)\)
cancels against the one from the product of \(\sinh\) in \(\CF(u)\)
and we get a simple power-law asymptotics \(\CF\sim u^{E}\), in
complete agreement with (\ref{T00largeuexpansion}).

\subsection{Fixing the normalization constants}
\label{sec:fixing-norm-const}
A freedom in the overall scale of the \(\bT\)-gauge was used to choose the normalization for \(f\) in (\ref{eqf}). There is no further freedom and the integration constants \(\L\) and \(\L_{\CF}\)  appearing in this article are  fixed once the normalization for \(f\) is chosen.

\(\L_\CF\) appearing in (\ref{Fexplicit}) is found from the comparison of the large \(u\) asymptotics of \(\CF=f\bar f\sqrt{\sT_{0,0}}\) and (\ref{Fexplicit}) and is given by
\begin{equation}
\label{eq:LambdaF}
        \L_{\CF}=\exp\left[-2\pi Mg+E \log2g+M\log2-
\int_{2g+i0}^{\infty+i0} {dv}\left(-i\log \left(Y_{1,1}(v)Y_{2,2}(v)\right)+\pi M-\frac Ev\right)
\right],
\end{equation}
where the branch of \(\log Y_{1,1}Y_{2,2}\) is the same as in (\ref{F0int}).

The comparison of  (\ref{Fexplicit})  and \(f\bar f\sqrt{\sT_{0,0}}\) at large \(u\) reveals also a practical expression for the energy,
\be
        E=M-\frac 12\int_{-\infty}^{\infty}dv\rho_B(v),
\ee
which is suitable for the numerical computation of \(E\).

\(\L\) appearing in (\ref{eq:U}) is fixed in two steps. First, we demand that\footnote{\(\qs_{\es}\) is regular in the upper half-plane above \(\hat Z_1\). This is the region where  \(\qs_{\emptyset}\) on mirror and magic sheets are identified. In contrary, mirror \(\wT_{1,1}\) and magic \(\hat \wT_{1,1}\) are identified inside their analyticity strip \(\CA_1\). Hence \(\qs_{\emptyset}\) and \(\wT_{1,1}\) are equal in magic but not in mirror.} \(\qs_{\emptyset}=\hat\wT_{1,1}\) is a real function on the magic sheet which allows us to fix the phase of \(\Lambda\). Indeed, by studying the equality \(\qs_{\es}=q_{\es}f^{++}f^{--}U^+U^-\) at large \(u\) we see that  \(\qs_{\es}\) is a real function, at least at large \(u\), if \(\Lambda^2 e^{\Psi*\rho_c}\) is a real function at large \(u\).  Hence:

\begin{equation}\label{ImNorm}
        \Im\log\L^2=-\frac 14\int_{-\infty}^{\infty}\rho_c(v)dv\;.
\end{equation}
The absolute value of \(\L\) is fixed from
\begin{equation}
       \sqrt{\sT_{0,0}^+\sT_{0,0}^-}=
       {\nU\bar
         \nU}\sT_{0,1}=
       {\nU\bar \nU}\frac{\rrho_2}{1-Y_{1,1}Y_{2,2}},
\label{eq:Normalization}
\end{equation}
which should be valid for \(-2g<u<2g\). This relation is an upper band analog of (\ref{discontinuitiyh}) and it is derived similarly.

Note that (\ref{eq:Normalization})  fixes \(U\bar U\) on \(\hat Z_0\) essentially from the knowledge of \(\rho_2\). Hence the nontrivial information about \(U\) is contained in the complementary region: \((-\infty, -2g]\cup[2g,\infty)\).

\bibliography{bibliography}
\bibliographystyle{bibstyle2}

\end{document}